\documentclass[aps,prd,groupedaddress,preprintnumbers]{revtex4}
\usepackage{axodraw,color}
\usepackage{pstricks}
\usepackage{epsfig}

\newcommand{\vs}[1]{\rule[- #1 mm]{0mm}{#1 mm}}
\newcommand{\be}{\vs{2}\begin{equation}}
\newcommand{\eq}{\vs{2}\begin{equation}}
\newcommand{\en}{\\[2mm]\end{equation}}
\newcommand{\ee}{\\[2mm]\end{equation}}
\newcommand{\bea}{\begin{eqnarray}}
\newcommand{\ena}{\end{eqnarray}}
\newcommand{\eea}{\end{eqnarray}}

\newlength{\dhatheight}

\begin{document}

\renewcommand{\theequation}{\Roman{section}.\arabic{equation}}

\preprint{CPT-P001-2013}
\preprint{LPT-ORSAY/13-35}

\title{Isospin breaking in the phases of the $K_{e4}$ form factors}

\author{V. Bernard}
\email{bernard@ipno.in2p3.fr}
\affiliation{Groupe de Physique Th\'eorique, Institut de Physique Nucl\'eaire\\
B\^at. 100, CNRS/IN2P3/Univ. Paris-Sud 11 (UMR 8608), 91405 Orsay Cedex, France}

\author{S. Descotes-Genon}
\email{sebastien.descotes-genon@th.u-psud.fr}
\affiliation{Laboratoire de Physique Th\'eorique, CNRS/Univ. Paris-Sud 11 
(UMR 8627), 91405 Orsay Cedex, France}

\author{M. Knecht}
\email{knecht@cpt.univ-mrs.fr} 
\affiliation{Centre de Physique Th\'eorique,
 CNRS/Aix-Marseille Univ./Univ. du Sud Toulon-Var (UMR 7332)\\
CNRS-Luminy Case 907, 13288 Marseille Cedex 9, France}



\date{\today}

\begin{abstract}
Isospin breaking in the $K_{\ell 4}$  form factors induced by the difference between charged and
neutral pion masses is studied. Starting from suitably subtracted dispersion representations,
the form factors are constructed in an iterative way up to two loops in the low-energy expansion by
implementing analyticity, crossing, and unitarity due to two-meson intermediate states. Analytical
expressions for the phases of the two-loop form factors of the $K^\pm\to\pi^+\pi^- e^\pm \nu_e$ channel
are given, allowing one to connect the difference of form-factor phase shifts measured experimentally (out
of the isospin limit) and the difference of $S$- and $P$-wave $\pi\pi$ phase shifts studied theoretically
(in the isospin limit). The isospin-breaking correction consists of the sum of a universal part, involving
only $\pi\pi$ rescattering, and a process-dependent contribution, involving the form factors in the coupled
channels. The dependence on the two $S$-wave scattering lengths $a_0^0$ and $a_0^2$ in the isospin limit
is worked out in a general way, in contrast to previous analyses based on one-loop chiral perturbation theory.
The latter is used only to assess the subtraction constants involved in the dispersive approach.
The two-loop universal and process-dependent contributions are estimated and cancel partially
to yield an isospin-breaking correction close to the one-loop case. The recent results on the phases of
$K^\pm\to\pi^+\pi^- e^\pm \nu_e$ form factors obtained by the NA48/2 collaboration at the CERN SPS are
reanalysed including this isospin-breaking correction to extract values for the scattering lengths $a_0^0$
and $a_0^2$, as well as for low-energy constants and order parameters of two-flavour $\chi$PT. 
\end{abstract}

\maketitle

\section{INTRODUCTION} \label{intro}
\setcounter{equation}{0}

One of the best tests of our understanding of low-energy QCD comes from $\pi\pi$ scattering,
as it probes the spontaneous breaking of chiral symmetry, responsible for the existence of light pions as Goldstone Bosons. 
As such, it provides a very stringent test of $N_f=2$ Chiral Perturbation Theory ($\chi$PT), the effective theory
for low-energy pion dynamics built on the chiral limit $m_u=m_d=0$, of its structure and of its range of 
validity~\cite{Gasser:1983yg,Bijnens:1997vq,Bijnens:2006zp}. In addition, an accurate determination
of the pattern of chiral symmetry breaking in this limit is important, as it can be compared
with studies of low-energy processes involving $K$ and/or $\eta$ mesons. The latter provide information on 
the pattern of chiral symmetry breaking in the $N_f=3$ chiral limit ($m_u=m_d=m_s=0$)~\cite{Gasser:1984gg}. 
Several studies indicate possibly significant differences of patterns between these two 
limits~\cite{Moussallam:1999aq,Moussallam:2000zf,DescotesGenon:2003cg,DescotesGenon:2007ta,DescotesGenon:2000di,Bernard:2010ex,Bernard:2012fw,Bernard:2012ci}. 
Such differences can be interpreted as a paramagnetic suppression of chiral order parameters when
the number of massless flavours in the theory increases, in relation with the role of $s\bar{s}$ vacuum  pairs in chiral 
dynamics -- which can be important, as suggested by the structure of scalar resonances~\cite{DescotesGenon:1999uh}.

Several experimental processes can be exploited to extract information on $N_f=2$ chiral symmetry breaking, each time using
final-state interactions to probe $\pi\pi$ (re)scattering~\footnote{Another interesting source of information comes from 
numerical simulations of QCD on a lattice, which are now able to determine scattering lengths and phase shifts in channels 
where rescattering does not involve disconnected diagrams~\cite{Beane:2005rj,Beane:2006gj}.}. This is for instance the case for 
the cusp in $K\to\pi\pi\pi$ at low $\pi\pi$ invariant mass~\cite{Colangelo:2006va}, 
for the energy levels of pionic atoms \cite{Sazdjian:1999jf,Gasser:2007zt}, as well as for $K_{\ell 4}$ decays.
Indeed, an angular analysis of $K_{\ell 4}$ data provides information on the interference between the $S$ and $P$ waves, 
as a function of the energy of the hadronic $\pi\pi$ pair~\cite{Cabibbo:1965zz,Berends:1968zz}. Dispersive methods, i.e. Roy 
equations, can then be used to reconstruct the low-energy $\pi\pi$ amplitude 
using unitarity, analyticity, and data at higher energies, with two subtraction parameters chosen conveniently as the
scattering lengths $a_0^0$ and $a_0^2$~\cite{Roy:1971tc,Ananthanarayan:2000ht}. The reconstructed amplitude can be checked 
against the prediction from $N_f=2$ $\chi$PT.
The dispersive constraints set by the Roy equations to match higher-energy data on $\pi\pi$ phase shifts do 
constrain the values of $(a_0^0,a_0^2)$ into a large so-called Universal 
Band, out of which the domain favoured by $\chi$PT represents only a small region.

Until 2001, the only available data on $K_{\ell 4}$ decay into two charged pions came from the old Geneva-Saclay 
experiment~\cite{Rosselet:1976pu} and from the 
more recent BNL-E865 experiment~\cite{Pislak:2001bf}. A first analysis using the Roy equations together with a theoretical estimate 
of the scalar radius of the pion led to a determination of the scattering lengths
in close agreement with the predictions from two-loop $\chi$PT~\cite{Colangelo:2000jc}. 
Another analysis of the data available at that time (including $I=2$ low-energy phase shifts) favoured a slightly larger value for 
$a_0^2$, 1 $\sigma$ away from the two-loop $\chi$PT prediction~\cite{DescotesGenon:2001tn}.
Recently, the  NA48/2 collaboration has performed a remarkable work in collecting high-statistics $K^\pm_{e 4}$ data
at the CERN SPS~\cite{Batley:2007zz,Batley:2010zza}. After the announcement of the preliminary results of NA48/2~\cite{BrigitteKaon07}, it was pointed out 
that the high level of accuracy 
reached by the experiments in extracting the $\pi\pi$ phase shifts required taking into account
isospin-breaking effects~\cite{GasserKaon07}. These effects stem from different sources. First, the contributions from real and 
virtual photons can be removed,
estimating the Coulomb exchanges and incorporating radiative processes through a Monte-Carlo treatment~\cite{Barberio:1990ms,Barberio:1993qi,Xu:2010sw,Davidson:2010ew}. Second, 
the effect of the mass difference between charged and neutral
pions on the one hand, which is also dominantly of electromagnetic origin, and between 
$up$ and $down$ quarks on the other hand, must be determined from a theoretical analysis. In the following,
we will focus on these remaining corrections, which we will call "isospin-breaking" 
for simplicity, being understood that the other photon effects mentioned above have been taken 
care of beforehand by appropriate means, or can otherwise be considered to be negligible. For more details
on this issue, we refer the reader to the corresponding discussions in Refs.~\cite{GasserKaon07,Colangelo:2008sm,DescotesGenon:2012gv}.

A computation of these corrections was performed using next-to-leading-order $\chi$PT~\cite{Colangelo:2007df,Colangelo:2008sm}, leading to a significant 
energy-dependent correction (of up to more than 10 milliradians) in the phase shifts.
This correction to the data, together with their analysis through the Roy equations, 
restored the agreement between the NA48/2 results and two-loop $\chi$PT.
However, this correction was evaluated in the framework of $\chi$PT, with a given set of counterterms 
with values corresponding to a rather narrow range of scattering lengths $a_0^0$ and $a_0^2$. The underlying assumption is that the 
correction remains the same for all values of $(a_0^0,a_0^2)$, including values that are reasonable from the dispersive 
point of view, i.e. consistent with Roy equations and higher-energy data, but cannot be accommodated from the chiral point 
of view, because they differ too much from the current-algebra results \cite{Weinberg66} 
and thus would constitute a breakdown of the very notion of perturbative 
expansion in powers of quark masses. If the correction had a strong dependence on $a_0^0$ and $a_0^2$, 
the latter would not be exhibited by the one-loop computation performed in the framework of $\chi$PT, but it could 
affect the outcome of the analysis of the data provided by the NA48/2 experiment. 

It is therefore necessary to develop a computational framework of isospin-breaking corrections
in the phases of the form factors
where the values of the scattering lengths are not unnecessarily restricted from the outset. In order to illustrate
the issue, let us consider a simple example, leaving aside violations of isospin symmetry for the sake of demonstration.
In the isospin limit, the one-loop expressions of the $K_{\ell 4}$ form factors are well documented in 
the literature \cite{Bijnens90,Bijnens94,Bijnens:1994me}, and one finds for one of the form
factors involved in the decay channel of interest, $K^+ \to \pi^+\pi^-\ell^+\nu_\ell$,
\begin{equation}
F^{\mbox{\tiny{$+-$}}} (s,t,u) = \frac{M_K}{\sqrt{2} F_\pi} \left[
1 + \cdots + \frac{2 s - M_\pi}{2 F_\pi^2} J_{\pi\pi}^r (s) + \cdots
\right],
\label{example_1}
\end{equation} 
where the ellipses stand for additional contributions, that need not be further specified at this stage,
$s$ denotes the square of the invariant mass of the dipion system, $F_\pi$ is the pion decay constant,
and  $J_{\pi\pi}^r$ is the renormalized one-loop two-point function \cite{Chew60,Gasser:1983yg,Gasser:1984gg}
[a more detailed account of the notation will be given below]. 
In this final expression of the one-loop form factor, no dependence on the scattering lengths is visible, neither
in this term nor in the omitted ones. However, in the computation of the form factors, the actual expression
in terms of the low-energy constants of the $\chi$PT Lagrangian reads
\begin{equation}
F^{\mbox{\tiny{$+-$}}} (s,t,u) = \frac{M_K}{\sqrt{2} F_\pi} \left[
1 + \cdots + \frac{2 s - 2 {\widehat m} B_0}{2 F_0^2}  J_{\pi\pi}^r (s) + \cdots
\right], \qquad\qquad [{\widehat m} = (m_u + m_d)/2]
\label{example_2}
\end{equation} 
which agrees with the previous expression (\ref{example_1}) if the leading-order relations $F_\pi = F_0$ and 
$M_\pi^2 = 2 {\widehat m} B_0$ are used [this is the appropriate order to consider in this example], explaining 
why the expression  (\ref{example_1})  is usually quoted. 
However, it is not straightforward to reinterpret the expression (\ref{example_2}) in terms of
the $\pi\pi$ scattering lengths $a_0^0$ and $a_0^2$: they are both proportional
to $2 {\widehat m} B_0$ at lowest order \cite{Gasser83bis,Gasser:1983yg}, but there are infinitely many combinations of 
$M_\pi^2$, $a_0^0$, and $a_0^2$ that sum up to $2 {\widehat m} B_0$.
Even if a contribution from the $I=2$ channel is forbidden  by the 
$\Delta I = 1/2$ rule of the corresponding weak charged current, the question still remains how to determine the combination
that gives the correct dependence on $a_0^0$. Obviously, the information provided
by Eq.~(\ref{example_2}) alone does not allow for an unambiguous answer.
As can be guessed easily, the missing link has to be provided by unitarity. The function $J_{\pi\pi}^r$ encodes the discontinuity
of the form factor $F^{\mbox{\tiny{$+-$}}} (s,t,u)$ along the positive real $s$-axis, which involves the $I=0$ 
$\pi\pi$ partial wave in the channel with zero angular momentum as a final-state interaction effect~\cite{Bijnens94,Bijnens:1994me}. 
A careful analysis, which will be detailed in the present article, shows that, at one-loop order, 
Eqs.~(\ref{example_1}) and (\ref{example_2}) actually read
\begin{equation}
F^{\mbox{\tiny{$+-$}}} (s,t,u) = \frac{M_K}{\sqrt{2} F_\pi} \left[
1 + \cdots + \left( \frac{s - 4 M_\pi^2}{F_0^2} + 16 \pi a_0^0 \right) J_{\pi\pi}^r (s) + \cdots
\right].
\label{example_3}
\end{equation} 
Let us stress that, barring higher-order contributions presently not under discussion,
the three representations are strictly identical. However, if one considers the scattering lengths $a_0^0$ 
and $a_0^2$ as free variables that have to be adjusted 
from a fit to experimental data, only the third form is actually suitable. It is certainly conceivable
to use the existing one-loop expressions of $K^+_{e4}$ form factors, now including isospin-violating
effects \cite{cuplov04,Cuplov:2003bj,Nehme:2003bz}, and to repeat the above analysis for each separate 
contribution. But this would represent a rather cumbersome exercise. 
Instead, we will develop a more global approach, where the relevant unitarity properties are put forward explicitly from the start.

The purpose of this article is therefore to reconsider the extraction of
information on low-energy $\pi\pi$ scattering 
from $K_{e4}^\pm$ decays using a representation of the form factors based on dispersive
properties, in order to check the validity of the implicit assumption that isospin-breaking corrections
are not sensitive to the values of the scattering lengths. Indeed, in presence of isospin breaking, several $\pi\pi$ channels 
can rescatter into a given final state, contributing to the isospin-breaking effects of interest here in direct link 
with the structure of the $\pi\pi$ amplitude itself. As shown in Refs.~\cite{Stern:1993rg,Knecht:1995tr}, the use of 
analyticity, unitarity and crossing is sufficient to reconstruct the $\pi\pi$ amplitude up to two loops in terms 
of a limited number of subtraction constants (subthreshold or threshold parameters). Refs.~\cite{Stern:1993rg,Knecht:1995tr}
considered the simpler situation where isospin symmetry holds. This analysis has recently been extended to the isospin-breaking 
case for pion form factors and scattering amplitudes
\cite{DescotesGenon:2012gv}. Here, we will explain how to set up a similar construction for the $K_{e4}$ form factors,
and we will use it in order to extract a more general expression for the isospin-breaking correction
 in the phases of the two-loop form factors, where the values of 
$a_0^0$ and $a_0^2$ remain as free parameters and are not fixed from the outset. 
Working at two loops will also allow us to address the issue of the dependence of the phases on
the invariant mass of the dilepton system. The isospin-symmetric situation is
recovered as the limit in which the values of the neutral pion and kaon masses tend towards the charged ones,
$M_{\pi^0} \to M_{\pi^\pm}$, $M_{K^0} \to M_{K^\pm}$, while keeping the latter fixed. This agrees with
the convention that we will follow here: all quantities without superscript refer to the charged case 
(taken as the default case, i.e. $M_\pi=M_{\pi^\pm}$ and $M_K=M_{K^\pm}$), and quantities involving neutral pions and kaons 
will carry an explicit $0$ superscript, e.g. $M_{\pi^0}$, $M_{K^0}$.

We close this introductory part with an outline of the article.
In Sec. \ref{sec:definitions} we define the form factors relevant for $K_{\ell 4}$ decays,
we discuss their properties regarding partial-wave expansions, crossing properties,
chiral counting, analyticity and unitarity properties, and we present a first discussion
concerning their phases.
In Sec.~\ref{sec:Dispersion_reps}, we construct a dispersive representation of the form factors
based on the previous properties that provides a two-step iterative construction of the
form factors up to and including the two-loop order in the low-energy expansion.
The general expressions of the $K_{\ell 4}$ form factors at one loop based on this representation
are presented in Sec.~\ref{sec:1-loop_ff}. Some issues related to the second iteration
are briefly discussed.
In Sec.~\ref{sec:IB_in_phases}, we
discuss the general structure of isospin breaking in the phases of the two-loop form factors,
and compute the relevant partial-wave projections of the one-loop form factors.
Sec.~\ref{sec:numerics} is devoted to a numerical analysis of the dependence of the
isospin correction in the difference between $S$- and $P$-wave phase shifts on
the values of the $S$-wave scattering lengths $a_0^0$ and $a_0^2$. In Sec.~\ref{sec:fits}, we reanalyze the NA48/2
data for this phase difference keeping the scattering lengths as free parameters.
Finally, we summarize our results and present our conclusions in Sec~\ref{sec:conclusion}.
Several appendices are devoted, in successive order,
to the leading-order expression of the mesonic scattering amplitudes involved
in the discontinuities of the form factors (App.~\ref{app:loampl}), to the determination
of the subtraction polynomials occurring in the dispersive representations of the form factors (App.~\ref{app:poly}),
to certain integrals of the loop functions required to perform the partial-wave projections
of the one-loop form factors (App.~\ref{app:indefinite}),
 and finally to a numerical approximate expression for the isospin-breaking correction to the
difference of phase shifts between the $S$ and $P$ waves (App.~\ref{app:approximate}).

\section{Definition and general properties of the $K_{\ell 4}$ form factors}\label{sec:definitions}
\setcounter{equation}{0}

In the Standard Model, the amplitudes corresponding to $K_{\ell 4}$ decays are defined from the 
matrix elements of the type $\langle \pi^a(p_a) \pi^b(p_b) \vert i A_{\mu}^{4-i5}(0) \vert K(k) \rangle$
involving the $\Delta S = \Delta Q = +1$ axial current \footnote{In the present study, we will not 
consider the matrix element of the vector current,
related to the axial anomaly, and described by a single form factor $H^{ab}(s,t,u)$. } between a (charged or
neutral) kaon state and the corresponding two-pion state, specifically
$(K,a,b)\in\{(K^+,+,-),(K^+,0,0),(K^0,0,-)\}$.
For our purposes, we also need to consider the matrix elements related to
$\langle \pi^a(p_a) \pi^b(p_b) \vert i A_{\mu}^{4-i5}(0) \vert K(k) \rangle$ through crossing, namely
$\langle \pi^a(p_a) {\bar K}(k) \vert iA_{\mu}^{4-i5}(0) \vert {\bar \pi}^b(p_b)  \rangle $ and
$\langle {\bar K}(k) \pi^b(p_b) \vert iA_{\mu}^{4-i5}(0) \vert {\bar \pi}^a(p_a) \rangle $.
In order to be able to treat these matrix elements 
simultaneously and on a common footing, we consider general matrix elements of the type
\begin{equation}
{\cal A}^{ab}_\mu (p_a, p_b ; p_c) = \langle a(p_a) \, b(p_b) \vert iA_\mu (0) \vert {\bar c}(p_c) \rangle
.
\label{A_ab}
\end{equation}
Here $a$, $b$, and $c$ denote spin-0 mesons with momenta $p_a$, $p_b$, and $p_c$, 
masses $M_a$, $M_b$, and $M_c$, and $A_\mu(x)$ may stand for $A_{\mu}^{4-i5} (x)$. 
We need actually not specify the other quantum numbers of these states and of this current,
but we only assume that they are such that the matrix element is not trivial.
It is convenient to denote the particle in the initial state as an anti-particle
${\bar c}$, while the final state contains particles. This structure is then maintained
under the operation of crossing, which allows us to simplify the notation. Likewise,
we do not mention the particle $c$ in ${\cal A}^{ab}_\mu$ explicitly, since the context and 
the particles $a$ and $b$ will specify it without ambiguity.
In practice the sets of interest are
$\{ a,b,c \} = \{ \pi^+, \pi^-, K^- \}$, $\{ \pi^0, \pi^0, K^- \}$ or $\{ \pi^0, \pi^-, {\bar K}_0 \}$.

The matrix element Eq.~(\ref{A_ab}) possesses the general decomposition into invariant form factors
\begin{equation}
{\cal A}^{ab}_\mu (p_a, p_b ; p_c) =
(p_a + p_b)_\mu F^{ab}(s,t,u) +
(p_a - p_b)_\mu G^{ab}(s,t,u) +
(p_c - p_a - p_b)_\mu R^{ab}(s,t,u) .
\label{decomp_A_mu}
\end{equation}
They depend on the variables
$s = (p_a + p_b)^2 ,\ t = (p_c - p_a)^2 ,\ u = (p_c - p_b)^2$, 
obeying the ``mass-shell'' condition
$ s + t + u \,=\, M_a^2 + M_b^2 + M_c^2 + s_{\ell} \equiv \Sigma_\ell$, 
with $\ s_{\ell} \equiv (p_c - p_a - p_b)^2$
being the square of the dilepton invariant mass.
In the physical region of the $K_{\ell 4}$ decay,
$s_{\ell}$ is strictly positive, $s_\ell \ge m_{\ell}^2$, and in what follows we will always assume this to
be the case. 
For reasons of simplicity, we use here a normalisation of the form factors
that differs from the one commonly adopted. 
There is no difficulty  in introducing 
any appropriate normalisation afterwards, through a simple rescaling of the form factors.

In the remainder of this section we discuss the general properties of these matrix elements
from the point of view of their partial-wave expansions, of their crossing properties,
of their low-energy expansions, and we briefly review the analyticity properties needed
in the following. We close this section with a general discussion
of the phases of the form factors at two loops in the low-energy expansion.

\subsection{Partial-wave expansion}\label{subsec:PW_exp}

The decomposition (\ref{decomp_A_mu}) leads to form factors which are free from kinematical singularities,
but which do not have simple decompositions into partial waves. For the latter, it is more
convenient to introduce another set of form factors. To this effect, adapting the method
of Ref.~\cite{Berends:1968zz} to the more general situation at hand, we define
\begin{eqnarray}
\Sigma^{ab}_\mu &=& (p_a + p_b)_\mu - \,\frac{M_c^2-s-s_\ell}{2 s_\ell}\, (p_c - p_a - p_b)_\mu
,
\nonumber\\
\Delta^{ab}_\mu &=& (p_a - p_b)_\mu + 2\,\frac{\lambda^{1\over 2}_{ab}(s)}{\lambda^{1\over 2}_{\ell c} (s)}\, 
\cos\theta_{ab}(p_c - p_a - p_b)_\mu
- \left[\frac{M_a^2 - M_b^2}{s} \,+\, \frac{M_c^2 - s - s_\ell}{s}\,\frac{\lambda^{1\over 2}_{ab}(s)}{\lambda^{1\over 2}_{\ell c} (s)}\, 
\cos\theta_{ab} \right] \! (p_a + p_b)_\mu
,
\nonumber\\
\Lambda^{ab}_\mu &=& (p_c - p_a - p_b)_\mu
.
\end{eqnarray}
These four-vectors are mutually orthogonal,
\begin{equation}
\Sigma^{ab} \cdot \Delta^{ab} = 0,\ \Sigma^{ab} \cdot \Lambda^{ab} = 0,\ \Delta^{ab} \cdot \Lambda^{ab} = 0,
\end{equation}
and $\Lambda^{ab} \cdot \Lambda^{ab} = s_\ell$. In these expressions, the functions $\lambda_{ab}(s)$
and $\lambda_{\ell c} (s)$ are defined in terms of K\"allen's function $\lambda(x,y,z)=x^2+y^2+z^2-2xy-2xz-2yz$
by $\lambda_{ab}(s)=\lambda(s,M_a^2,M_b^2)$ and $\lambda_{\ell c} (s) = \lambda(s,s_\ell,M_c^2)$,
respectively. Furthermore,
$\theta_{ab}$ denotes the angle made by the line of flight of particle $a$ in the $(a,b)$ rest frame with the
direction of ${\vec p}_a + {\vec p}_b$ in the rest frame of particle ${\bar c}$,
\begin{equation}
\cos\theta_{ab}\,=\, \frac{(M_a^2 - M_b^2)(s_\ell -M_c^2) - s(t-u)}{\lambda^{1\over 2}_{ab}(s) \lambda^{1\over 2}_{\ell c} (s)}
\,=\, \frac{(M_a^2 - M_b^2)(s_\ell -M_c^2) + s(\Sigma_\ell - s - 2t)}{\lambda^{1\over 2}_{ab}(s) \lambda^{1\over 2}_{\ell c} (s)}
.
\label{cos_theta}
\end{equation}
We can then write down another decomposition of the matrix element, in terms of transverse and longitudinal components,
\begin{equation}
{\cal A}^{ab}_\mu (p_a, p_b ; p_c) =
\Sigma^{ab}_\mu {\cal F}^{ab}(s,t,u) +
\Delta^{ab}_\mu {\cal G}^{ab}(s,t,u) +
\Lambda^{ab}_\mu {\cal R}^{ab}(s,t,u) .
\label{decomp_A_mu_cal}
\end{equation}
There exists a one-to-one correspondence between the two sets of form factors,
\begin{eqnarray}
{\cal F}^{ab}(s,t,u) &=& F^{ab}(s,t,u) + \left[\frac{M_a^2 - M_b^2}{s} \,+\, \frac{M_c^2 - s - s_\ell}{s}\,
\frac{\lambda^{1\over 2}_{ab}(s)}{\lambda^{1\over 2}_{\ell c} (s)}\, \cos\theta_{ab} \right] G^{ab}(s,t,u)
,
\nonumber\\
{\cal G}^{ab}(s,t,u) &=& G^{ab}(s,t,u)
,
\nonumber\\
{\cal R}^{ab}(s,t,u) &=& R^{ab}(s,t,u) +\,\frac{M_c^2-s-s_\ell}{2 s_\ell}\,F^{ab}(s,t,u)
\nonumber\\
&&
 +\, \frac{1}{2 s s_\ell} \left[
(M_a^2 - M_b^2)(M_c^2 - s - s_\ell) + \lambda^{1\over 2}_{ab}(s) \lambda^{1\over 2}_{\ell c} (s) \cos\theta_{ab}
\right] G^{ab}(s,t,u).
\label{F_G_R_cal}
\end{eqnarray}
Notice that the form factor ${\cal R}^{ab}(s,t,u)$ describes the matrix element of the
divergence of the current $A^\mu (x)$,
\begin{equation}
\langle a(p_a) \, b(p_b) \vert \partial^\mu A_\mu (0) \vert {\bar c}(p_c) \rangle \,=\, -s_\ell {\cal R}^{ab}(s,t,u)
\label{calR:div_A}
.
\end{equation}
In the center-of-mass frame of the $(a,b)$ pair of particles, one obtains [the metric we are using has
signature $(+,-,-,-)$]
\begin{equation}
{\cal A}^{ab}_\mu (p_a, p_b ; p_c) =
\left(
\begin{tabular}{c}
$ {\displaystyle{\frac{1}{2 \sqrt{s}}}} \left[ 
-\, {\displaystyle{\frac{\lambda_{\ell c}(s)}{s_\ell}}}\,{\cal F}^{ab}(s,t,u) + (M_c^2 - s - s_\ell) {\cal R}^{ab}(s,t,u)
\right] $
\\
$ - {\displaystyle{\frac{\lambda^{1\over 2}_{ab}(s)}{\sqrt{s}}}}  \,{\cal G}^{ab}(s,t,u) \,\sin\theta_{ab} $
\\
$ 0 $
\\
$ -\, {\displaystyle{\frac{\lambda^{1\over 2}_{\ell c} (s)}{2 \sqrt{s}}}}
\left[
{\displaystyle{\frac{M_c^2 - s - s_\ell}{2 s_\ell}}}\,{\cal F}^{ab}(s,t,u) - {\cal R}^{ab}(s,t,u)
\right] $
\end{tabular}
\right)
.
\end{equation}
This entails the partial-wave decompositions \cite{Berends:1968zz}
\begin{eqnarray} 
{\cal F}^{ab}(s,t,u) &=& \sum_{l \ge 0} f_{l}^{ab}(s,s_{\ell}) P_{l}(\cos\theta_{ab}) 
,
\nonumber\\
{\cal G}^{ab}(s,t,u) &=& \sum_{l \ge 1} g_{l}^{ab}(s,s_{\ell}) P_{l}^{\prime}(\cos\theta_{ab})
,
\nonumber\\
{\cal R}^{ab}(s,t,u) &=& \sum_{l \ge 0} r_{l}^{ab}(s,s_{\ell}) P_{l}(\cos\theta_{ab})
.
\label{FandG_PW_decomp}
\end{eqnarray}
The partial waves are obtained upon projection
of the form factors,
\bea
f_{l}^{ab}(s,s_{\ell}) &=& \frac{2 l +1}{2}\,\int_{-1}^{+1}d(\cos\theta_{ab}){\cal F}^{ab}(s,t,u)
P_{l}(\cos\theta_{ab})
,
\nonumber\\
g_{l}^{ab}(s,s_{\ell}) &=& \frac{2 l +1}{2}\,\frac{1}{l(l+1)}
\int_{-1}^{+1}d(\cos\theta_{ab}) {\cal G}^{ab}(s,t,u) \sin\theta_{ab} 
P_{l}^1(\cos\theta_\pi)
,
\nonumber\\
r_{l}^{ab}(s,s_{\ell}) &=& \frac{2 l +1}{2}\,\int_{-1}^{+1}d(\cos\theta_{ab}){\cal R}^{ab}(s,t,u)
P_{l}(\cos\theta_{ab})
,
\label{formfactorpartialwave}
\eea
where we have used the definition
$P_{l}^1(\cos\theta)=\sin\theta P_{l}^{\prime}(\cos\theta)$
and the orthogonality properties
\bea
\int_{-1}^{+1}d(\cos\theta) P_{l}(\cos\theta)
P_{l^\prime}(\cos\theta) &=& \frac{2}{2 l + 1}\,\delta_{l l^\prime}
,
\nonumber\\
\int_{-1}^{+1}d(\cos\theta) P_{l}^1(\cos\theta)
P_{l^\prime}^1(\cos\theta) &=& \frac{2 l(l + 1)}{2 l + 1}\,\delta_{l l^\prime}
.
\eea 
Since $\{F ; G ; R\}^{ab} (s , t , u) = \{F ; - G ; R\}^{ba} (s , u , t)$ and $\cos\theta_{ab} = - \cos\theta_{ba}$,
one has the symmetry properties
\begin{equation}
f_{l}^{ba} (s,s_{\ell}) = (-1)^l f_{l}^{ab} (s,s_{\ell})\,, 
\ g_{l}^{ba} (s,s_{\ell}) = (-1)^{l} g_{l}^{ab} (s,s_{\ell}) \,, 
\ r_{l}^{ba} (s,s_{\ell}) = (-1)^l r_{l}^{ab} (s,s_{\ell}) 
.
\label{sym_PV}
\end{equation}
The set of form factors $(F,G,R)$
is free of kinematical singularities, but does not exhibit a simple expansion in partial waves,
while the opposite holds for the other set $({\cal F},{\cal G},{\cal R})$, which admits simple partial-wave decompositions, 
but is  plagued with kinematical singularities. According to which aspect one wishes to 
emphasise, one set is more adapted than the other, which explains why we sometimes need to
switch back and forth between these two sets in the following.

\subsection{Crossing properties}\label{subsec:crossing}

The crossing properties are expressed through the relations
\bea
{\cal A}_{\mu}^{ac}(p_a,p_c;p_b) \,=\, \lambda_b \lambda_c {\cal A}_{\mu}^{ab}(p_a,-p_b;-p_c)
\, ,\quad
{\cal A}_{\mu}^{cb}(p_c,p_b;p_a) \,=\, \lambda_a \lambda_c {\cal A}_{\mu}^{ab}(-p_a,p_b;-p_c)
,
\eea
where the matrix elements on the right-hand sides are related through analytic 
continuations
to the original matrix element ${\cal A}_{\mu}^{ab}(p_a,p_b;p_c)$,
assuming that the usual analyticity properties hold.
The coefficients $\lambda_{a,b,c}$ are crossing phases, which 
are chosen such as to reduce to the Condon-Shortley phase convention in the isospin limit,
\be
\lambda_{K^\pm}\,=\,\lambda_{\pi^\pm} \,=\, -1,
\ \lambda_{\pi^0} \,=\, \lambda_{K^0} \,=\, \lambda_{{\bar K}^0} \,=\, +1 .
\ee
At the level of the form factors themselves, these crossing relations become
\be
{\bf A}^{ac}(s,t,u)\,=\,\lambda_b \lambda_c\, {\cal C}_{st} {\bf A}^{ab}(t,s,u)\,,
\ {\bf A}^{cb}(s,t,u)\,=\,\lambda_a \lambda_c\, {\cal C}_{us} {\bf A}^{ab}(u,t,s)\,,
\ {\bf A}^{ba}(s,t,u)\,=\,{\cal C}_{tu} {\bf A}^{ab}(s,u,t)
,
\\
\label{crossing}
\ee
with
\bea
{\bf A}^{X}(s,t,u) \,=\, \left(
\begin{array}{c}
F^{X}(s,t,u) \\
G^{X}(s,t,u) \\
R^{X}(s,t,u)
\end{array}
\right)
,
\label{A^X_three}
\eea
where $X$ stands for any one of the couples of indices $ab$ (and, in the present case, also $ba$),
$ac$, or $cb$,
and
\bea
{\cal C}_{st} \,=\,  \left(
\begin{array}{ccc}
-\frac{1}{2} & +\frac{3}{2} & 0 \\
+\frac{1}{2} & +\frac{1}{2} & 0 \\
-1           &    +1        & +1\\
\end{array}
\right)
\,,\ {\cal C}_{us} \,=\,  \left(
\begin{array}{ccc}
-\frac{1}{2} & -\frac{3}{2} & 0 \\
-\frac{1}{2} & +\frac{1}{2} & 0 \\
-1           &    -1        & +1\\
\end{array}
\right)
\,,\ {\cal C}_{tu} \,=\, \left(
\begin{array}{ccc}
+1 & 0 & 0 \\
0 & -1 & 0 \\
0           &    0        & +1\\
\end{array}
\right)
\,.
\eea
Each of these crossing matrices squares to the identity matrix. In addition, they satisfy  the relations
\be
{\cal C}_{st}{\cal C}_{us} \,=\, {\cal C}_{us}{\cal C}_{tu},
\ {\cal C}_{us}{\cal C}_{st} \,=\, {\cal C}_{st}{\cal C}_{tu},
\ {\cal C}_{st}{\cal C}_{tu} \,=\, {\cal C}_{tu}{\cal C}_{us}
.
\ee
It is useful to notice that under crossing the form factors 
$F^X$ and $G^X$ transform into form factors $F^Y$ and $G^Y$, without
mixing with the form factors $R^Y$. In the following, we will omit the form factors  $R^X$ from the discussion most of the time, writing
\bea
{\bf A}^{X}(s,t,u) \,=\, \left(
\begin{array}{c}
F^{X}(s,t,u) \\
G^{X}(s,t,u) 
\end{array}
\right)\,,
\label{A^X_two}
\eea
instead of Eq.~(\ref{A^X_three}). When it is the case, it is understood that the crossing matrices 
are reduced to their upper-left $2\times 2$ blocks.
All the previous relations between these matrices remain unaffected by this truncation.
As can be seen from the crossing properties of the form factors $F$, $G$, and $R$,
the type-${\cal F}$ and ${\cal G}$ form factors transform among themselves under crossing.
On the other hand, and in contrast with the form factor $R$,
the type-${\cal R}$ form factors transform into themselves, without mixing with ${\cal F}$ and ${\cal G}$,
\bea
{\cal R}^{ac}(s,t,u) \,=\, \lambda_b\lambda_c {\cal R}^{ab}(t,s,u)\,,
\ {\cal R}^{cb}(s,t,u) \,=\, \lambda_a\lambda_c {\cal R}^{ab}(u,t,s)
.
\eea
This result follows from the  relationship between the ${\cal R}$ form factors and
the matrix elements of the divergence of the current $A_{\mu}(x)$,
as shown in Eq.~(\ref{calR:div_A}), so that they cannot mix under crossing with the other form factors,
which correspond to the transverse components of the same current.

\subsection{Chiral counting}\label{subsec:chi_count}

The next ingredient is provided by the low-energy behaviour of the partial waves~\cite{Colangelo:1994qy},
based on the chiral counting $M_P \sim {\cal O}(E)$, $s,t,u,s_\ell \sim {\cal O}(E^2)$, where $M_P$ stands for
the mass of any of the light pseudoscalar states. This
singles out the $S$ and $P$ waves as dominant at low energies, and makes them the central subject of study for $K_{\ell 4}$ decays.
Note that we treat here  $s_\ell$ on the same footing as one of the masses squared, which is
compatible with its allowed range inside the $K_{\ell 4}$ phase space.
We emphasise that this treatment is mandatory for the chiral counting of the partial waves to be 
correct.
At this stage, we should recall that $K_{\ell 4}$ form factors are traditionally normalized with 
$1/M_K$ factored out on the right-hand side of Eq.~(\ref{decomp_A_mu}), which
makes the form factors artificially proportional to $M_K$. Here we deal with
form factors normalized as in (\ref{decomp_A_mu}), which are of order ${\cal O}(E^0)$ at tree level.
Additional normalisation factors are not to be taken into account in the discussion of the chiral behaviour.
With this proviso, we have the following chiral counting of the partial waves \cite{Colangelo:1994qy}:
\bea
&{\mbox{Re}}f_0^{ab}(s, s_\ell),\ {\mbox{Re}}f_1^{ab}(s,s_\ell),\ {\mbox{Re}\,}g_1^{ab}(s,s_\ell) \sim {\cal O}(E^0)),\quad
&{\mbox{Im}}f_0^{ab}(s,s_\ell),\ {\mbox{Im}}f_1^{ab}(s,s_\ell),\ {\mbox{Im}\,}g_1^{ab}(s,s_\ell)  \sim {\cal O}(E^2)
\nonumber
\\
&\!\!\!\!\!\!\!\!\!\!\!\!
{\mbox{Re}}f_l^{ab}(s,s_\ell),\ {\mbox{Re}}\,g_l^{ab}(s,s_\ell) \sim {\cal O}(E^2) ,\ l\ge 2 ,\quad
&{\mbox{Im}}f_l^{ab}(s,s_\ell),\ {\mbox{Im}}\,g_l^{ab}(s,s_\ell)\sim {\cal O}(E^6) ,\ l\ge 2
.
\eea

In terms of the form factors $F^{ab}(s,t,u)$ and $G^{ab}(s,t,u)$, and thanks to Eqs.~(\ref{F_G_R_cal}) 
and (\ref{formfactorpartialwave}), the chiral counting of the partial
waves translates into the decompositions
\bea
F^{ab}(s,t,u) &=& F_S^{ab}(s,s_\ell)\,+\,F_P^{ab} (s,s_\ell) \cos\theta_{ab} \,+\,F^{ab}_>(s,\cos\theta_{ab},s_\ell)
,
\nonumber
\\
G^{ab}(s,t,u) &=& G_P^{ab}(s,s_\ell)  \,+\,G^{ab}_>(s,\cos\theta_{ab},s_\ell)
.
\label{decomp_F_G}
\eea
The contributions of the partial waves with $\ell \ge 2$ are collected in $F^{ab}_>(s,\cos\theta_{ab},s_\ell)$ and in 
$G^{ab}_>(s,\cos\theta_{ab},s_\ell)$, with the counting
${\mbox{Re}} F^{ab}_>(s,\cos\theta_{ab},s_\ell)$, ${\mbox{Re}} G^{ab}_>(s,\cos\theta_{ab},s_\ell) \sim {\cal O}(E^2)$ and
${\mbox{Im}} F^{ab}_>(s,\cos\theta_{ab},s_\ell)$, ${\mbox{Im}} G^{ab}_>(s,\cos\theta_{ab},s_\ell)$ $\sim {\cal O}(E^6)$, 
while the contributions from $S$ and $P$ waves are collected in 
\bea
F_S^{ab}(s,s_\ell) &=& f_0^{ab}(s,s_\ell) \,-\,\frac{M_a^2 - M_b^2}{s}\,g_1^{ab}(s,s_\ell)
,
\nonumber\\
F_P^{ab}(s,s_\ell) &=& f_1^{ab}(s,s_\ell) \,-\, \frac{M_c^2 - s - s_{\ell}}{s}\,
\frac{\lambda^{\frac{1}{2}}_{ab}(s)}{\lambda^{\frac{1}{2}}_{\ell c}(s)}\,g_1^{ab}(s,s_\ell)\,,
\nonumber
\\
G_P^{ab}(s,s_\ell) &=& g_1^{ab}(s,s_\ell)
.
\label{def_F_S_and_F_P}
\eea
These equations provide the bridge at the level of the lowest partial waves between the two representations 
of the matrix elements, in terms of the form factors $\cal F$ and $\cal G$, or $F$ and $G$.
Similar relations hold between the form factors and partial-wave projections in the $ac$ and $cb$
channels, and are obtained from the above upon performing cyclic permutations of the labels $a,b,c$,
while replacing the variables $s$ and $\cos \theta_{ab}$ by their appropriate counterparts.

\subsection{Analyticity and unitarity properties}\label{subsec:Ana_Uni}

We now assume that the form factors $F^{ab}(s,t,u)$ and $G^{ab}(s,t,u)$ have the usual analyticity properties
with respect to the variable $s$, for fixed values of $t$ and of $u$ (and of $s_\ell \geq 0$), with a cut
on the positive $s$-axis, whose discontinuity is fixed by unitarity, and a cut on the negative
$s$-axis generated by unitarity in the crossed channel. The form factors
are regular and real in the interval between $s=0$ and the positive value of $s$ corresponding
to the lowest-lying intermediate state. This singularity structure 
is transmitted to the partial waves (along, possibly, with other singularities produced by the
projection procedure itself \cite{PWanalyticity1,PWanalyticity2,PWanalyticity3,PWanalyticity4}). 
In the following, we will only need to know the discontinuities
of the lowest partial waves along the positive $s$-axis at low energies, which are fixed by unitarity. 
Up to and including two loops in the chiral
counting discussed in the previous subsection, these discontinuities (with respect to $s$ at a fixed $s_\ell$)
originate from mesonic two-particle intermediate states, and are therefore given by 
\bea 
{\mbox{Im}}\,f^{ab}_l (s,s_\ell) 
&=&
\sum_{\{ a^\prime , b^\prime \}} \frac{1}{{\cal S}_{a^\prime b^\prime }}\,\frac{\lambda^{\frac{1}{2}}_{a^\prime b^\prime }(s)}{s}\,
{\mbox{Re}}\,\left\{t_l^{a^\prime b^\prime ; a b}(s) 
\left[ f^{a^\prime b^\prime }_l (s,s_\ell )\right]^{\star} 
\right\} \theta (s - s_{a^\prime b^\prime })
 + {\cal O}(E^8)
 ,
\nonumber\\
{\mbox{Im}}\,g^{ab}_l (s,s_\ell) 
&=&
\sum_{\{ a^\prime , b^\prime \}} \frac{1}{{\cal S}_{a^\prime b^\prime }}\,\frac{\lambda^{\frac{1}{2}}_{a^\prime b^\prime }(s)}{s}\,
\frac{\lambda^{\frac{1}{2}}_{a^\prime b^\prime }(s)}{\lambda^{\frac{1}{2}}_{ab}(s)}\,
{\mbox{Re}}\,\left\{t_l^{a^\prime b^\prime ; a b}(s) 
\left[ g^{a^\prime b^\prime }_l (s,s_\ell)\right]^{\star}
\right\} \theta(s - s_{a^\prime b^\prime })
 + {\cal O}(E^8)
,
\label{Imf_Img}
\eea
where $l=0,1$, and $t_l^{a^\prime b^\prime ; a b}(s)$ denotes the $l$-th partial wave of the
$a^\prime b^\prime \to a b$ scattering amplitude. $s_{a^\prime b^\prime }$ stands for 
the lowest invariant mass squared of the corresponding intermediate state, 
$s_{a^\prime b^\prime } = (M_{a^\prime} + M_{b^\prime})^2$ in terms of the masses
$M_{a^\prime}, M_{b^\prime}$ of the particles in the intermediate state. The symmetry factor reads
${\cal S}_{a^\prime b^\prime } = 1$ in all cases of interest, except for 
$\{ a^\prime , b^\prime \}= \{\pi^0 , \pi^0 \}$ or $\{\eta , \eta\}$, 
where ${\cal S}_{a^\prime b^\prime } = 2$. 
Notice that we do not need to include the factor ${\cal S}_{a^\prime b^\prime } $ in the unitarity sum concerning
partial waves with odd values of $l$, since the amplitudes with two identical particles
either in the initial or final state will only produce partial waves $t_l^{a^\prime b^\prime ; a b}(s)$
with even values of $l$.

The partial waves $t_l^{a^\prime b^\prime ; a b} (s)$ of the mesonic scattering amplitudes
$A^{a^\prime b^\prime ; a b} (s,{\hat t})$, ${\hat t} = (p_a - p_{a^\prime})^2$, are defined as usual,
\be
A^{a^\prime b^\prime ; a b} (s,{\hat t}) \,=\, 16 \pi \sum_l (2l+1) t_l^{a^\prime b^\prime ; a b}(s) P_l(\cos{\hat \theta})
,
\label{PW_decomp}   
\ee
where ${\hat\theta}$ is the corresponding scattering angle in the center-of-mass frame of the reaction ${a^\prime b^\prime \to a b}$,
and $s + {\hat t} + {\hat u} = M_a^2 + M_b^2 + M_{a^\prime}^2 + M_{b^\prime}^2$, ${\hat u} = (p_a - p_{b^\prime})^2$.
As far as their chiral counting is concerned, the partial-wave projections of these scattering amplitudes behave as~\cite{Stern:1993rg} 
\bea
&{\mbox{Re}}\,t^{a^\prime b^\prime ; a b}_{l}(s)  \sim {\cal O}(E^2),\ l = 0,1 , \quad
&{\mbox{Re}}\,t^{a^\prime b^\prime ; a b}_{l}(s)  \sim {\cal O}(E^4),\ l \ge 2 ,\nonumber
\\
&{\mbox{Im}}\,t^{a^\prime b^\prime ; a b}_{l}(s)  \sim {\cal O}(E^4),\ l = 0,1 , \quad
&{\mbox{Im}}\,t^{a^\prime b^\prime ; a b}_{l}(s)  \sim {\cal O}(E^8),\ l  \ge 2 . 
\label{PW_chiral_count}
\eea  

\subsection{Phases of the form factors}\label{subsec:phases}

As discussed in the introduction, we are eventually interested in the phases of 
the $F_S$, $F_P$, and $G_P$ components of the $F$ and $G$ form factors 
corresponding to the decay channel 
$K^+ \to \pi^+ \pi^- \ell^+ \nu_\ell$.
More precisely, we will consider the differences
of these phases that are observable in the interferences occurring in the differential decay distribution~\footnote{In this paper, we consider only CP-even ``strong'' phases, and discard any CP-odd ``weak'' phases.}.
The channel $K^\pm \to \pi^+ \pi^- \ell^\pm \nu_\ell$ is the only one where both $F$ and $G$ occur already at lowest order (in the $K_{\ell 4}$ decay mode of the neutral kaon, a tiny form factor $F$ is
 induced already at lowest order in the presence of isospin breaking only).
In order to simplify the notation in this subsection, we have
suppressed the $+-$ superscript whenever no confusion can arise.
The generic low-energy structure of the form factors can be written as in Eq.~(\ref{decomp_F_G}),
\begin{eqnarray}
F(s,t,u) &=& {\widehat F}_S(s,s_\ell) e^{i\delta_S(s, s_\ell)} + {\widehat F}_P(s,s_\ell) e^{i\delta_P(s,s_\ell)} \cos\theta
+ {\mbox{Re}} F_>(s,\cos\theta ,s_\ell) + {\cal O}(E^6)
,
\nonumber\\
G(s,t,u) &=& {\widehat G}_P(s,s_\ell) e^{i\delta_P(s,s_\ell)}
+ {\mbox{Re}} G_>(s,\cos\theta ,s_\ell) + {\cal O}(E^6)
,
\label{F_S_hat_and_G_P_hat}
\end{eqnarray} 
where we have introduced the real functions ${\widehat F}_S(s,s_\ell)$ ($\equiv {\widehat f}_0(s,s_\ell)$ for $M_a = M_b$), 
${\widehat F}_P(s,s_\ell)$, and ${\widehat G}_P(s,s_\ell)$ ($\equiv {\widehat g}_1(s,s_\ell))$, 
which correspond to the quantities appearing in Eq.~(\ref{decomp_F_G}), but with
their phases removed, ${\widehat F}_S(s,s_\ell) =  e^{-i\delta_S(s,s_\ell)} F_S(s+i0,s_\ell)$, etc.
Notice that we have assumed these phases to depend on $s_\ell$, and that 
we have assigned the same phase to $F_P(s,s_\ell)$ and $G_P(s,s_\ell)$. We will comment on this feature below.

From the point of view of 
the chiral counting discussed in Sec.~\ref{subsec:chi_count}, these quantities behave as follows:
\begin{eqnarray}
{\widehat F}_S(s,s_\ell),\, {\widehat G}_P(s,s_\ell)
& \sim & {\cal O}(E^0) + {\cal O}(E^2) + {\cal O}(E^4) + \cdots
,
\nonumber\\
{\widehat F}_P(s,s_\ell)
& \sim & {\cal O}(E^2) + {\cal O}(E^4) + \cdots
,
\nonumber\\
{\mbox{Re}} F_>(s,\cos\theta ,s_\ell), \, {\mbox{Re}} G_>(s,\cos\theta ,s_\ell) & \sim & {\cal O}(E^2) + \cdots
,
\end{eqnarray}
where we have shown the orders relevant up to two loops, the higher orders (not discussed here)
being denoted by the ellipses. Notice that the behaviour of ${\widehat F}_P(s,s_\ell)$, which
receives its first contribution at next-to-leading order only, is different from 
the cases of ${\widehat F}_S(s,s_\ell)$ and of ${\widehat G}_P(s,s_\ell)$. The form factors $F(s,t,u)$ and $G(s,t,u)$
being both constant at lowest order, a dependence with respect to $\cos \theta$ cannot appear at this order.
This observation can also be made directly from the definition of ${F}_P(s,s_\ell)$ in Eq.~(\ref{def_F_S_and_F_P}):
since $F(s,t,u)$ is constant at lowest order, the contribution to $f_1(s,s_\ell)$ comes entirely from the last
term in the expression for ${\cal F}(s,t,u)$ in Eq.~(\ref{F_G_R_cal}), which exactly cancels the
contribution proportional to $g_1(s,s_\ell)$ in (\ref{def_F_S_and_F_P}). 

In order to obtain the expression of the phases order by order, we now consider the chiral expansion of the 
real parts of the lowest partial-wave projections of the meson-meson scattering amplitudes,
\begin{equation}
{\mbox{Re}}\,t_l^{a^\prime b^\prime ; {\mbox{\tiny{$+-$}}}} (s) = \varphi_l^{a^\prime b^\prime ; {\mbox{\tiny{$+-$}}}} (s) + 
\psi_l^{a^\prime b^\prime ; {\mbox{\tiny{$+-$}}}} (s) + {\cal O}(E^6) ,
\end{equation}
for $l=0,1$, and with $\varphi_{0,1}^{a^\prime b^\prime ; {\mbox{\tiny{$+-$}}}} (s) \sim {\cal O}(E^2)$ and
$\psi_{0,1}^{a^\prime b^\prime ; {\mbox{\tiny{$+-$}}}} (s) \sim {\cal O}(E^4)$. Writing a similar expansion for the
form factors themselves, e.g.
\begin{equation}
{\mbox{Re}}\, F_{S}(s,s_\ell) = F_{S[0]} + F_{S[2]}(s,s_\ell) + {\cal O}(E^4) ,
\qquad {\mbox{Re}}\, G_{P}(s,s_\ell) = G_{P[0]} + G_{P[2]}(s,s_\ell) + {\cal O}(E^4)
,
\end{equation}
and using the unitarity condition Eq.~(\ref{Imf_Img}) for the imaginary parts, we obtain the expressions
\begin{equation}
\delta_S(s,s_\ell) = 
\sum_{\{ a^\prime , b^\prime \}} \frac{1}{{\cal S}_{a^\prime b^\prime}}\,\frac{\lambda^{\frac{1}{2}}_{a^\prime b^\prime}(s)}{s}\,
\left[
\varphi_0^{a^\prime b^\prime ; {\mbox{\tiny{$+-$}}}} (s) \,
\frac{F_{S[0]}^{a^\prime b^\prime} +  F_{S[2]}^{a^\prime b^\prime} (s,s_\ell)}{F_{S[0]} + F_{S[2]} (s,s_\ell)}
\,
+
\,
\psi_0^{a^\prime b^\prime ; {\mbox{\tiny{$+-$}}}}(s) \,
\frac{F_{S[0]}^{a^\prime b^\prime}}{F_{S[0]}}
\right]
\theta (s - s_{a^\prime b^\prime} ) 
+
{\cal O}(E^6)
,
\label{delta_S}
\end{equation}
and
\begin{equation}
\delta_P(s,s_\ell) = 
\sum_{\{ a^\prime , b^\prime \}} 
\frac{\lambda^{\frac{1}{2}}_{a^\prime b^\prime}(s)}{s}\,
\frac{\lambda^{\frac{1}{2}}_{a^\prime b^\prime}(s)}{\lambda^{\frac{1}{2}}_{ab}(s)}
\left[
\varphi_1^{a^\prime b^\prime ; {\mbox{\tiny{$+-$}}}} (s) \,
\frac{G_{P[0]}^{a^\prime b^\prime} +  G_{P[2]}^{a^\prime b^\prime} (s,s_\ell)}{G_{P[0]} + G_{P[2]} (s,s_\ell)}
\,
+
\,
\psi_1^{a^\prime b^\prime ; {\mbox{\tiny{$+-$}}}}(s) \,
\frac{G_{P[0]}^{a^\prime b^\prime}}{G_{P[0]}}
\right]
\theta (s - s_{a^\prime b^\prime} ) 
+
{\cal O}(E^6)
.
\label{delta_P}
\end{equation}
We see that the phases $\delta_S(s,s_\ell)$ and $\delta_P(s,s_\ell)$ depend on $s_\ell$
through the order ${\cal O}(E^2)$ corrections to the form factors, as soon as a second
intermediate state $a^\prime b^\prime \neq +-$ is involved.
In the case of the $P$-wave phase shift,
there can be no contribution from states with two identical particles due to Bose symmetry,
explaining the absence of 
the factor $1/{{\cal S}_{a^\prime b^\prime}}$ in $\delta_P(s,s_\ell)$.
Hence, for $\delta_P$ in the specific case $ab = +-$ and for $s\le M_K^2$, the sum boils down to the single $\pi^+ \pi^-$ 
intermediate state, the contribution from form factors drops out altogether, and there is no $s_\ell$ dependence left. In other words, while Watson's theorem does not apply to the case of the $\delta_S(s , s_\ell)$ phase
shift due to the occurrence of two distinct possible intermediate states 
[$\pi^0 \pi^0$ and $\pi^+ \pi^-$ for $s\le M_K^2$], it
is still operative in the $l=1$ channel. This explains both why the phases of $F_P(s,s_\ell)$
and of $G_P(s , s_\ell)$ are identical, and why this common phase $\delta_P(s)$ actually does not depend on $s_\ell$.

In the isospin limit, the dependence on $s_\ell$ also drops out from $\delta_S(s,s_\ell)$, and Watson's theorem is recovered, 
i.e. the phases tend towards
\begin{equation}
\delta_S(s,s_\ell) \to \, \delta_0(s),
\ \delta_P(s) \to \, \delta_1(s)
\end{equation}
where $\delta_0(s)$ and $\delta_1(s)$ denote the $\pi\pi$ phases in the $l=0$, $I=0$ and $l=1$, $I=1$ channels, respectively.
The quantity that is determined from experiment is the difference $\delta_S(s,s_\ell) - \delta_P(s)$
and our aim is to compute its deviation from the difference $\delta_0(s) - \delta_1(s)$. 
Let us stress that the dependence on $s_\ell$ is also not present
in $\delta_S(s,s_\ell)$ at lowest order (i.e. the case considered in Ref.~\cite{Colangelo:2008sm}), where the expression
for $\delta_S$ reduces to
\begin{equation}
\delta_S(s) = 
\sum_{\{ a^\prime , b^\prime \}} \frac{1}{{\cal S}_{a^\prime b^\prime}}\,\frac{\lambda^{\frac{1}{2}}_{a^\prime b^\prime}(s)}{s}\,
\varphi_0^{a^\prime b^\prime ; {\mbox{\tiny{$+-$}}}} (s) \,
\frac{F_{S[0]}^{a^\prime b^\prime}}{F_{S[0]}}
\, \theta (s - s_{a^\prime b^\prime} ) 
+
{\cal O}(E^4)
.
\end{equation}
It appears that the available statistics has not allowed the NA48/2 experiment to identify a dependence of the phases on $s_\ell$ \cite{Batley:2007zz,Batley:2010zza}.  Another one of our aims will be to 
check that, from the theoretical side, the dependence on $s_\ell$ is indeed sufficiently small, as compared to other sources of error.

\indent

\section{Two-loop representation of $K_{\ell 4}$ form factors}\label{sec:Dispersion_reps}
\setcounter{equation}{0}

In this section, we derive a representation
of the $K_{\ell 4}$ form factors $F^{ab}(s,s_\ell)$ and $G^{ab}(s,s_\ell)$
that holds up to and including two loops in the low-energy expansion. From this 
representation, we can then obtain the various quantities
that enter the expressions of the phases of the form factors.
The idea is to proceed as in the case of the $\pi \pi$ amplitude in
Ref.~\cite{Stern:1993rg}, or as discussed for the scalar form factor of the pion
in Ref.~\cite{GasserMeissner91} (in the isospin limit) and in Ref.~\cite{DescotesGenon:2012gv} (with
isospin breaking included). As compared to the latter case, one has to deal
with some additional kinematic complexities when addressing the $K_{\ell 4}$ form factors. 

As a starting point, we assume fixed-$t$ dispersion relations for all form factors, in all three channels.
In order to obtain low-energy representations for these form factors accurate at 
two loops, it is convenient to write dispersion relations with two subtractions.
Notice that if the convergence of the dispersive integrals was the only issue, a lesser number
of subtractions would probably be sufficient (for the application of dispersion relations to the
$K_{e4}$ form factors in a somewhat different context, see Ref.~\cite{DispRelKe4_1,DispRelKe4_2}). 
Assuming the usual
analyticity properties for the form factors, with a first cut extending to infinity along part of the real positive $s$-axis, 
and a similar second cut along the real negative $s$-axis, due to the $u$-channel singularities,
we obtain the following dispersion relations [momentarily omitting the dependence with respect to $u$ and/or $s_\ell$]
\begin{equation}
{\bf A}^{ab}(s,t) = {\bf P}^{ab}(t \vert s,u) + \frac{s^2}{\pi} \int \frac{d x}{x^2} \frac{1}{x-s-i0} \, {\mbox{Im}} {\bf A}^{ab}(x,t)
+ \frac{u^2}{\pi} \int \frac{d x}{x^2} \frac{1}{x-u-i0} \, \lambda_a \lambda_c {\cal C}_{us} {\mbox{Im}} {\bf A}^{cb}(x,t)
.
\label{Disp_Rel}
\end{equation}
Each integral runs slightly above or below the corresponding cut in the complex $s$-plane, 
from the relevant threshold, $s_{ab}$ or $u_{ab}$, to infinity.
${\bf P}^{ab}(t \vert s,u)$ denotes a pair of subtraction functions that are polynomials
of the first degree in $s$ and $u$, with coefficients given by arbitrary functions of $t$.
Using the decompositions Eqs.~(\ref{cos_theta}) and (\ref{decomp_F_G}), we may write
\begin{equation}
 {\mbox{Im}} {\bf A}^{ab}(s,t) \,=\, \left(
\begin{array}{l}
{\mbox{Im}} F^{ab}_S(s) + 
{\displaystyle{\frac{s(\Sigma_\ell -s - 2t)-(M_a^2 - M_b^2)(M_c^2 - s_\ell)}{\lambda^{1\over 2}_{ab}(s) \lambda^{1\over 2}_{\ell c} (s)}}}
\, {\mbox{Im}} F^{ab}_P(s)\\
{\mbox{Im}} g_1^{ab}(s)
\end{array}
\right)
+ {\mbox{Im}} {\bf A}^{ab}(s,t)_{l\ge 2}
,
\end{equation}
where $F^{ab}_S(s)$ and $F^{ab}_P(s)$ were expressed in terms of the lowest partial waves in Eq.~(\ref{def_F_S_and_F_P}).
Furthermore, ${\mbox{Im}} {\bf A}^{ab}(s,t)_{l\ge 2}$ collects the contributions of the higher ($l\ge 2$)
partial-wave projections in (\ref{FandG_PW_decomp}), so that at low energies, ${\mbox{Im}} {\bf A}^{ab}(s,t)_{l\ge 2} = {\cal O}(E^6)$,
as discussed in Sec.~\ref{subsec:chi_count}. 
The last property is relevant as long as $s$ and $u$ remain below a typical hadronic scale $\Lambda_H \sim 1$ GeV,
but one should remember that the
integrals involving ${\mbox{Im}} {\bf A}^{ab}(s,t)_{l\ge 2}$ run up to infinity. 
However, in the range above $\Lambda_H$, ${\mbox{Im}} {\bf A}^{ab}(s,t)_{l\ge 2} = {\cal O}(E^0)$,
so that (see the similar discussion in Ref.~\cite{Stern:1993rg})
\begin{equation}
\frac{s^2}{\pi} \int \frac{d x}{x^2} \frac{1}{x-s-i0} \, {\mbox{Im}} {\bf A}^{ab}(x,t)_{l\ge 2}
\,=\, \left(\frac{s}{\Lambda_H}\right)^2 {\bf H}^{ab} \,+\, {\cal O}(E^6)
,
\label{HPV_exp}
\end{equation}
where ${\bf H}^{ab}$ denotes a set of constants, whose precise definitions need not concern us here.
We thus obtain the expression
\begin{eqnarray}
 {\bf A}^{ab}(s,t,u) &=&  {\bf P}^{ab}(t \vert s,u) +
\left[ {\bf \Phi}_{\!{\mbox{\tiny$+$}}}^{ab} (s) - (t-u) {\bf \Phi}_{\!{\mbox{\tiny$-$}}}^{ab} (s) \right]
 + \lambda_a \lambda_c {\cal C}_{us} \left[ {\bf \Phi}_{\!{\mbox{\tiny$+$}}}^{cb} (u) - (t-s) {\bf \Phi}_{\!{\mbox{\tiny$-$}}}^{cb} (u) \right]   
\nonumber\\
&& + \,
\lambda_b \lambda_c {\cal C}_{st} \left[ {\bf \Phi}_{\!{\mbox{\tiny$+$}}}^{ac} (t) - (s-u) {\bf \Phi}_{\!{\mbox{\tiny$-$}}}^{ac} (t) \right]
\,+\, {\cal O}(E^6)
.
\label{Disp_Rep}
\end{eqnarray}
In this expression, ${\bf P}^{ab}(t \vert s,u)$ denotes a pair of polynomials in $s$ and $u$ with
coefficients given by arbitrary functions of $t$ as before, but it differs 
from the one introduced initially in Eq.~(\ref{Disp_Rel}) in two respects. First, it contains
a contribution that compensates the fourth term on the right-hand side of Eq.~(\ref{Disp_Rep}),
which has been introduced for convenience as will become clear shortly. 
Second, the terms of Eq.~(\ref{HPV_exp}) generated by the higher partial waves have also
been absorbed into these polynomials. Therefore,  ${\bf P}^{ab}(t \vert s,u)$ in Eq.~(\ref{Disp_Rep})
still represents a pair of arbitrary polynomials of at most second order in $s$ and $u$,  whose
coefficients are functions of $t$. As for the functions ${\bf \Phi}_{\!{\mbox{\tiny$\pm $}}}^{ab} (s)$,
they are given in terms of the lowest partial waves
\begin{eqnarray}
{\bf \Phi}_{\!{\mbox{\tiny$+$}}}^{ab} (s) &=& \frac{s^2}{\pi} \int \frac{d x}{x^2} \frac{1}{x-s-i0}
\left(
\begin{array}{l}
{\mbox{Im}} F^{ab}_S(x) - 
{\displaystyle{\frac{(M_a^2 - M_b^2)(M_c^2 - s_\ell)}{\lambda^{1\over 2}_{ab}(x) \lambda^{1\over 2}_{\ell c} (x)}}}
{\mbox{Im}} F^{ab}_P(x)\\
{\mbox{Im}} g^{ab}_1(x) 
\end{array}
\right)
\nonumber\\
{\bf \Phi}_{\!{\mbox{\tiny$-$}}}^{ab} (s) &=& \frac{s}{\pi} \int {d x} \frac{1}{x-s-i0} \,
{\displaystyle{\frac{1}{\lambda^{1\over 2}_{ab}(x) \lambda^{1\over 2}_{\ell c} (x)}}}
\left(
\begin{array}{l}
{\mbox{Im}} F^{ab}_P(x) \\
0
\end{array}
\right)
,
\end{eqnarray}
where the integrals run over the right-hand cuts only. From these definitions, 
and according to the symmetry properties (\ref{sym_PV}), it follows that
\begin{equation}
 {\cal C}_{tu} {\bf \Phi}_{\!{\mbox{\tiny$\pm$}}}^{ba} (s) \,=\, \pm {\bf \Phi}_{\!{\mbox{\tiny$\pm$}}}^{ab} (s)\,.
\end{equation}
Alternatively, the functions ${\bf \Phi}_{\!{\mbox{\tiny$\pm $}}}^{ab} (s)$ can be defined
by specifying their analyticity properties in the complex $s$-plane, where their singularities 
are restricted to a cut along the positive real axis, with discontinuities along this cut expressed
in terms of the lowest partial waves as 
\begin{eqnarray}
 {\mbox{Im}} {\bf \Phi}_{\!{\mbox{\tiny$+$}}}^{ab} (s) &=& 
\left(
\begin{array}{l}
{\mbox{Im}} f^{ab}_0(s) -  
{\displaystyle{\frac{(M_a^2 - M_b^2)}{\lambda_{\ell c} (s)}}}
\left[
(s - M_c^2 - 3 s_\ell) {\mbox{Im}} g^{ab}_1(s) + 
(M_c^2 - s_\ell) {\displaystyle{\frac{\lambda^{1\over 2}_{\ell c} (s)}{\lambda^{1\over 2}_{ab}(s)}}}{\mbox{Im}} f^{ab}_1(s)
\right]
\\
{\mbox{Im}} g^{ab}_1(s) 
\end{array}
\right)\,,
\nonumber\\
{\mbox{Im}} {\bf \Phi}_{\!{\mbox{\tiny$-$}}}^{ab} (s) &=&
\frac{s}{\lambda^{\frac{1}{2}}_{ab}(s) \lambda^{\frac{1}{2}}_{\ell c}(s)} 
\left(
\begin{array}{l}
{\mbox{Im}} f_1^{ab}(s) \,-\, 
{\displaystyle{\frac{M_c^2 - s - s_\ell}{s}\,\frac{\lambda^{1\over 2}_{ab}(s)}{\lambda^{1\over 2}_{\ell c} (s)}}} {\mbox{Im}} g_1^{ab}(s) \\
0
\end{array}
\right)
,
\end{eqnarray}
supplemented by ${\bf \Phi}_{\!{\mbox{\tiny$\pm $}}}^{ab} (0) = 0$ and by the asymptotic conditions  
\begin{equation}
 \lim_{\vert s \vert \to \infty} \, s^{-3 + \frac{1}{2}(1\mp 1) } \, {\bf \Phi}_{\!{\mbox{\tiny$\pm$}}}^{ab} (s) = 0
.
\end{equation}
These conditions define ${\bf \Phi}_{\!{\mbox{\tiny$+$}}}^{ab} (s)$ (${\bf \Phi}_{\!{\mbox{\tiny $-$}}}^{ab} (s) $)
only up to a polynomial ambiguity, which is of second (first) order in $s$. The contributions of these polynomials
to ${\bf A}^{ab}(s,t,u)$ can then be absorbed by the arbitrary subtraction functions ${\bf P}^{ab}(t|s,u)$
already at hand.
Let us stress once more that the functions ${\bf \Phi}_{\!{\mbox{\tiny$\pm $}}}^{ab} (s)$
only possess right-hand cuts, with discontinuities specified in terms of those
of the partial waves, whereas the partial-wave projections themselves
in general have a more complicated analytical structure \cite{PWanalyticity1,PWanalyticity2,PWanalyticity3,PWanalyticity4}.

Finally, it remains to enforce the crossing relations (\ref{crossing}). One easily
checks that the three terms in Eq.~(\ref{Disp_Rep}) involving the {functions}
${\bf \Phi}_{\!{\mbox{\tiny$\pm $}}} (s)$ satisfy these relations among themselves, so that
(\ref{crossing}) need only be enforced on the contributions involving ${\bf P}^{ab}(t \vert s,u)$.
Following the same argument as in \cite{Stern:1993rg}, this 
means that the latter boil down to a pair  of polynomials  ${\bf P}^{ab}(s,t,u)$ of at most second order in 
all three variables $s$, $t$, and $u$, with arbitrary {\it constant} coefficients. These coefficients may 
depend on the masses and on $s_\ell$, in a way that is compatible with the chiral counting. 
The polynomials in the different channels are then related by 
\begin{equation}
{\bf P}^{ac}(s,t,u)\,=\, \lambda_b \lambda_c {\cal C}_{st} {\bf P}^{ab}(t,s,u)\,,
\ {\bf P}^{cb}(s,t,u)\,=\, \lambda_a \lambda_c {\cal C}_{us} {\bf P}^{ab}(u,t,s)\,,
\ {\bf P}^{ba}(s,t,u)\,=\,{\cal C}_{tu} {\bf P}^{ab}(s,u,t)
.
\label{P_crossing}
\end{equation}

\begin{figure}[t]
\begin{center}
\begin{picture}(150,100)(50,100)
\SetWidth{1}
\Text(60,160)[]{$K^+$}
\Text(135,175)[]{$\pi^+$}
\Text(135,130)[]{$\pi^-$}
\Line(50,150)(100,150)
\Photon(100,150)(100,200){3}{4}
\SetColor{Blue}
\CBoxc(100,150)(10,10){Blue}{Blue}
\Line(100,150)(150,200)
\Line(100,150)(150,100)
\end{picture}
\begin{picture}(150,100)(50,100)
\SetWidth{1}
\Text(60,160)[]{$K^+$}
\Text(135,182)[]{$\pi^+$}
\Text(135,123)[]{$\pi^-$}
\Text(180,182)[]{$\pi^+$}
\Text(180,123)[]{$\pi^-$}
\Line(50,150)(100,150)
\Photon(100,150)(100,200){3}{4}
\SetColor{Blue}
\CBoxc(100,150)(10,10){Blue}{Blue}
\Oval(127,150)(25,25)(0)
\Line(152,150)(175,200)
\Line(152,150)(175,100)
\SetColor{Black}
\Vertex(152,150){3}
\end{picture}
\begin{picture}(150,100)(50,100)
\SetWidth{1}
\Text(60,160)[]{$K^+$}
\Text(135,182)[]{$\pi^0$}
\Text(135,122)[]{$\pi^0$}
\Text(180,182)[]{$\pi^+$}
\Text(180,123)[]{$\pi^-$}
\Line(50,150)(100,150)
\Photon(100,150)(100,200){3}{4}
\SetColor{Red}
\Oval(127,150)(25,25)(0)
\CBoxc(100,150)(10,10){Red}{Red}
\SetColor{Blue}
\Line(152,150)(175,200)
\Line(152,150)(175,100)
\SetColor{Black}
\Vertex(152,150){3}
\end{picture}
\end{center}
\caption{$K^+_{e 4}$ form factors: leading-order contribution,
and contribution at one loop from unitarity in the $s$-channel. \label{fig:schannel}}
\end{figure}
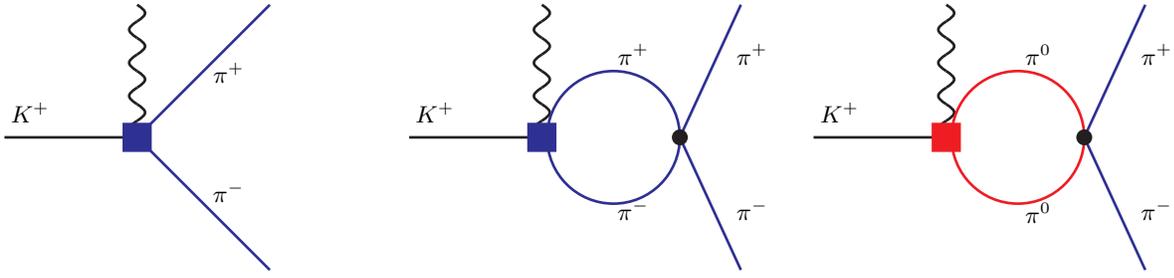

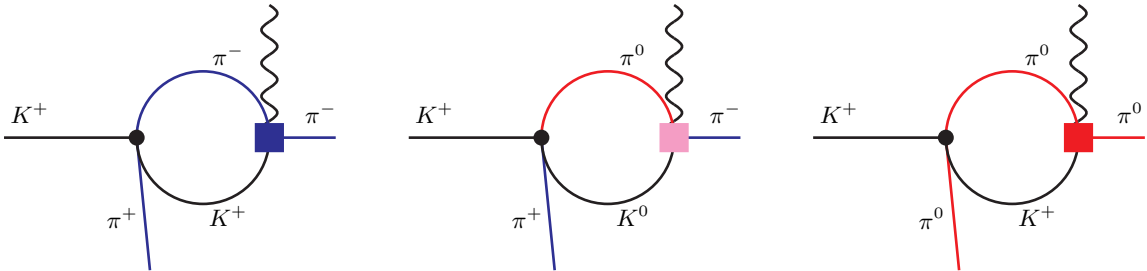
\begin{figure}[b]
\begin{center}
\begin{picture}(150,100)(60,100)
\SetWidth{1}
\Text(60,160)[]{$K^+$}
\Text(135,182)[]{$\pi^-$}
\Text(135,121)[]{$K^+$}
\Text(170,160)[]{$\pi^-$}
\Text(95,120)[]{$\pi^+$}\Line(50,150)(100,150)
\Photon(150,150)(150,200){3}{4}
\SetColor{Blue}
\CArc(125,150)(25,0,180)
\SetColor{Blue}
\Line(152,150)(175,150)
\Line(100,150)(105,100)
\SetColor{Black}
\Vertex(100,150){3}
\CArc(125,150)(25,180,0)
\CBoxc(150,150)(10,10){Blue}{Blue}
\end{picture}
\begin{picture}(150,100)(60,100)
\SetWidth{1}
\Text(60,160)[]{$K^+$}
\Text(135,182)[]{$\pi^0$}
\Text(135,121)[]{$K^0$}
\Text(170,160)[]{$\pi^-$}
\Text(95,120)[]{$\pi^+$}\Line(50,150)(100,150)
\Photon(150,150)(150,200){3}{4}
\SetColor{Red}
\CArc(125,150)(25,0,180)
\SetColor{Blue}
\Line(152,150)(175,150)
\Line(100,150)(105,100)
\SetColor{Black}
\Vertex(100,150){3}
\CArc(125,150)(25,180,0)
\CBoxc(150,150)(10,10){Lavender}{Lavender}
\end{picture}
\begin{picture}(150,100)(60,100)
\SetWidth{1}
\Text(60,160)[]{$K^+$}
\Text(135,182)[]{$\pi^0$}
\Text(135,121)[]{$K^+$}
\Text(170,160)[]{$\pi^0$}
\Text(95,120)[]{$\pi^0$}
\Line(50,150)(100,150)
\Photon(150,150)(150,200){3}{4}
\SetColor{Red}
\CArc(125,150)(25,0,180)
\SetColor{Red}
\Line(152,150)(175,150)
\Line(100,150)(105,100)
\SetColor{Black}
\Vertex(100,150){3}
\CArc(125,150)(25,180,0)
\CBoxc(150,150)(10,10){Red}{Red}
\end{picture}
\caption{$K^+_{e 4}$ form factors: typical rescattering diagrams involved in the reconstruction of the $K_{e4}^+$ form 
factors in the $t$- and $u$-channels. 
\label{fig:tuchannels}}
\end{center}
\end{figure}

In the kinematical range of interest, the discontinuities
in both form factors and relevant meson-meson scattering amplitudes are limited
to two-meson intermediate states [cf. examples of typical diagrams at one loop shown
in Figs.~\ref{fig:schannel} and \ref{fig:tuchannels}] up to and including two-loop order.
Discontinuities generated by states made of more than two mesons contribute
only at higher orders, while discontinuities due to non-Goldstone intermediate
states occur only at higher energies.

\begin{figure}[t]
\begin{center}
\begin{picture}(375,130)(0,130)

\SetWidth{1.5}

\Boxc(0,200)(70,20)
\Text(0,200)[]{$A$ at order {$E^{2k}$}}

\ArrowLine(35,200)(90,200)

\Text(62.5,240)[]{projection}
\Text(62.5,230)[]{over partial waves}

\Boxc(125,200)(70,20)
\Text(125,200)[]{$f$ at order {$E^{2k}$}}

\ArrowLine(160,200)(202.5,200)

\Text(180,230)[]{unitarity}

\Boxc(250,200)(95,20)
\Text(250,200)[]{Im $f$ at order {$E^{2k+2}$}}

\ArrowLine(297.5,200)(335,200)

\Text(317.5,230)[]{dispersion relation}

\Boxc(375,200)(80,20)
\Text(375,200)[]{$A$ at order {$E^{2k+2}$}}

\ArrowLine(375,190)(375,150)
\ArrowLine(375,150)(0,150)
\ArrowLine(0,150)(0,190)

\end{picture}
\end{center}
\caption{Recursive construction for two-loop representations of the $K_{e4}^+$ form factors and $\pi\pi$ scattering 
amplitudes in the low-energy regime. $A$ denotes the amplitude of interest, whereas $f$ corresponds to partial waves. \label{iterconst}}
\end{figure}
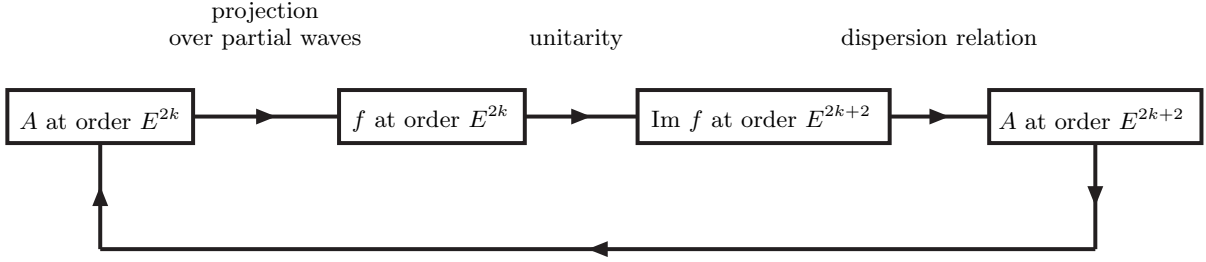

The above analysis provides an iterative set-up that may be used to construct the $K_{e4}$
form factors at two loops through a two-step process, as illustrated schematically in Fig.~\ref{iterconst}.
The starting point is provided by the form factors and amplitudes at lowest order. Since these are
given by at most first order polynomials in the corresponding Lorentz invariant
kinematical variables, the computation of the lowest partial waves required for the
one-loop discontinuities is a simple exercise. Likewise,
finding the appropriate explicit representation of the one-loop functions with the prescribed discontinuities
presents no particular difficulties. Things become less tractable with the implementation of the second iteration,
which requires the partial-wave projections of the one-loop form factors and scattering amplitudes.
For a discussion of some of the technical aspects related to the second iteration, we refer the interested
reader to Ref.~\cite{DescotesGenon:2012gv}.
For our present aims, we fortunately need not go through the whole second iteration. Getting the real parts of the one-loop
partial wave projections will be enough, and this part is still tractable in an analytic way. 

\indent

\section{First iteration: one-loop form factors}\label{sec:1-loop_ff}
\setcounter{equation}{0}

Using the results displayed in the previous section, we give the general structure
of the form factors $F^{ab}(s,t,u)$ and $G^{ab}(s,t,u)$ at one loop in terms of the lowest-order partial-wave
projections of the various amplitudes involved. We then discuss the two instances
of direct interest in greater detail, namely $ab = +-$ and $ab = 00$, corresponding to the channels
$K^+ \to \pi^+ \pi^- \ell^+ \nu_\ell$ and $K^+ \to \pi^0 \pi^0 \ell^+ \nu_\ell$, respectively.

\subsection{The general case}\label{subsec:1-loop_ff_general}

At one loop, only the lowest-order expressions of the form factors and of the partial waves 
are required in the unitarity condition (\ref{Imf_Img}). 
As far as the form factors are concerned, they reduce to ${\cal O} (E^0)$ constants,
\begin{equation}
\left. F^{ab}(s,t,u) \right\vert_{LO} = F^{ab}_{S [0]},
\ \left. G^{ab}(s,t,u) \right\vert_{LO} = G^{ab}_{P [0]}
,
\label{F_G_LO}
\end{equation}
so that the corresponding lowest partial-wave projections are real and given by
\begin{eqnarray}
f_0^{ab}(s,s_\ell) &=& F^{ab}_{S [0]} \left[ 1 \,+\, \frac{G^{ab}_{P [0]}}{F^{ab}_{S [0]}} \frac{M_a^2 - M_b^2}{s}  \right] \,+\, {\cal O}(E^2)
, 
\nonumber\\
f_1^{ab}(s,s_\ell) &=&  G^{ab}_{P [0]} \, \frac{M_c^2 - s - s_\ell}{s}
\cdot \frac{\lambda_{ab}^{\frac{1}{2}} (s)}{\lambda_{\ell c}^{\frac{1}{2}} (s)} \,+\, {\cal O}(E^2) 
,
\nonumber\\
g_1^{ab}(s,s_\ell) &=& G^{ab}_{P [0]} \,+\, {\cal O}(E^2) 
.
\label{F_G_pw_LO}
\end{eqnarray}
We write for the lowest partial waves of the scattering amplitudes $A^{ab;a^\prime b^\prime}(s,{\hat t},{\hat u})$,
${\hat t} = (p_a - p_{a^\prime})^2$, ${\hat u} = (p_a - p_{b^\prime})^2$,
\begin{equation}
{\mbox{Re}}\, t_l^{ab;a^\prime b^\prime}(s) = \varphi_l^{ab;a^\prime b^\prime}(s) + {\cal O}(E^4) ,
\ {\mbox{Im}}\, t_l^{ab;a^\prime b^\prime}(s) = 0 + {\cal O}(E^4) ,
\ l=0,1.
\end{equation}
In contrast to the case of the $\pi\pi$ amplitude in the isospin limit \cite{Stern:1993rg,Knecht:1995tr},
the lowest-order partial waves $\varphi_0^{ab;a^\prime b^\prime}(s)$ and $\varphi_1^{ab;a^\prime b^\prime}(s)$ 
are in general not first-order polynomials
in the variable $s$, due to the possible occurrence of unequal masses. But, considering that the
corresponding scattering amplitudes are polynomials
of at most first orders in the Mandelstam variables $s,{\hat t},{\hat u}$ at lowest order, and that the general expressions
of the variables ${\hat t}$ and ${\hat u}$ in terms of $s$ and of the scattering angle ${\hat\theta}$ are given by
[$\Delta_{ab} \equiv M_a^2 - M_b^2$, $\Delta_{a^\prime b^\prime} \equiv M_{a^\prime}^2 - M_{b^\prime}^2$, 
$s + {\hat t} + {\hat u} = M_a^2 + M_b^2 + M_{a^\prime}^2 + M_{b^\prime}^2$]
\begin{eqnarray}
{\hat t} = \frac{1}{2} \left( M_a^2 + M_b^2 + M_{a^\prime}^2 + M_{b^\prime}^2 -s \right)
- \frac{\Delta_{ab} \Delta_{a^\prime b^\prime}}{2 s} +
\frac{1}{2} \frac{\lambda^{\frac{1}{2}}_{ab} (s) \lambda^{\frac{1}{2}}_{a^\prime b^\prime} (s)}{s}\,
\cos {\hat\theta}
\nonumber\\
{\hat u} = \frac{1}{2} \left( M_a^2 + M_b^2 + M_{a^\prime}^2 + M_{b^\prime}^2 -s \right)
+ \frac{\Delta_{ab} \Delta_{a^\prime b^\prime}}{2 s} -
\frac{1}{2} \frac{\lambda^{\frac{1}{2}}_{ab} (s) \lambda^{\frac{1}{2}}_{a^\prime b^\prime} (s)}{s}\,
\cos {\hat\theta}
,
\end{eqnarray}
we deduce that
\begin{equation}
\varphi^{ab;a^\prime b^\prime}_P \equiv \varphi^{ab;a^\prime b^\prime}_1(s) \cdot 
\frac{s}{\lambda^{\frac{1}{2}}_{ab} (s) \lambda^{\frac{1}{2}}_{a^\prime b^\prime} (s)}
\label{varphi_P}
\end{equation}
is a constant, while
\begin{equation}
\varphi^{ab;a^\prime b^\prime}_S(s) \equiv \varphi^{ab;a^\prime b^\prime}_0(s)  + 
3\, \frac{\Delta_{ab} \Delta_{a^\prime b^\prime}}{s} \cdot  \varphi^{ab;a^\prime b^\prime}_P
\label{varphi_S}
\end{equation}
is a polynomial of at most first order in $s$. 

In order to build the functions $\Phi_+$ and $\Phi_-$ at this order, it is useful to notice that 
not only does the combination $F^{ab}_P(s)$, defined in Eq.~(\ref{def_F_S_and_F_P}),
vanish at lowest order, cf. Eq.~(\ref{F_G_pw_LO}), but so does also its discontinuity along the positive real axis
at order ${\cal O}(E^2)$.  This is due to the fact that $F^{ab}(s,t,u)$ is
constant at lowest order, and thus gives no contribution to $f_1^{ab}(s)$. The latter
is entirely produced by the (constant) form factor $G^{ab}(s,t,u)$ at this order, with
the coefficient required so that the combination in (\ref{def_F_S_and_F_P}) vanishes, cf. Eq.~(\ref{F_G_R_cal}).
At order ${\cal O}(E^2)$, the discontinuity of $F^{ab}_P(s)$ involves the lowest-order expressions
of $f_1^{a^\prime b^\prime}(s)$ and $g_1^{a^\prime b^\prime}(s)$, both multiplied by 
$\varphi^{ab;a^\prime b^\prime}_1(s)$, and thus vanishes for the same reason.
Consequently, ${\mbox{Im}}  {\bf \Phi}_{\!{\mbox{\tiny$-$}}}^{ab} (s) = 0 + {\cal O}(E^4)$.
Of course, similar statements hold for ${\bf \Phi}_{\!{\mbox{\tiny$-$}}}^{cb} (s)$
and for ${\bf \Phi}_{\!{\mbox{\tiny$-$}}}^{ac} (s)$, and the contributions of these functions 
at one loop are thus entirely absorbed by the subtraction polynomials. 
We therefore end up with the following one-loop expressions of the form factors,
\begin{eqnarray}
F^{ab}(s,t,u) &=& F^{ab}_{S [0]}  + \frac{1}{F_0^2}
\left[
P_F^{ab}(s,t,u) + \Delta^{ab}_F (s) + A_F^{ab}(t) + B_F^{ab}(t) - B_F^{ab}(u)
\right]
,
\nonumber\\
G^{ab}(s,t,u) &=& G^{ab}_{P [0]}  + \frac{1}{F_0^2}
\left[
P_G^{ab}(s,t,u) + \Delta^{ab}_G (s) + A_G^{ab}(t) + B_G^{ab}(t) - B_G^{ab}(u)
\right]
,
\label{ff_gen_exp}
\end{eqnarray}
where we have normalized the one-loop corrections by $F_0^2$, the square of the
pion decay constant in the $N_f=3$ chiral limit~\cite{Gasser:1984gg}.
The unitarity (i.e. non polynomial) contributions are given by
\begin{eqnarray}
A_F^{ab}(s) &=& - \frac{1}{2} \lambda_b \lambda_c 
\left[
\Delta^{ac}_F (s) - 3 \Delta^{ac}_G (s)
\right]
- \frac{1}{2} \lambda_a \lambda_c 
\left[
\Delta^{cb}_F (s) + 3 \Delta^{cb}_G (s)
\right]
,
\nonumber\\
B_F^{ab}(s) &=&
+ \frac{1}{2} \lambda_a \lambda_c 
\left[
\Delta^{cb}_F (s) + 3 \Delta^{cb}_G (s)
\right]
,
\nonumber\\
A_G^{ab}(s) &=& + \frac{1}{2} \lambda_b \lambda_c 
\left[
\Delta^{ac}_F (s) + \Delta^{ac}_G (s)
\right]
- \frac{1}{2} \lambda_a \lambda_c 
\left[
\Delta^{cb}_F (s) - \Delta^{cb}_G (s)
\right]
,
\nonumber\\
B_G^{ab}(s) &=&
+ \frac{1}{2} \lambda_a \lambda_c 
\left[
\Delta^{cb}_F (s) - \Delta^{cb}_G (s)
\right]
,
\label{As_and_Bs}
\end{eqnarray}
with 
\begin{eqnarray}
{\mbox{Im}} \Delta^{ab}_F (s) &=& F_0^2 \sum_{\{a^\prime , b^\prime\}}
\frac{1}{{\cal S}_{a^\prime b^\prime}}
\frac{\lambda^{\frac{1}{2}}_{a^\prime b^\prime} (s)}{s}
\left[
F_{S [0]}^{a^\prime b^\prime} 
 +
G_{P [0]}^{a^\prime b^\prime} 
 \cdot \frac{\Delta_{a^\prime b^\prime}}{s}
\right]
\varphi^{ab;a^\prime b^\prime}_0 \!(s)
\times \theta(s-s_{a^\prime b^\prime})
- 
\frac{\Delta_{ab}}{s} \cdot {\mbox{Im}} \Delta^{ab}_G (s)
,
\nonumber\\
{\mbox{Im}} \Delta^{ab}_G (s) &=& F_0^2 \sum_{\{a^\prime , b^\prime\}}
G_{P [0]}^{a^\prime b^\prime} \cdot \varphi^{ab;a^\prime b^\prime}_1 \!(s) 
\cdot
\frac{\lambda^{\frac{1}{2}}_{a^\prime b^\prime} (s)}{\lambda^{\frac{1}{2}}_{ab} (s)}
\cdot
\frac{\lambda^{\frac{1}{2}}_{a^\prime b^\prime} (s)}{s}
\times \theta(s-s_{a^\prime b^\prime})
.
\label{Im_Delta_FG}
\end{eqnarray}

There exist many possibilities to express the functions $\Delta_F^{ab} (s)$ and $\Delta_G^{ab} (s)$ themselves,
the differences being compensated by the polynomial parts $P_F^{ab}(s,t,u)$ and $P_G^{ab}(s,t,u)$. To make contact
with the structure of the form factors at one loop as displayed in Ref.~\cite{Bijnens:1994me}, we may for instance write
\begin{eqnarray}
\Delta_F^{ab} (s) &=&
\frac{F_0^2}{F_\pi^2} \sum_{\{ a^\prime b^\prime\}}
\frac{1}{{\cal S}_{a^\prime b^\prime}}
\bigg\{
F_{S [0]}^{a^\prime b^\prime}
\left[
16 \pi F_\pi^2 \varphi^{ab;a^\prime b^\prime}_S \!(s) 
- 3 \Delta_{a^\prime b^\prime} \gamma^{ab;a^\prime b^\prime}_P
\right]
{\bar J}_{a^\prime b^\prime} (s)
+ 6 F_{S [0]}^{a^\prime b^\prime} \gamma^{ab;a^\prime b^\prime}_P
\left[
s - \Delta_{ab} 
\right]
K_{a^\prime b^\prime} (s)
\nonumber\\
&&\!\!\!\!
+ G_{P [0]}^{a^\prime b^\prime} \beta^{ab;a^\prime b^\prime}
\Delta_{a^\prime b^\prime}
{\bar J}_{a^\prime b^\prime} (s)
+ 2 G_{P [0]}^{a^\prime b^\prime} 
\left[
16 \pi F_\pi^2 \varphi^{ab;a^\prime b^\prime}_S \!(0) K_{a^\prime b^\prime} (s)
- 6 \gamma^{ab;a^\prime b^\prime}_P \!\Delta_{ab}
{M}_{a^\prime b^\prime} (s)
\right]
\bigg\}\,,
\nonumber\\
\Delta_G^{ab} (s) &=&
\frac{F_0^2}{F_\pi^2} \sum_{\{ a^\prime b^\prime\}}
12 G_{P [0]}^{a^\prime b^\prime}
\gamma^{ab;a^\prime b^\prime}_P
\left[
s {M}_{a^\prime b^\prime} (s) - L_{a^\prime b^\prime} (s)
\right]
,
\label{Delta_F_and_Delta_G}
\end{eqnarray}
where
\begin{equation}
\beta^{ab;a^\prime b^\prime} \equiv  16 \pi {F_\pi^2} \, \frac{\varphi^{ab;a^\prime b^\prime}_S (s) - \varphi^{ab;a^\prime b^\prime}_S (0)}{s}  
, \quad
\gamma^{ab;a^\prime b^\prime}  \equiv  16 \pi {F_\pi^2} \, \varphi^{ab;a^\prime b^\prime}_P
,
\label{beta_and_gamma_def}
\end{equation}
and
\begin{eqnarray}
K_{ab} (s) &=& \frac{\Delta_{ab}}{2 s } {\bar J}_{ab}(s)
,
\nonumber\\
L_{ab} (s) &=& \frac{\Delta_{ab}^2}{4 s } {\bar J}_{ab}(s)
,
\nonumber\\
{M}_{ab} (s) &=& \frac{1}{12 s} \left( s - 2 M_a^2 - 2 M_b^2 \right) {\bar J}_{ab}(s)
+ \frac{\Delta_{ab}^2}{3 s^2 } \ {\bar{\!\!{\bar J}}}_{ab}(s)
+ \frac{1}{288 \pi^2}
.
\label{loop_functions}
\end{eqnarray}
Here ${\bar J}_{ab}(s)$ denotes the two-point scalar loop function subtracted at $s=0$,
which can be written in a dispersive form,
\begin{equation}
{\bar J}_{ab}(s) = \frac{s}{16\pi^2} \int_{(M_a + M_b)^2}^\infty \frac{dx}{x} \frac{1}{x-s} 
\frac{\lambda_{ab}^{\frac{1}{2}}(x)}{x}
,
\end{equation}
and
\begin{equation}
{{\bar{\!\!{\bar J}}}}_{ab}(s) = {\bar J}_{ab}(s) - s {\bar J}_{ab}'(0)  
= \frac{s^2}{16\pi^2} \int_{(M_a + M_b)^2}^\infty \frac{dx}{x^2} \frac{1}{x-s} 
\frac{\lambda_{ab}^{\frac{1}{2}}(x)}{x}
,
\end{equation}
with
\begin{equation}
{\bar J}_{ab}'(0) = \frac{1}{32 \pi^2}
\left[
\frac{M_a^2 + M_b^2}{\Delta_{ab}^2} + 2 \frac{M_a^2 M_b^2}{\Delta_{ab}^3} \ln \frac{M_b^2}{M_a^2}
\right] 
.
\end{equation}
Notice that the formulas given in Ref.~\cite{Bijnens:1994me} were expressed in terms of 
the scale-dependent renormalized functions
$J^r_{ab} (s) = {\bar J}_{ab} (s) - 2 k_{ab}(\mu)$ and $M^r_{ab} (s) = {M}_{ab} (s)  - \frac{1}{6} k_{ab}(\mu)$,
with
\begin{equation}
k_{ab}(\mu) \equiv \frac{1}{32 \pi^2} \, \frac{1}{\Delta_{ab}}
\left[ M_a^2 \ln {\displaystyle\frac{M_a^2}{\mu^2}} - M_b^2 \ln {\displaystyle\frac{M_b^2}{\mu^2}} \right]
.
\end{equation}
The difference that results from using one set of functions rather than the other 
is a polynomial of at most first order in the variables $s$, $t$, and $u$ that can be absorbed
in the (so far arbitrary) subtraction polynomials $P_F^{ab}(s,t,u)$ and $P_G^{ab}(s,t,u)$.

\subsection{The case $K^+ \to \pi^+ \pi^- \ell^+ \nu_\ell$}\label{subsec:1-loop_ff_+-}

For the process $K^+ \to \pi^+ \pi^- \ell^+ \nu_\ell$ the general structure of the
form factors at one loop, as displayed in the equations (\ref{ff_gen_exp}) and (\ref{As_and_Bs}) 
given above, leads to the following expressions (notice that we chose a somewhat different normalization from the general formulas given previously, by factorizing 
$F^{\mbox{\tiny{$+-$}}}_{S [0]} = G^{\mbox{\tiny{$+-$}}}_{P [0]} = M_K/(\sqrt{2} F_0)$).

\begin{eqnarray}
F^{\mbox{\tiny{$+-$}}} (s,t,u) &=& \frac{M_K}{\sqrt{2} F_0}
\left\{
1 +
\frac{1}{F_0^2}
\left[
P_F^{\mbox{\tiny{$+-$}}} (s,t,u) + \Delta_F^{\mbox{\tiny{$+-$}}} (s) +  
A_F^{\mbox{\tiny{$+-$}}} (t) + B^{\mbox{\tiny{$+-$}}} (t) - B^{\mbox{\tiny{$+-$}}} (u)
\right]
\right\}
,
\nonumber\\
\label{FG_+-_1loop}
\\
G^{\mbox{\tiny{$+-$}}} (s,t,u) &=& \frac{M_K}{\sqrt{2} F_0}
\left\{
1 +
\frac{1}{F_0^2}
\left[
P_G^{\mbox{\tiny{$+-$}}} (s,t,u) + \Delta_G^{\mbox{\tiny{$+-$}}} (s) +  
A_G^{\mbox{\tiny{$+-$}}} (t) + B^{\mbox{\tiny{$+-$}}} (t) - B^{\mbox{\tiny{$+-$}}} (u)
\right]
\right\}
,
\nonumber
\end{eqnarray}
with
\begin{eqnarray}
\Delta_F^{\mbox{\tiny{$+-$}}} (s) &=&
\frac{F_0^2}{F_\pi^2} \sum_{\{ a^\prime b^\prime\}} \frac{1}{{\cal S}_{a^\prime b^\prime}}
\frac{F_{S [0]}^{a^\prime b^\prime}}{F_{S [0]}^{\mbox{\tiny{$+-$}}}}
\left[
16 \pi F_\pi^2 \varphi^{\mbox{\tiny{$+-$}};a^\prime b^\prime}_S (0) +
\beta^{\mbox{\tiny{$+-$}};a^\prime b^\prime} s \right]
{\bar J}_{a^\prime b^\prime} (s)
,
\nonumber\\
A_F^{\mbox{\tiny{$+-$}}} (t) + B^{\mbox{\tiny{$+-$}}} (t) &=&
- \frac{1}{2}
\frac{F_0^2}{F_\pi^2} \sum_{\{ a^\prime b^\prime\}} 
\frac{F_{S [0]}^{a^\prime b^\prime}}{F_{S [0]}^{\mbox{\tiny{$+-$}}}}
\bigg\{
\left[
16 \pi F_\pi^2 \varphi^{\mbox{\tiny{$+K^-$}};a^\prime b^\prime}_S (0) + \beta^{\mbox{\tiny{$+K^-$}};a^\prime b^\prime} t
\right]
{\bar J}_{a^\prime b^\prime} (t)
\nonumber\\
&&
+ \Delta_{a^\prime b^\prime} 
\left(
\frac{G_{P [0]}^{a^\prime b^\prime}}{F_{S [0]}^{a^\prime b^\prime}}
\beta^{\mbox{\tiny{$+K^-$}};a^\prime b^\prime} - 3 \gamma^{\mbox{\tiny{$+K^-$}};a^\prime b^\prime}
\right)
{\bar J}_{a^\prime b^\prime} (t)
\nonumber\\
&&
+ 2
\left[
\frac{G_{P [0]}^{a^\prime b^\prime}}{F_{S [0]}^{a^\prime b^\prime}}
16 \pi F_\pi^2 \varphi^{\mbox{\tiny{$+K^-$}};a^\prime b^\prime}_S (0)
+3 \gamma^{\mbox{\tiny{$+K^-$}};a^\prime b^\prime} (t - M_\pi^2 + M_K^2) 
\right]
K_{a^\prime b^\prime} (t)
\nonumber\\
&&
- 12 \frac{G_{P [0]}^{a^\prime b^\prime}}{F_{S [0]}^{a^\prime b^\prime}} \gamma^{\mbox{\tiny{$+K^-$}};a^\prime b^\prime}
\left[
(3 t + M_\pi^2 - M_K^2) M_{a^\prime b^\prime} (t) - 3 L_{a^\prime b^\prime} (t)
\right]
\bigg\}
,
\nonumber\\
B^{\mbox{\tiny{$+-$}}} (u) &=&
\frac{1}{2}
\frac{F_0^2}{F_\pi^2} \frac{F_{S [0]}^{\mbox{\tiny{$K^- -$}}}}{F_{S [0]}^{\mbox{\tiny{$+-$}}}}
\bigg\{
\left[
16 \pi F_\pi^2 \varphi^{\mbox{\tiny{$K^- -$}};\mbox{\tiny{$K^- -$}}}_S (0) + \beta^{\mbox{\tiny{$K^- -$}};\mbox{\tiny{$K^- -$}}} u
\right]
{\bar J}_{K \pi} (u)
\nonumber\\
&& 
- 3 \gamma^{\mbox{\tiny{$K^- -$}};\mbox{\tiny{$K^- -$}}} (M_K^2 - M_\pi^2 )
{\bar J}_{K \pi} (u)
+ 6 \gamma^{\mbox{\tiny{$K^- -$}};\mbox{\tiny{$K^- -$}}} (u + M_\pi^2 - M_K^2) 
K_{K \pi} (u)
\bigg\}
,
\end{eqnarray}
and
\begin{eqnarray}
\Delta_G^{\mbox{\tiny{$+-$}}} (s) &=& 12
\frac{F_0^2}{F_\pi^2} \sum_{\{ a^\prime b^\prime\}}  
\frac{G_{P [0]}^{a^\prime b^\prime}}{G_{P [0]}^{\mbox{\tiny{$+-$}}}}
\gamma^{\mbox{\tiny{$+-$}};a^\prime b^\prime} 
\left[ s M_{a^\prime b^\prime} (s) - L_{a^\prime b^\prime} (s)
\right]
,
\nonumber\\
A_G^{\mbox{\tiny{$+-$}}} (t) + B^{\mbox{\tiny{$+-$}}} (t) &=&
- A_F^{\mbox{\tiny{$+-$}}} (t) - B^{\mbox{\tiny{$+-$}}} (t)
+ 24
\frac{F_0^2}{F_\pi^2} \sum_{\{ a^\prime b^\prime\}}  
\frac{G_{P [0]}^{a^\prime b^\prime}}{F_{S [0]}^{\mbox{\tiny{$+-$}}}} \gamma^{\mbox{\tiny{$+K^-$}};a^\prime b^\prime}
\left[
t  M_{a^\prime b^\prime} (t) -  L_{a^\prime b^\prime} (t)
\right]
.
\end{eqnarray}
\begin{table}[t]
\begin{center}
\begin{tabular}{c|l}
$\pi^+\pi^-$  & $\ \pi^+ \pi^-$, $\pi^0 \pi^0$, $K^+ K^-$, $K^0 {\bar K}^0$, ~$\eta\, \eta$~, $\pi^0 \eta$~
\\
$\pi^+ K^-$ & $\ \pi^+ K^-$, $\pi^0 {\bar K}^0$, ~$\eta {\bar K}^0$
\\
$K^- \pi^-$  & $\ K^- \pi^-$  
\end{tabular}
\caption{The pairs $\{a,b\}$ (on the left) and the corresponding possible intermediate states $\{a^\prime , b^\prime \}$
(on the right) relevant for the form factors $F^{\mbox{\tiny{$+-$}}} (s,t,u)$ and $G^{\mbox{\tiny{$+-$}}} (s,t,u)$.}
\label{tab:intermediate_+-}
\end{center}
\end{table}
The various intermediate states that occur in these expressions are listed in
Tab.~\ref{tab:intermediate_+-} and the associated form factors at lowest order
are collected in Tab.~\ref{tab:FF_LO}. Notice that a single intermediate state, $K^- \pi^-$,
contributes to $B^{\mbox{\tiny{$+-$}}} (t)$, and that $G_{LO}^{\mbox{\tiny{$K^- -$}}} = 0$
has already been taken into account in the expression given above. We also recall
that the absence of superscript on pion and kaon masses refers to the charged case.
Likewise, ${\bar J}_{K \pi} (s)$ stands for ${\bar J}_{K^\pm \pi^\pm} (s)$, and so on.
The lowest-order scattering amplitudes, together with the corresponding parameters
$\beta^{ a b ;a^\prime b^\prime}$, $\gamma^{ a b ;a^\prime b^\prime}$ and $\varphi^{ a b ;a^\prime b^\prime}_S (0)$,
are given in App.~\ref{app:loampl}. In the superscripts, the pion states are simply
mentioned by their charges, e.g 
$\beta^{\mbox{\tiny{$+-$}};a^\prime b^\prime} \equiv \beta^{\mbox{\tiny{$\pi^+ \pi^- $}};a^\prime b^\prime}$,
or $\gamma^{\mbox{\tiny{$+ K^- ; 0 {\bar K}^0$}}} \equiv \gamma^{\mbox{\tiny{$\pi^ +K^- ; \pi^0 {\bar K}^0$}}}$, and so on.

Finally, at this order, the subtraction polynomials
have the following structure:
\begin{eqnarray}
{P}_F^{\mbox{\tiny{$+-$}}}(s,t,u) &=& \pi^{\mbox{\tiny{$+-$}}}_{0,F} \,+\, \pi^{\mbox{\tiny{$+-$}}}_{1,F} s 
\,+\,\pi^{\mbox{\tiny{$+-$}}}_{2,F} s_{\ell} \,+\,\pi^{\mbox{\tiny{$+-$}}}_{3,F} (t-u) 
,
\nonumber\\
{P}_G^{\mbox{\tiny{$+-$}}}(s,t,u) &=& \pi^{\mbox{\tiny{$+-$}}}_{0,G} \,+\, \pi^{\mbox{\tiny{$+-$}}}_{1,G} s 
\,+\,\pi^{\mbox{\tiny{$+-$}}}_{2,G} s_{\ell} \,+\,\pi^{\mbox{\tiny{$+-$}}}_{3,G} (t-u) 
.
\label{P_F_and_P_G_+-}
\end{eqnarray}
The subtraction constants $\pi_{i, F/G}^{\mbox{\tiny{$+-$}}}$ are discussed in App.~ \ref{app:poly}.

\begin{table}[b]
\begin{center}
\begin{tabular}{|c|c|c|}
\hline
  ~$a^\prime b^\prime$~  & ~$F_{S [0]}^{a^\prime b^\prime} / F_{S [0]}^{\mbox{\tiny{$+-$}}}$~ & 
~$G_{P [0]}^{a^\prime b^\prime} / F_{S [0]}^{\mbox{\tiny{$+-$}}}$~
\\ \hline\hline
$\pi^+ \pi^-$  & $1$  &  $1$
\\ \hline
$\pi^0 \pi^0$  & $-\left( 1 + 2 \sqrt{3} \epsilon_2 \right)$  &  $0$
\\ \hline
$K^+ K^-$  & $2$  &  $2$
\\ \hline
$K^0 {\bar K}^0$  & $-1$  &  $-1$
\\ \hline
$\eta \eta$  & $-3 \left( 1 - 2 \frac{\epsilon_1}{\sqrt{3}} \right)$  &  $0$
\\ \hline
$\pi^0 \eta$  & $- \sqrt{3} \left( 1 - \frac{\epsilon_1}{\sqrt{3}} + \sqrt{3} \epsilon_2 \right)$  &  $0$
\\ \hline\hline
$\pi^+ K^-$  & $1$  &  $1$
\\ \hline
$\pi^0 {\bar K}^0$  & $- \frac{3}{\sqrt{2}} \left( 1 + \frac{\epsilon_2}{\sqrt{3}} \right)$  &  
$- \frac{1}{\sqrt{2}} \left( 1 - \sqrt{3} \epsilon_2 \right)$
\\ \hline
$\eta {\bar K}^0$  & $- \sqrt{\frac{3}{2}} \left( 1 - \sqrt{3} \epsilon_1 \right)$  &  
$\sqrt{\frac{3}{2}} \left( 1 + \frac{\epsilon_1}{\sqrt{3}} \right) $
\\ \hline
$\pi^- K^-$  & $-2$  &  $0$
\\ \hline\hline
$\pi^0 K^-$  & $- \frac{1}{2} \left( 1 + 2 \sqrt{3} \epsilon_2 \right)$  &  $\frac{1}{2} \left( 1 + 2 \sqrt{3} \epsilon_2 \right)$
\\ \hline
$\pi^- {\bar K}^0$  & $- \frac{3}{\sqrt{2}} \left( 1 - \frac{\epsilon_2}{\sqrt{3}} \right)$  & 
$- \frac{1}{\sqrt{2}} \left( 1 + \sqrt{3} \epsilon_2 \right)$
\\ \hline
$\eta K^-$  & $- \frac{\sqrt{3}}{2} \left( 1 - \frac{\epsilon_1}{\sqrt{3}} + \sqrt{3} \epsilon_2 \right)$  &  
$\frac{\sqrt{3}}{2} \left( 1 - \frac{\epsilon_1}{\sqrt{3}} + \sqrt{3} \epsilon_2 \right)$
\\ \hline
\end{tabular}
\caption{The intermediate states that contribute to the one-loop unitarity parts of the
form factors $F^{\mbox{\tiny{$+-$}}} (s,t,u)$ and $G^{\mbox{\tiny{$+-$}}} (s,t,u)$ (the first 10 lines), 
or $F^{\mbox{\tiny{$00$}}} (s,t,u)$ and $G^{\mbox{\tiny{$00$}}} (s,t,u)$ (the first 6 lines and the 3 last ones),
together with the associated form factors at leading order, normalized by
$F_{S [0]}^{\mbox{\tiny{$+-$}}} = M_K/(\sqrt{2} F_0)$, where $F_0$ is the pion
decay constant in the three-flavour chiral limit. In these expressions,
$\epsilon_1$ and $\epsilon_2$ denote the two $\pi^0 - \eta$ mixing angles,
defined in App.~\ref{app:loampl}.}\label{tab:FF_LO}
\end{center}
\end{table}

\subsection{The case $K^+ \to \pi^0 \pi^0 \ell^+ \nu_\ell$}\label{subsec:1-loop_ff_00}

For the decay mode $K^+ \to \pi^0 \pi^0 \ell^+ \nu_\ell$, the expressions of the 
form factors at one loop follow
from the equations (\ref{ff_gen_exp}) and (\ref{As_and_Bs}) up to a change of normalization
\begin{eqnarray}
F^{\mbox{\tiny{$00$}}} (s,t,u) &=& - \frac{M_K}{\sqrt{2} F_0}
\left\{
1 + 2 \sqrt{3} \epsilon_2 +
\frac{1}{F_0^2}
\left[
P_F^{\mbox{\tiny{$00$}}} (s,t,u) + \Delta^{\mbox{\tiny{$00$}}} (s) + 
A^{\mbox{\tiny{$00$}}} (t) + A^{\mbox{\tiny{$00$}}} (u)
\right]
\right\}
,
\nonumber\\
\label{FG_00_1loop}
\\
G^{\mbox{\tiny{$00$}}} (s,t,u) &=& - \frac{M_K}{\sqrt{2} F_0}
\left\{
0 +
\frac{1}{F_0^2}
\left[
P_G^{\mbox{\tiny{$00$}}} (s,t,u) +  
B^{\mbox{\tiny{$00$}}} (t) - B^{\mbox{\tiny{$00$}}} (u) 
\right]
\right\}
,
\nonumber
\end{eqnarray}
with
\begin{eqnarray}
\Delta^{\mbox{\tiny{$00$}}} (s) &=&
- \frac{F_0^2}{F_\pi^2} \sum_{\{ a^\prime b^\prime\}} \frac{1}{{\cal S}_{a^\prime b^\prime}}
\frac{F_{S [0]}^{a^\prime b^\prime}}{F_{S [0]}^{\mbox{\tiny{$+-$}}}}
\left[
16 \pi F_\pi^2 \varphi^{\mbox{\tiny{$00$}};a^\prime b^\prime}_S (0) +
\beta^{\mbox{\tiny{$00$}};a^\prime b^\prime} s + \right]
{\bar J}_{a^\prime b^\prime} (s)
,
\nonumber\\
A^{\mbox{\tiny{$00$}}} (t)  &=&
- \frac{1}{2}
\frac{F_0^2}{F_\pi^2} \sum_{\{ a^\prime b^\prime\}} 
\frac{F_{S [0]}^{a^\prime b^\prime}}{F_{S [0]}^{\mbox{\tiny{$+-$}}}}
\bigg\{
\left[
16 \pi F_\pi^2 \varphi^{\mbox{\tiny{$0K^-$}};a^\prime b^\prime}_S (0) + \beta^{\mbox{\tiny{$0K^-$}};a^\prime b^\prime} t
\right]
{\bar J}_{a^\prime b^\prime} (t)
\nonumber\\
&&
+ \Delta_{a^\prime b^\prime} 
\left(
\frac{G_{P [0]}^{a^\prime b^\prime}}{F_{S [0]}^{a^\prime b^\prime}}
\beta^{\mbox{\tiny{$0K^-$}};a^\prime b^\prime} - 3 \gamma^{\mbox{\tiny{$0K^-$}};a^\prime b^\prime}
\right)
{\bar J}_{a^\prime b^\prime} (t)
\nonumber\\
&&
+ 2
\left[
\frac{G_{P [0]}^{a^\prime b^\prime}}{F_{S [0]}^{a^\prime b^\prime}}
16 \pi F_\pi^2 \varphi^{\mbox{\tiny{$0K^-$}};a^\prime b^\prime}_S (0)
+ 3 \gamma^{\mbox{\tiny{$0K^-$}};a^\prime b^\prime} (t - M_{\pi^0}^2 + M_K^2) 
\right]
K_{a^\prime b^\prime} (t)
\nonumber\\
&&
- 12 \frac{G_{P [0]}^{a^\prime b^\prime}}{F_{S [0]}^{a^\prime b^\prime}} \gamma^{\mbox{\tiny{$0K^-$}};a^\prime b^\prime}
\left[
(3 t + M_{\pi^0}^2 - M_K^2) M_{a^\prime b^\prime} (t) - 3 L_{a^\prime b^\prime} (t)
\right]
\bigg\}\,,
\nonumber\\
B^{\mbox{\tiny{$00$}}} (t)  &=&
- A^{\mbox{\tiny{$00$}}} (t) 
+ 24
\frac{F_0^2}{F_\pi^2} \sum_{\{ a^\prime b^\prime\}}  
\frac{G_{P [0]}^{a^\prime b^\prime}}{F_{S [0]}^{\mbox{\tiny{$+-$}}}} \gamma^{\mbox{\tiny{$0K^-$}};a^\prime b^\prime}
\left[
t  M_{a^\prime b^\prime} (t) -  L_{a^\prime b^\prime} (t)
\right]
.
\end{eqnarray}
\begin{table}[t]
\begin{center}
\begin{tabular}{c|l}
$\pi^0\pi^0$ & $\ \pi^+ \pi^-$, $\pi^0 \pi^0$, $K^+ K^-$, $K^0 {\bar K}^0$, ~$\eta\, \eta$~, $\pi^0 \eta$~
\\
$\pi^0 K^-$ & $\ \pi^0 K^-$, $\pi^- {\bar K}^0$, ~$\eta K^-$
\end{tabular}
\caption{The pairs $\{a,b\}$ (on the left) and the corresponding possible intermediate states $\{a^\prime , b^\prime \}$
(on the right) relevant for the form factors $F^{\mbox{\tiny{$00$}}} (s,t,u)$ and $G^{\mbox{\tiny{$00$}}} (s,t,u)$.}
\label{tab:intermediate_00}
\end{center}
\end{table}

The structure (\ref{FG_00_1loop}) of the form factors, which involve only three
functions $\Delta^{\mbox{\tiny{$00$}}} (s)$, $A^{\mbox{\tiny{$00$}}} (t)$, and $B^{\mbox{\tiny{$00$}}} (t)$,
is a consequence of Bose symmetry [in the notation of Eq.~(\ref{ff_gen_exp}), we 
have $A_F^{\mbox{\tiny{$00$}}} (t) = 2 A^{\mbox{\tiny{$00$}}} (t)$,
$B_F^{\mbox{\tiny{$00$}}} (t) = - A^{\mbox{\tiny{$00$}}} (t)$, $A_F^{\mbox{\tiny{$00$}}} (t) = 0$,
and $B_G^{\mbox{\tiny{$00$}}} (t) = B^{\mbox{\tiny{$00$}}} (t)$, up to the change of
normalization].
Tab.~\ref{tab:intermediate_00} lists the various intermediate states that can contribute
in this case and the associated form factors at lowest order are given in Tab.~\ref{tab:FF_LO}. 
For the lowest-order scattering amplitudes, together with the corresponding parameters
$\beta^{ a b ;a^\prime b^\prime}$, $\gamma^{ a b ;a^\prime b^\prime}$ and $\varphi^{ a b ;a^\prime b^\prime}_S (0)$,
we again refer the reader to App.~\ref{app:loampl}.
The form of the polynomials $P_F^{\mbox{\tiny{$00$}}} (s,t,u)$ and $P_F^{\mbox{\tiny{$00$}}} (s,t,u)$
is also restricted by Bose symmetry, and they have a somewhat simpler structure than in the channel
with two charged pions:
\begin{eqnarray}
{P}_F^{\mbox{\tiny{$00$}}}(s,t,u) &=& \pi^{\mbox{\tiny{$00$}}}_{0,F} \,+\, \pi^{\mbox{\tiny{$00$}}}_{1,F} s 
\,+\,\pi^{\mbox{\tiny{$00$}}}_{2,F} s_{\ell}  
,
\nonumber\\
{P}_G^{\mbox{\tiny{$00$}}}(s,t,u) &=& \pi^{\mbox{\tiny{$00$}}}_{3,G} (t-u) 
.
\label{P_F_and_P_G_00}
\end{eqnarray}
The  subtraction constants $\pi_{i, F/G}^{\mbox{\tiny{$00$}}}$ are discussed in App.~\ref{app:poly}.

\indent

\section{Isospin breaking in the phases of the two-loop form factors}\label{sec:IB_in_phases}
\setcounter{equation}{0}

In this section we address the issue of isospin breaking in the phases of the form factors
$F^{\mbox{\tiny{$+-$}}} (s,t,u)$ and $G^{\mbox{\tiny{$+-$}}} (s,t,u)$, building on the 
results obtained in the previous section, and on the discussion in Sec.~\ref{subsec:phases}.
Since the low-energy $\pi\pi$ scattering amplitudes play a central role in this discussion, we first simplify the
notation. Quantities related to the process $\pi^+\pi^- \to \pi^+\pi^-$ ($\pi^0\pi^0 \to \pi^0\pi^0$) 
will be distinguished by a $+-$ ($00$) superscript or subscript, e.g. 
$\varphi_0^{\mbox{\tiny{$+-$}}}(s) \equiv \varphi_0^{\mbox{\tiny{$+- ; +-$}}}(s)$. For the inelastic
channel $\pi^+\pi^- \to \pi^0\pi^0$, we use the superscript/subscript $x$, so that 
$\varphi_0^{x}(s) \equiv \varphi_0^{\mbox{\tiny{$+- ; 00$}}}(s)$, for instance.
These changes are not only meant to make the notation more compact, but also to
make contact with Ref.~\cite{DescotesGenon:2012gv}, to which we will often refer 
in this section.

In the particular case we study here, the general formulas (\ref{delta_S}) and (\ref{delta_P}) read,
for $4 M_\pi^2 \le s \le (M_K - m_\ell)^2$,
\begin{eqnarray}\label{eq:deltaS}
\delta_S(s,s_\ell) - \delta_0(s) &=&
\sigma(s) \left\{
\left[
\varphi_0^{\mbox{\tiny{$+-$}}}(s) - \stackrel{{ }_{\mbox{\scriptsize{$o$}}}}{\varphi} \stackrel{{\mbox{\tiny{$+-$}}}}{_{\!\!\!\! 0}} \! \! (s)
\right]
+
\left[
\psi_0^{\mbox{\tiny{$+-$}}}(s) - \stackrel{{ }_{\mbox{\scriptsize{$o$}}}}{\psi} \stackrel{{\mbox{\tiny{$+-$}}}}{_{\!\!\!\! 0}} \! \! (s)
\right]
\right\}
\nonumber\\
&&
+
\frac{1}{2}
\sigma_0 (s) \left[
\varphi_0^{x}(s) \, \frac{F_{S[0]}^{00} + F_{S[2]}^{00} (s,s_\ell)}{F_{S[0]}^{\mbox{\tiny{$+-$}}} + F_{S[2]}^{\mbox{\tiny{$+-$}}} (s,s_\ell)} +
\psi_0^{x}(s)\, \frac{F_{S[0]}^{00}}{F_{S[0]}^{\mbox{\tiny{$+-$}}}}
\right]
\nonumber\\
&&
+
\frac{1}{2}
\sigma_0 (s) \left[
\stackrel{{ }_{\mbox{\scriptsize{$o$}}}}{\varphi} \stackrel{x}{_{0}} (s) + 
\stackrel{{ }_{\mbox{\scriptsize{$o$}}}}{\psi} \stackrel{x}{_{0}} (s) \right]
+ {\cal O} (E^6)
,
\end{eqnarray}
and
\begin{eqnarray}\label{eq:deltaP}\label{eq:isospinbreakingP}
\delta_P(s) - \delta_1(s) &=&
\sigma(s) \left\{
\left[
\varphi_1^{\mbox{\tiny{$+-$}}}(s) - \stackrel{{ }_{\mbox{\scriptsize{$o$}}}}{\varphi} \stackrel{{\mbox{\tiny{$+-$}}}}{_{\!\!\!\! 1}} \! \! (s)
\right]
+
\left[
\psi_1^{\mbox{\tiny{$+-$}}}(s) - \stackrel{{ }_{\mbox{\scriptsize{$o$}}}}{\psi} \stackrel{{\mbox{\tiny{$+-$}}}}{_{\!\!\!\! 1}} \! \! (s)
\right]
\right\}
+ {\cal O} (E^6)
.
\end{eqnarray}
Here $\sigma (s)$ and $\sigma_0 (s)$ stand for the phase-space factors for two charged or two
neutral pions, respectively:

\begin{equation}
\sigma (s) = \sqrt{1 - \frac{4 M_\pi^2}{s}} \,, 
\quad
\sigma_0 (s) = \sqrt{1 - \frac{4 M_{\pi^0}^2}{s}}
,
\end{equation}
and for any quantity $A$, $\stackrel{{ }_{\mbox{\scriptsize{$o$}}}}{\!A}$ denotes its counterpart in the isospin limit.
Using $F_{S [0]}^{\mbox{\tiny{$00$}}}/F_{S [0]}^{\mbox{\tiny{$+-$}}} = - (1 + 2 \sqrt{3} \epsilon_2)$,
cf. Tab.~\ref{tab:FF_LO}, and writing
\begin{equation}
F_{S[2]}^{\mbox{\tiny{$+-$}}} (s,s_\ell) = F_{S [0]}^{\mbox{\tiny{$+-$}}} \cdot {\mbox{\large{f}}}_0^{\mbox{\tiny{$+-$}}}(s,s_\ell)\,,\qquad
F_{S[2]}^{00} (s,s_\ell) = - F_{S [0]}^{\mbox{\tiny{$+-$}}} \cdot {\mbox{\large{f}}}_0^{00}(s,s_\ell)
,
\end{equation}
one obtains 
\begin{eqnarray}\label{eq:isospinbreakingS}
\delta_S(s,s_\ell) - \delta_0(s) &=&
\sigma(s) \left\{
\left[
\varphi_0^{\mbox{\tiny{$+-$}}}(s) - \stackrel{{ }_{\mbox{\scriptsize{$o$}}}}{\varphi} \stackrel{{\mbox{\tiny{$+-$}}}}{_{\!\!\!\! 0}} \! \! (s)
\right]
+
\left[
\psi_0^{\mbox{\tiny{$+-$}}}(s) - \stackrel{{ }_{\mbox{\scriptsize{$o$}}}}{\psi} \stackrel{{\mbox{\tiny{$+-$}}}}{_{\!\!\!\! 0}} \! \! (s)
\right]
\right.
\nonumber\\
&&
\left.
-
\frac{1}{2} \left[ \varphi_0^{x}(s) - \stackrel{{ }_{\mbox{\scriptsize{$o$}}}}{\varphi} \stackrel{x}{_{0}} (s) \right] -
\frac{1}{2} \left[ \psi_0^{x}(s) - \stackrel{{ }_{\mbox{\scriptsize{$o$}}}}{\psi} \stackrel{x}{_{0}} (s) \right]
\right\}
\nonumber\\
&&
+
\frac{1}{2} \left[ \sigma (s) - (1 + 2 \sqrt{3} \epsilon_2) \sigma_0 (s) \right]
\left[ \varphi_0^{x}(s) + \psi_0^{x} (s)  \right]
\nonumber\\
&&
+
\frac{1}{2}
\sigma_0 (s) \varphi_0^{x}(s) 
\left[ (1 + 2 \sqrt{3} \epsilon_2) {\mbox{\large{f}}}_0^{\mbox{\tiny{$+-$}}}(s,s_\ell)  - {\mbox{\large{f}}}_0^{00}(s,s_\ell) \right]
+ {\cal O} (E^6)
,
\end{eqnarray}
with
\begin{equation}\label{eq:partialwaveprojectionf}
{\mbox{\large{f}}}_0^{ab}(s,s_\ell) = \frac{1}{2}
\int_{-1}^{+1} d(\cos\theta_{ab}) \, \frac{1}{F_0^2} 
\left[
{\mbox{Re}} U_F^{ab} (s,t,u) + 
\frac{M_K^2 - s - s_\ell}{s} \frac{\lambda^{\frac{1}{2}}_{ab} (s)}{\lambda^{\frac{1}{2}}_{\ell K} (s)} 
\cos \theta_{ab} \, {\mbox{Re}}U_G^{ab} (s,t,u)
\right]
,
\end{equation}
for $ab=+-,00$. $U_{F/G}^{\mbox{\tiny{$+-$}}} (s,t,u)$ 
represent the combinations of terms $\Delta_{F/G}^{\mbox{\tiny{$+-$}}} (s) +  
A_{F/G}^{\mbox{\tiny{$+-$}}} (t) + B^{\mbox{\tiny{$+-$}}} (t) - B^{\mbox{\tiny{$+-$}}} (u)$
in the two expressions of Eq.~(\ref{FG_+-_1loop}), and 
likewise the case $ab=00$ involves the two combinations $\Delta^{\mbox{\tiny{$00$}}} (s) + 
A^{\mbox{\tiny{$00$}}} (t) + A^{\mbox{\tiny{$00$}}} (u)$ (for $U_F^{\mbox{\tiny{$00$}}}$) and  
$B^{\mbox{\tiny{$00$}}} (t) - B^{\mbox{\tiny{$00$}}} (u)$ (for $U_G^{\mbox{\tiny{$00$}}}$)
in the expressions of Eq.~(\ref{FG_00_1loop}). Notice also that 
$\lambda^{\frac{1}{2}}_{\mbox{\tiny{$+-$}}} (s) = s \sigma (s)$, while $\lambda^{\frac{1}{2}}_{\mbox{\tiny{$00$}}} (s) = s \sigma_0 (s)$.
The variables $s,t,u$ and $\cos \theta_{ab}$ are related, in each case, by a relation given in Eq.~(\ref{cos_theta}),
where $M_a = M_b$ $( = M_\pi$ or $M_{\pi^0}$).

In agreement with Sec.~V in Ref.~\cite{DescotesGenon:2012gv}, we see that isospin-breaking effects take place in the $S$-wave phase shift through two types of contributions: the first two lines in Eq.~(\ref{eq:isospinbreakingS}) are universal as they depend only on $\pi\pi$ (re)scattering, whereas the last two are process-dependent as  they involve isospin-breaking in the $K_{\ell 4}$ form factors. For the third term, this dependence is not as explicit as for the last one, but one should recall that the factor $- (1 + 2 \sqrt{3} \epsilon_2)$ originates from the ratio
$F_{S [0]}^{\mbox{\tiny{$00$}}}/F_{S [0]}^{\mbox{\tiny{$+-$}}}$, as shown in Tab.~\ref{tab:FF_LO}.
This is in contrast with the scalar form factors considered in Ref.~\cite{DescotesGenon:2012gv},
where the corresponding ratio was equal to unity at lowest order. On the other
hand,  isospin breaking in the $P$-wave phase shift Eq.~(\ref{eq:deltaP}) is universal, in relation with our comments in Sec.~\ref{subsec:phases} concerning the presence of a single intermediate $\pi\pi$ state at low energy in this channel, even in presence of isospin breaking.

In order to relate the data from $K^\pm_{e4}$ decays to the $\pi\pi$ phases shifts $\delta_0 (s) - \delta_1 (s)$ in the isospin
limit, we need to evaluate the isospin-breaking correction
\begin{equation}
 \Delta_{\mbox{\scriptsize{IB}}} (s, s_\ell) = \left[ \delta_S(s,s_\ell) - \delta_0(s) \right] - \left[ \delta_P(s) - \delta_1(s) \right]\,,
\label{Delta_IB}
\end{equation}
at next-to-leading order. This requires the determination of the partial-wave projections ${\mbox{\large{f}}}_0^{\mbox{\tiny{$+-$}}}(s,s_\ell)$
and ${\mbox{\large{f}}}_0^{\mbox{\tiny{$00$}}}(s,s_\ell)$ of the $K_{\ell 4}$ form factors on the one hand, and on the other hand, the
$\pi\pi$ partial waves $\varphi_{0,1}^{\mbox{\tiny{$+-$}}}(s)$, $\varphi_0^{x}(s) $, $\psi_{0,1}^{\mbox{\tiny{$+-$}}}(s)$, and $\psi_0^{x}(s) $.
We can rely on the results obtained in Ref.~\cite{DescotesGenon:2012gv},
using in particular App.~F therein, to express the latter in terms of the scattering lengths $a_{ab}$ and slope parameters $b_{ab},c_{ab}$ defined in App.~\ref{app:loampl} below.

We thus discuss
the partial-wave projections ${\mbox{\large{f}}}_0^{\mbox{\tiny{$+-$}}}(s,s_\ell)$
and ${\mbox{\large{f}}}_0^{\mbox{\tiny{$00$}}}(s,s_\ell)$ in greater detail. 
Starting with the former,
we rewrite the integration over the angle $\theta_{\mbox{\tiny{$+-$}}}$  in Eq.~(\ref{eq:partialwaveprojectionf}) as an integration over the
variable $t$, so that
\begin{eqnarray}  
F_0^2 \! \cdot {\mbox{\large{f}}}_0^{\mbox{\tiny{$+-$}}}(s,s_\ell) &=& \pi^{\mbox{\tiny{$+-$}}}_{0,F} 
+ \pi^{\mbox{\tiny{$+-$}}}_{1,F} s + \pi^{\mbox{\tiny{$+-$}}}_{2,F} s_{\ell} + {\mbox{Re}}\,\Delta^{\mbox{\tiny{$+-$}}}_F(s)
- \frac{1}{3}\,\left(1 - \frac{4M_{\pi}^2}{s}\right) 
(M_{K}^2 - s - s_\ell) \pi^{\mbox{\tiny{$+-$}}}_{3,G}
\nonumber\\
&&
+
\frac{1}{\sigma (s)\lambda_{\ell K}^{1/2}(s) }
\int_{t_-^c}^{t_+^c} \!\!\! dt\ \Bigg\{
{\mbox{Re}}\,A_{F}^{\mbox{\tiny{$+-$}}}(t) 
+
\frac{( M_{K}^2-s-s_\ell ) (M_{K}^2 + 2 M_{\pi}^2 + s_\ell - s - 2t )}{\lambda_{\ell K} (s)}
\nonumber\\
&&
\qquad\qquad\qquad \times
\left[ {\mbox{Re}}\,A_{G}^{\mbox{\tiny{$+-$}}}(t) + 2 {\mbox{Re}}\,B^{\mbox{\tiny{$+-$}}}(t)  \right]
\Bigg\}
.
\label{f_0_+-}
\end{eqnarray}
The range of integration is given by ${t_-^c} (s,s_\ell) \le t\le {t_+^c} (s,s_\ell)$, where
\begin{equation}
{t_\pm^c} (s,s_\ell) = \frac{1}{2}\,(M_{K}^2 + 2 M_{\pi}^2 + s_\ell -s) \pm
\frac{1}{2}\,\sigma (s)\lambda_{\ell K}^{1/2}(s)
.
\end{equation}

In the channel with two neutral pions, we obtain a similar expression,
\begin{eqnarray}   
F_0^2 \! \cdot {\mbox{\large{f}}}_0^{00}(s,s_\ell) &=& \pi^{\mbox{\tiny{$00$}}}_{0,F} + \pi^{\mbox{\tiny{$00$}}}_{1,F} s 
+\pi^{\mbox{\tiny{$00$}}}_{2,F} s_{\ell} + {\mbox{Re}}\,\Delta^{\mbox{\tiny{$00$}}}(s)
- \frac{1}{3}\,\left(1 - \frac{4M_{\pi^0}^2}{s}\right) 
(M_{K}^2 - s - s_\ell) \pi^{\mbox{\tiny{$00$}}}_{3,G}
\\
&&
+
\frac{2}{\sigma_{0}(s)\lambda_{\ell K}^{1/2}(s) }
\int_{t_-^n}^{t_+^n} \!\!\! dt \Bigg\{
{\mbox{Re}}\,A^{\mbox{\tiny{$00$}}}(t) \,+\,
\frac{( M_{K}^2-s-s_\ell ) ( M_{K}^2 + 2 M_{\pi^0}^2 + s_\ell - s - 2t )}{\lambda_{\ell K}(s)}
\, {\mbox{Re}}\,B^{\mbox{\tiny{$00$}}}(t)  
\Bigg\} 
,
\quad\ { }
\nonumber
\label{f_0_00}
\end{eqnarray}
The range of integration is now given by ${t_-^n} (s,s_\ell) \le t \le {t_+^n} (s,s_\ell)$, where
\bea
{t_\pm^n} (s,s_\ell) = \frac{1}{2}\,(M_{K}^2 + 2 M_{\pi^0}^2 + s_\ell -s) \pm
\frac{1}{2}\,\sigma_{0}(s)\lambda_{\ell K}^{1/2}(s)
.
\eea

In order to perform the partial-wave projection, we need to evaluate the integrals
in Eqs.~(\ref{f_0_+-}) and in (\ref{f_0_00}). They involve the function ${\bar J}_{ab}(s)$ and the other
loop functions defined in Eq.~(\ref{loop_functions}). The indefinite integrals necessary
in order to perform this step have been collected in App.~\ref{app:indefinite}.
We do not give the resulting explicit expressions for ${\mbox{\large{f}}}_0^{\mbox{\tiny{$+-$}}}(s,s_\ell)$
and for ${\mbox{\large{f}}}_0^{00}(s,s_\ell)$, since they are rather lengthy and do not present any
particular interest as such.

\section{Numerical analysis}\label{sec:numerics}
\setcounter{equation}{0}

\subsection{Isospin breaking in the one-loop phases}

Before collecting all the above elements to assess the isospin-breaking correction 
$\Delta_{\rm IB} (s, s_\ell)$  for the phases
up to two loops, it is interesting to study the lowest-order case first from Eqs.~(\ref{eq:isospinbreakingP}) and (\ref{eq:isospinbreakingS})
 \begin{eqnarray}
 \Delta_{\mbox{\scriptsize{IB}}}^{\mbox{\scriptsize{LO}}} (s) &=&
\sigma(s) \left\{
\left[
\varphi_0^{\mbox{\tiny{$+-$}}}(s) - \stackrel{{ }_{\mbox{\scriptsize{$o$}}}}{\varphi} \stackrel{{\mbox{\tiny{$+-$}}}}{_{\!\!\!\! 0}} \! \! (s)
\right]
-
\frac{1}{2} \left[ \varphi_0^{x}(s) - \stackrel{{ }_{\mbox{\scriptsize{$o$}}}}{\varphi} \stackrel{x}{_{0}} (s) \right]
\right\}
\nonumber\\
&&
+
\frac{1}{2} \left[ \sigma (s) - (1 + 2 \sqrt{3} \epsilon_2) \sigma_0 (s) \right]
\varphi_0^{x}(s)
 -\sigma(s)
\left[
\varphi_1^{\mbox{\tiny{$+-$}}}(s) - \stackrel{{ }_{\mbox{\scriptsize{$o$}}}}{\varphi} \stackrel{{\mbox{\tiny{$+-$}}}}{_{\!\!\!\! 1}} \! \! (s)
\right]
+ {\cal O} (E^4)
.
\label{Delta_IB_LO}
\end{eqnarray}
At this order, there is no isospin breaking in the $P$ wave,
$\varphi_1^{\mbox{\tiny{$+-$}}}(s) = \stackrel{{ }_{\mbox{\scriptsize{$o$}}}}{\varphi} \stackrel{{\mbox{\tiny{$+-$}}}}{_{\!\!\!\! 1}} \! \! (s)$. Using the expressions given in Eq.~(\ref{IB_in_a_and_b_LO}), one has simply
%
%

\begin{figure}[t]
\center\epsfig{figure=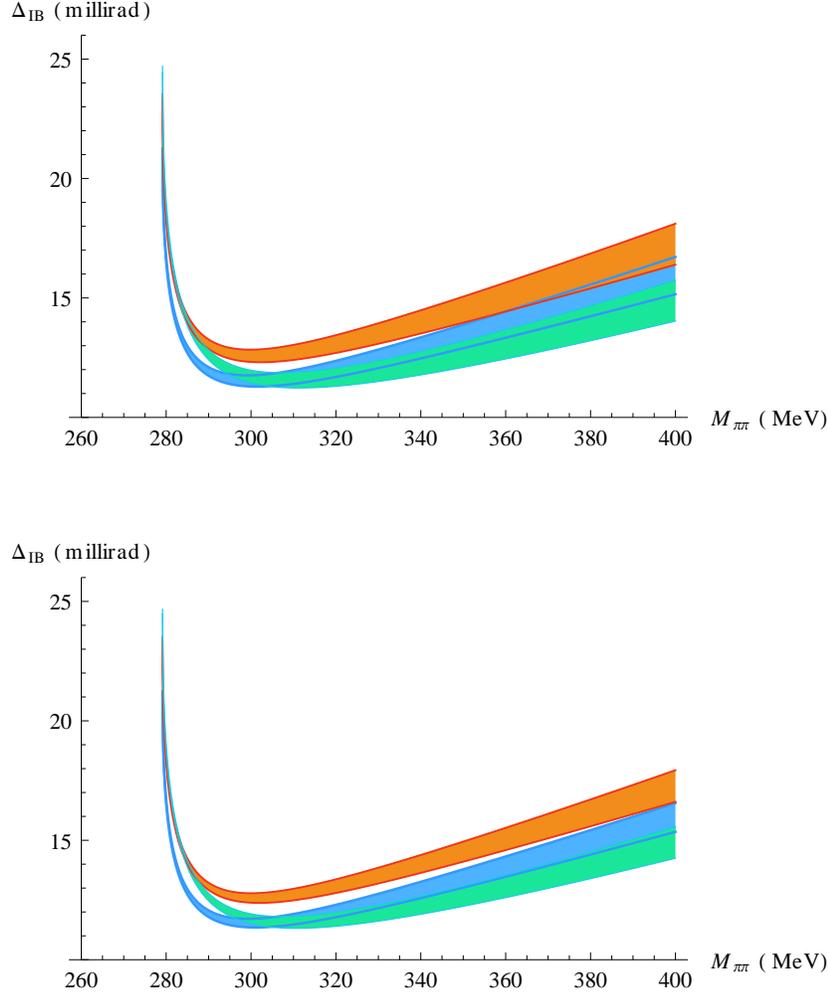,height=14.0cm}
\caption{Isospin breaking in the phase of the one-loop form factors, $\Delta_{\mbox{\scriptsize{IB}}}^{\mbox{\scriptsize{LO}}} (s)$,
as a function of the dipion invariant mass $M_{\pi\pi}=\sqrt{s}$, for different values of the scattering lengths.
The middle (light-blue) band corresponds to $(a_0^0,a_0^2) = ({\bar a}_0^0,{\bar a}_0^2)$
for which $\Delta_{\mbox{\scriptsize{IB}}}^{\mbox{\scriptsize{LO}}} (s) = {\bar\Delta}_{\mbox{\scriptsize{IB}}}^{\mbox{\scriptsize{LO}}} (s)$ (we have used
$F = 86.2$ MeV, i.e. $({\bar a}_0^0,{\bar a}_0^2)=(0.182,-0.052)$),
whereas the other two cases shown correspond to $(a_0^0,a_0^2) = (0.205, -0.055)$ (upper orange band)
and to $(a_0^0,a_0^2) = (0.24, -0.035)$ (lower green band). The widths of these bands result from the
uncertainty on the quark mass ratio $R$. In the upper part of the figure, we have taken the value $R=37 \pm 5$
considered in Ref.~\cite{Colangelo:2008sm}, while the lower part corresponds to the more recent determination $R = 36.6 \pm 3.8$ from a review of lattice results~\cite{Colangelo:2010et}.
\label{fig_IB_in_phase_LO}}
\end{figure}

\begin{table}[t]
\begin{center}
\begin{tabular}{c|c|c}
$M_{\pi\pi}$ (MeV) & $\bar\Delta^{\mbox{\scriptsize{LO}}}_{\mbox{\scriptsize{IB}}}$ for $F=86.2$ MeV & $\bar\Delta^{\mbox{\scriptsize{LO}}}_{\mbox{\scriptsize{IB}}}$ for $F=F_\pi$\\
\hline
286.06 & 12.53 & 10.91\\
295.95 & 11.55 & 10.05\\
304.88 & 11.54 & 10.04\\
313.48 & 11.75 & 10.23\\
322.02 & 12.06 & 10.50\\
330.80 & 12.43 & 10.82\\
340.17 & 12.86 & 11.19\\
350.94 & 13.37 & 11.64\\
364.57 & 14.04 & 12.22\\
389.95 & 15.31 & 13.33
\end{tabular}
\end{center}
\caption{Leading-order value of the isospin-breaking correction $\bar\Delta^{\mbox{\scriptsize{LO}}}_{\mbox{\scriptsize{IB}}}(s)$ (obtained for $a_0^0=\bar{a}_0^0,a_0^2=\bar{a}_0^2$) as a function of the dipion invariant mass $M_{\pi\pi}=\sqrt{s}$, for two choices of $F$ and for $R=37$. All phase shifts are given in milliradians.}
\label{tab:DeltaLO}
\end{table}

\begin{figure}[b]
\center\epsfig{figure=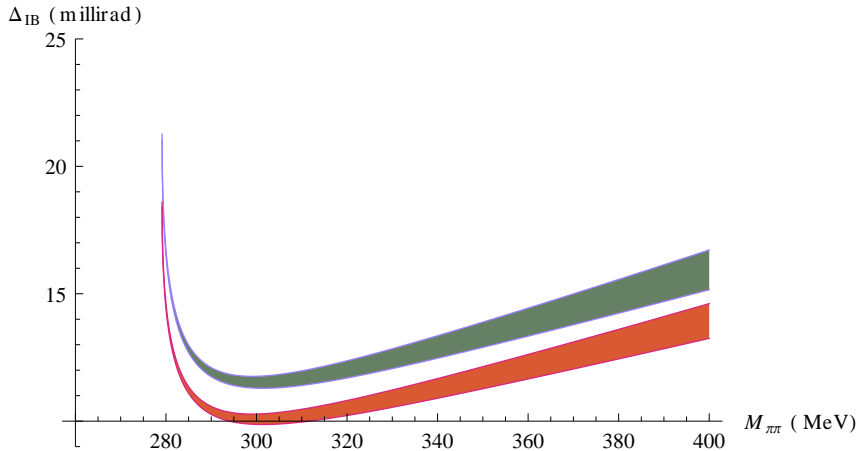,height=6.0cm}
\caption{Isospin breaking in the phase of the one-loop form factors, ${\bar\Delta}_{\mbox{\scriptsize{IB}}}^{\mbox{\scriptsize{LO}}} (s)$ for $(a_0^0,a_0^2) = ({\bar a}_0^0,{\bar a}_0^2)$,
but with $F = 82.2$ MeV (upper dark green band) and $F=92.2$ MeV (lower red band),
as a function of the dipion invariant mass $M_{\pi\pi}=\sqrt{s}$. The widths of these bands results from the
uncertainty on the quark mass ratio $R=37 \pm 5$.
\label{fig_IB_in_phase_LO_Fpi}}
\end{figure}

%
%
\begin{eqnarray}
 \Delta_{\mbox{\scriptsize{IB}}}^{\mbox{\scriptsize{LO}}} (s) &=&
\frac{1}{2} \sigma(s)
\left[ - \frac{2}{3} a_0^0 + \frac{2}{3} a_0^2 - 4 a_0^2 \frac{\Delta_\pi}{M_\pi^2} -\frac{1}{12} (2a_0^0-5a_0^2) 
\frac{s - 4 M_\pi^2}{ M_\pi^2} \right]
\nonumber\\
&&
- \frac{1}{2} \sigma_0(s) (1 + 2 \sqrt{3} \epsilon_2)
\left[ - \frac{2}{3} a_0^0 + \frac{2}{3} a_0^2 + a_0^2 \frac{\Delta_\pi}{M_\pi^2}-\frac{1}{12} (2a_0^0-5a_0^2)\frac{s - 4 M_\pi^2}{ M_\pi^2}\right]
.
\label{Delta_IB_LO_expl}
\end{eqnarray}
We may contrast this expression with the result of Ref.~\cite{Colangelo:2008sm}. Our expression agrees with
the result obtained there provided that we consider Eq.~(\ref{Delta_IB_LO_expl}) and replace the scattering
lengths $a_0^0$, and $a_0^2$ with ${\bar a}_0^0$, and ${\bar a}_0^2$, their tree-level values according to $N_f=2$ $\chi$PT,
\begin{equation}
{\bar a}_0^0 = \frac{7 M_\pi^2}{32 \pi F^2} \, , \ {\bar a}_0^2 = - \frac{M_\pi^2}{16 \pi F^2}
,
\end{equation}
and furthermore take $\epsilon_2 = \sqrt{3}/(4R)$ (see App.~\ref{app:poly}). Explicitly, this gives
\begin{equation}\label{eq:DeltaIBCGR}
 {\bar\Delta}_{\mbox{\scriptsize{IB}}}^{\mbox{\scriptsize{LO}}} (s) =
\frac{1}{32 \pi F^2} \left[
\sigma (s) \left( M_\pi^2 + 4 \Delta_\pi - s \right) - \sigma_0 (s) \left(M_\pi^2 - \Delta_\pi -s \right)
\left(1 + \frac{3}{2R} \right)
\right].
\end{equation}
In the notation of \cite{Colangelo:2008sm}, this last formula corresponds to the difference
$\psi_0 - \delta_0$, with $\psi_0$ given by Eq.~(6.1) of \cite{Colangelo:2008sm}.
In these expressions, $F$ is the value of the pion decay constant in the $N_f=2$ chiral limit, which is expected to be smaller than $F_\pi$ on general grounds~\cite{DescotesGenon:1999uh,DescotesGenon:1999zj},
and for which estimates from lattice simulations are available~\cite{Colangelo:2010et}.
For the sake of comparison with Ref.~\cite{Colangelo:2008sm}, we adopt the values given there: $F=86.2 \pm 0.5$ MeV
and $R = 37 \pm 5$. In Eq.~(\ref{eq:DeltaIBCGR}), we notice the presence
of the ratio of normalisations $F_{S [0]}^{\mbox{\tiny{$00$}}}/F_{S [0]}^{\mbox{\tiny{$+-$}}} =-(1+3/(2R))$, which in Refs.~\cite{Colangelo:2008sm,Colangelo:2007df}
is seen as arising from $\pi^0 - \eta$ mixing induced by the difference $m_d - m_u$.

In Fig.~\ref{fig_IB_in_phase_LO} we show the value of $\Delta_{\mbox{\scriptsize{IB}}}^{\mbox{\scriptsize{LO}}}$
as a function of the dipion invariant mass $M_{\pi\pi} = \sqrt{s}$ for different values of $a_0^0$ and of $a_0^2$,
with the uncertainty coming from the ratio of quark masses $R$. Even taking into account this uncertainty
we see differences in the isospin-breaking corrections for different values of $a_0^0$ and $a_0^2$.
For the tree-level $N_f=2$ $\chi$PT values $(a_0^0,a_0^2) = ({\bar a}_0^0,{\bar a}_0^2)$, we obtain ${\bar\Delta}_{\mbox{\scriptsize{IB}}}^{\mbox{\scriptsize{LO}}} (s)$ in agreement
with Ref.~\cite{Colangelo:2008sm} if we adopt the same choice for $F$. Let us however emphasise that this  choice has a significant impact on the correction at high dipion invariant mass $M_{\pi\pi}$, as shown in Fig.~\ref{fig_IB_in_phase_LO_Fpi}
and in Tab.~\ref{tab:DeltaLO} for the central values of the bins discussed in Ref.~\cite{Batley:2010zza}
(as pointed out in \cite{Colangelo:2008sm}, ${\bar\Delta}_{\mbox{\scriptsize{IB}}}^{\mbox{\scriptsize{LO}}} (s) \propto s/F^2$
for $s\gg M_\pi^2$).

\subsection{Isospin breaking in the two-loop phases}\label{sec:IBtwoloops}

Let us now discuss the two-loop phases.
The previous sections have provided all the elements needed to compute the isospin-breaking corrections
encoded in Eqs.~(\ref{eq:isospinbreakingP}) and (\ref{eq:isospinbreakingS}).
In these expressions, we distinguish universal contributions coming from $\pi\pi$ (re)scattering, 
denoted by the functions $\varphi$ and $\psi$,
and process-dependent contributions coming from $K_{\ell 4}$ form factors. Following App.~F in 
Ref.~\cite{DescotesGenon:2012gv}, the former are expressed in 
terms of scattering parameters $a_{ab},b_{ab},c_{ab}$ and higher-order subthreshold parameters
 $\lambda^{(1,2)}_{ab}$. The latter are expressed in terms of subtraction constants for $K_{\ell 4}$ form factors 
and of projection integrals of the parts generated by the unitarity condition, which are  expressed  in turn in terms of 
ratios of leading-order form factors $F_{S[0]}/F^{+-}_{S[0]}$ and $G_{S[0]}/F^{+-}_{S[0]}$, and parameters of scattering amplitudes $\beta^{ab;a'b'}$, $\gamma^{ab;a'b'}$ and $\varphi_S^{ab;a'b'}(0)$ corresponding to the various two-meson intermediate states (other than $\pi\pi$) allowed by unitarity at low energy.
 
Taken together, this leaves us with a rather large set of parameters not fixed by the general properties
(unitarity, analyticity, chiral counting) on which we have built our approach. With a very large set of data in the various decay channels for both the moduli and the phases of the form factors, one could 
in principle contemplate the possibility to determine all the various parameters (subtraction constants,
scattering lengths\ldots) involved in these expressions. However, despite the large statistics of the experimental data collected so far, this still remains out of reach at present. Our goal is however more modest, since we mainly wish to keep the dependence on the scattering lengths $a_0^0$ and $a_0^2$ in $\Delta_{\mbox{\scriptsize{IB}}} (s, s_\ell)$. Additional information must be provided on the remaining parameters appearing in this quantity. We briefly describe how we have proceeded in each case.
\begin{itemize}
\item
Concerning the quantities related to the $\pi\pi$ partial-wave projections, 
$\varphi_{0,1}^{\mbox{\tiny{$+-$}}}(s), \varphi_0^{x}(s)$ or $\psi_{0,1}^{\mbox{\tiny{$+-$}}}(s), \psi_0^{x}(s)$, 
their expressions in terms of the corresponding
threshold or subthreshold parameters (e.g. scattering lengths) are given in Sec.~4 and App.~F of Ref.~\cite{DescotesGenon:2012gv}.
We have next related the relevant scattering lengths and slope parameters appearing in these expressions 
to the two $S$-wave scattering $a_0^0$ and $a_0^2$ in the isospin limit using the results of Ref.~\cite{DescotesGenon:2012gv}.
The details of this calculation and the resulting expressions can be found in App.~\ref{app:loampl}.
 \item
 For the other lowest-order two-meson scattering amplitudes contributing to the real parts
of the form factors at one-loop, we have used the expressions shown in Tabs.~\ref{tab:beta_gamma} (in App.~\ref{app:loampl}) and \ref{tab:varphi(0)} (in App.~\ref{app:poly}), 
performing the identifications
\begin{equation}
2\hat{m}B_0 = M_{\pi^0}^2\,, \qquad 
m_sB_0 =M_{K}^2-M_{\pi^0}^2-\Delta_\pi -\frac{\Delta _K}{2}\,, \qquad \Delta_K=M_{K}^2 - M_{K^0}^2\,.
\end{equation}
 This might not look quite at the same level of
generality as in the case of the $\pi\pi$ amplitudes. In some cases, like for instance
$\pi K$ scattering, we could have used instead existing phenomenological information~\cite{Buettiker:2003pp}. However, the
numerical weight of all these contributions is quite small, well below the level of the uncertainties
generated by the other terms, as as we will see below when discussing the numerical results.
Therefore, the necessity to look for more elaborate treatments is not compelling.
 
\item
Finally, as far as the subtraction constants $\pi^{ab}_{i,F/G}$ are concerned,
we assume that $N_f=3$ $\chi$PT provides a reasonably accurate framework
for their determination. In order to proceed in this direction, we first have
to work out the relation of the constants $\pi^{ab}_{i,F/G}$ to low-energy constants, tadpole and unitarity contributions
coming from the one-loop expressions of the form factors. This identification
is described in App.~\ref{app:poly}, where we also provide explicit expressions
of the subtraction constants in terms of low-energy constants and of the 
same parameters of scattering amplitudes $\beta^{ab;a'b'}$, $\gamma^{ab;a'b'}$ and $\varphi_S^{ab;a'b'}(0)$
mentioned above. 
\end{itemize}

The numerical values  for the various low-energy constants that occur in all these
expressions are given in Tab.~\ref{tab:LECs}. 
For the strong low-energy constants $L_i$, we take as estimates the 
central values of the so-called $O(p^4)$ fit in Refs.~\cite{Bijnens:2002hp,Bijnens:2011tb}, and 
assign an uncertainty of $\pm 0.5\cdot 10^{-3}$ to each of the low-energy constants.
For the electromagnetic counterterms $\widehat{k}_i$ and $\widehat{K}_i$, we use resonance estimates obtained in $N_f=2$ and $N_f=3$ $\chi$PT~\cite{Haefeli:2007ey,Ananthanarayan:2004qk}. We recall that we assume 
(as already done in Refs.~\cite{Colangelo:2008sm,DescotesGenon:2012gv}) that the 
low-energy constants $\widehat{k}_i$ and $\widehat{K}_i$
 involved in the theory without virtual photons are identical
to those in the full theory, ${k}_i$ and ${K}_i$. For a discussion of this aspect, we refer the reader
to Ref.~\cite{DescotesGenon:2012gv}. This identification induces a systematic theoretical error whose size
is difficult to assess, but which will be assumed to be small compared to the other sources of uncertainties. 

\begin{table}[t]
\begin{center}\begin{tabular}{c|c|c|c|c|c}
$i$ & $\hat{k}_i(M_\rho)\cdot 10^3$~\cite{DescotesGenon:2012gv}& 
$i$ & $L_i(M_\rho)\cdot 10^{3}$~\cite{Bijnens:2002hp,Bijnens:2011tb} & $i$ &$\widehat{K} _i(M_\rho)\cdot 10^{3}$~\cite{Moussallam:1997xx,Ananthanarayan:2004qk}  \\
\hline
1 & $8.4\pm 2.8$ & 1 &$1.12\pm 0.5$ & 1 &$-2.71\pm 0.90$\\
2 & $3.4\pm 1.2$ & 2 &$1.23\pm 0.5$& 2& $0.69\pm 0.23$\\ 
3 & $2.7\pm 0.9$& 3 & $-3.98\pm 0.5$ & 3& $2.71\pm 0.90$\\
4 & $1.4\pm 0.5$& 4 & $1.50\pm 0.5$& 4 & $1.38\pm 0.46$\\
5 & $-0.8 \pm 6.3$& 9 & $7.0\pm 0.5$ & 5& $11.59\pm 3.86$\\
6 & $3.9 \pm 6.3$&  &  & 6& $2.77\pm 0.92$\\ 
7 & $3.7\pm 6.3$ &  & & 12 & $-4.2\pm 1.4$\\
8 & $-1.3\pm 2.5$ & & & & \\
14 & $-0.4 \pm 6.3$& &&\\
\end{tabular}
\caption{Values of the low-energy constants used for the estimate of the subtraction constants.
For $\widehat{K} _i(M_\rho)$, we take the central values from Refs.~\cite{Moussallam:1997xx,Ananthanarayan:2004qk}, and 
we follow Ref.~\cite{DescotesGenon:2012gv} in assigning a 33\% relative uncertainty to these numbers.  \label{tab:LECs}}
\end{center}
\end{table}

In addition to these inputs, we take for the isospin-breaking quantities~\cite{Colangelo:2010et}
\begin{equation}
R=36.6\pm 3.8\,, \qquad \epsilon_1=\epsilon_2=\frac{\sqrt{3}}{4R}\,,
\end{equation}
where the replacement of $\epsilon_{1,2}$ by their leading-order expression is justified by the smallness
of higher-order corrections~\cite{Gasser:1984gg}.
We also need the values of the $\pi\pi$ subthreshold parameters $\lambda_1$ and $\lambda_2$,
which we take from Ref.~\cite{DescotesGenon:2001tn}
\begin{equation}\label{eq:lambda12}
\lambda_1=(-4.18\pm 0.63) \cdot 10^{-3}\,, \qquad \lambda_2=(8.96\pm 0.12)\cdot 10^{-3}\,.
\end{equation}
In Sec.~\ref{sec:fits}, after having performed a new analysis of NA48/2 data including isospin-breaking corrections,
we will come back to these values.

The above expressions also involve the pseudoscalar decay constant in the $N_f=3$ chiral limit $m_u=m_d=m_s=0$. Despite extensive studies, using either phenomenological information~\cite{Bijnens:2011tb} or lattice simulations~\cite{Colangelo:2010et}, its value remains poorly known. General arguments based on the paramagnetic behaviour of the spectrum of the Dirac operator~\cite{DescotesGenon:1999uh,DescotesGenon:1999zj} indicate that it should be smaller than its counterpart $F$ in $N_f=2$ $\chi$PT, which is evaluated around 85 MeV by lattice simulations. On the other hand, recent fits of data to next-to-next-to-leading-order $\chi$PT suggest values of $F_0$ as low as 65 MeV~\cite{Bijnens:2011tb}, in agreement with results obtained by considering lattice data in the framework of resummed $\chi$PT designed to cope with such low $N_f=3$ chiral order parameters~\cite{Bernard:2010ex,Bernard:2012fw,Bernard:2012ci}. In order to cover the span of possible values, we take
\begin{equation}
F_0=75\pm 10\ {\rm MeV}\,.\label{F_0_value}
\end{equation}
Finally, we have neglected the electron mass in our numerical evaluations ($m_e=0$).

Before considering the full correction $\Delta_{\mbox{\scriptsize{IB}}}$, it is instructive to separate the different contributions to 
\begin{equation}\label{eq:deltafroman0}
\Delta {\mbox{\large{f}}}_0(s,s_\ell)= (1 + 2 \sqrt{3} \epsilon_2) {\mbox{\large{f}}}_0^{\mbox{\tiny{$+-$}}}(s,s_\ell)  - {\mbox{\large{f}}}_0^{00}(s,s_\ell) \,,
\end{equation}
given in Tab.~\ref{tab:Deltafroman} for the central values of the above parameters and taking $a_0^0=0.22$, $a_0^2=-0.045$ and $s_\ell =(M_{K^+}-\sqrt{s})^2/2$ (middle of the allowed range for the dilepton invariant mass). 
The values of the energies
reported in the first column correspond to the central values of the bins in the dipion invariant mass discussed 
in Ref.~\cite{Batley:2010zza}.
One can see that $\Delta {\mbox{\large{f}}}_0$ is dominated by the subtraction constants $\Delta\pi_{0,F}$, $\Delta\pi_{1,F}$ [where $\Delta\pi_{i,F/G}=(1 + 2 \sqrt{3} \epsilon_2)\pi_{i,F/G}^{\mbox{\tiny{$+-$}}} - \pi_{i,F/G}^{00}$] and by
the combination of $s$-dependent functions 
$\Delta_S (s) = (1 + 2 \sqrt{3} \epsilon_2) {\mbox{Re}}\,\Delta^{\mbox{\tiny{$+-$}}}_F(s) - {\mbox{Re}}\,\Delta^{\mbox{\tiny{$00$}}}(s)$, whereas the cumbersome integrals over $t$ are negligible.
The main contributions to $\Delta_S (s)$
come from the $\pi^0\pi^0$ and $\pi^+\pi^-$ channels, and the $\pi^0\eta$ channel, to a lesser extent. 
The structure of the subtraction constants $\Delta\pi_{0,F}$ and $\Delta\pi_{1,F}$ can be examined thoroughly through their analytic expressions in App.~\ref{app:poly}, and 
numerical investigation shows that they are mainly sensitive to the input values of $L_1$, $L_2$ and $L_3$. 
For the tadpole and unitarity contributions,  in the case of $\Delta\pi_{0,F}$,
the largest contribution comes from $\gamma^{{\mbox{\tiny{$+-$}}}; {\mbox{\tiny{$K^+K^-$}}}}$, with several others (related to $\pi\pi\to\pi\pi$ and $\pi\pi\to \pi\eta$ scattering) smaller but of comparable size. In the case of $\Delta\pi_{1,F}$, the main individual contributions
come from $\beta^{{\mbox{\tiny{$+K^-$}}};{\mbox{\tiny{$\eta K^0$}}}}$ and $\varphi_S^{{\mbox{\tiny{$0 K^+$}}}; {\mbox{\tiny{$\eta K^+$}}}}(0)$, largely canceling each other.

A comment is in order concerning the sensitivity of $\Delta {\mbox{\large{f}}}_0$ on the dilepton invariant mass $s_\ell$, as this is the only place where this kinematic variable will enter the isospin-breaking correction to the phases.
A first dependence comes from the polynomial  terms involving $\Delta \pi_{2,F}$ and $\Delta\pi_{3,G}$. The contribution proportional to $\Delta \pi_{2,F}$  vanishes  at $s_\ell=0$ and remains very small all over phase space.
For the values of $a_0^0$, $a_0^2$ and the inputs considered here, one has 
$\Delta\pi_{2,F}\simeq (-1.5\ {\rm GeV}^{-2})\times \Delta\pi_{0,F} $, which indicates that for the physical range of the dilepton invariant mass $0\leq s_\ell\leq (M_{K^+}-\sqrt{s})^2$, 
the contribution from $\Delta \pi_{2,F}$ is necessarily much smaller in magnitude than that from $\Delta \pi_{0,F}$.
The contribution proportional to $\Delta\pi_{3,G}$ is almost constant over the whole phase space (the values in Tab.~\ref{tab:Deltafroman} would be identical for $s_\ell=0$ at the level of accuracy taken here).
A second dependence on $s_\ell$ arises from the integrals in Eqs.~(\ref{f_0_+-}) and (\ref{f_0_00}), with an apparent singularity for $s_\ell\to(M_{K^+}-\sqrt{s})^2$  due to negative powers of $\lambda_{\ell K}(s)$. However, if we expand the integrands around 
$t=(t_+^c+t_-^c)/2$ and $t=(t_+^n+t_-^n)/2$ before performing
the integrals in Eqs.~(\ref{f_0_+-}) and (\ref{f_0_00}) respectively,
we can show easily that only positive powers of $\lambda_{\ell K}(s)$ arise in $\Delta {\mbox{\large{f}}}_0$, with no singularities in $s_\ell$.
The dependence of $\Delta {\mbox{\large{f}}}_0$ on $s_\ell$ turns out to be very mild:
in the following, we will always set $s_\ell=0$ -- taking $s_\ell=(M_{K^+}-\sqrt{s})^2$ would affect our result for the isospin-breaking correction by much less than 1\%, far less than the uncertainties induced by our inputs.
The general situation described here remains unchanged for other values of the scattering lengths $a_0^0$ and $a_0^2$
taken in a reasonable range (see below) around the values $a_0^0=0.22$, $a_0^2=-0.045$ considered for this analysis.

\begin{table}[t]
\begin{center}\begin{tabular}{c|c|c|c|c|c|c|c}
$M_{\pi\pi}$ (MeV) & $\Delta \pi_{0,F}$ & $\Delta\pi_{1,F}$& $\Delta \pi_{2,F}$
& $\Delta_S$ & $\Delta \pi_{3,G}$ & $\Delta \int_{t_-}^{t_+}$ & $\Delta {\mbox{\large{f}}}_0$\\
\hline
286.06 & 0.0195 & 0.0218 & -0.0006 & 0.0144 & 0.0025 & 0.0046 & 0.0618\\
295.95 & 0.0195 & 0.0233 & -0.0006 & 0.0137 & 0.0022 & 0.0046 & 0.0624\\
304.88 & 0.0195 & 0.0247 & -0.0005 & 0.0130 & 0.0019 & 0.0046 & 0.0630\\
313.48 & 0.0195 & 0.0261 & -0.0005 & 0.0124 & 0.0017 & 0.0046 & 0.0636\\
322.02 & 0.0195 & 0.0275 & -0.0004 & 0.0118 & 0.0015 & 0.0046 & 0.0643\\
330.80 & 0.0195 & 0.0291 & -0.0004 & 0.0111 & 0.0013 & 0.0046 & 0.0650\\
340.17 & 0.0195 & 0.0307 & -0.0003 & 0.0104 & 0.0011 & 0.0046 & 0.0659\\
350.94 & 0.0195 & 0.0327 & -0.0003 & 0.0095 & 0.0009 & 0.0046 & 0.0669\\
364.57 & 0.0195 & 0.0353 & -.00002 & 0.0084 & 0.0007 & 0.0047 & 0.0683\\
389.95 & 0.0195 & 0.0404 & -0.0002 & 0.0062 & 0.0004& 0.0047 & 0.0710\\
\end{tabular}\end{center}
\caption{Contributions to $\Delta {\mbox{\large{f}}}_0(s,s_\ell)$ (in milliradians) at $s_\ell=(M_{K^+}-\sqrt{s})^2/2$, $a_0^0=0.22$, $a_0^2=-0.045$, 
as a function of the dipion invariant mass $M_{\pi\pi}=\sqrt{s}$,
separated into the contributions from the (isospin-breaking difference of) the various subtraction parameters $\pi_{i}$, the purely $s$-dependent functions $\Delta_S$ and the remaining projection integrals. With a slight abuse of notation,
the labels of the intermediate columns denote the terms contributing to Eq.~(\ref{eq:deltafroman0}), which add up
to the final column, i.e. $\Delta {\mbox{\large{f}}}_0$.}
\label{tab:Deltafroman}
\end{table}

We now turn to $\Delta_{\mbox{\scriptsize{IB}}}$, the isospin-breaking corrections to the difference of phase shifts, and focus on the corrections in the $S$-wave, $\delta_s(s,0)-\delta_0(s)$, which can be split into different contributions:
\begin{eqnarray}
\Delta_S[\varphi_0^{\mbox{\tiny{$+-$}}}]&=&\sigma(s)\left[
\varphi_0^{\mbox{\tiny{$+-$}}}(s) - \stackrel{{ }_{\mbox{\scriptsize{$o$}}}}{\varphi} \stackrel{{\mbox{\tiny{$+-$}}}}{_{\!\!\!\! 0}} \! \! (s)
\right]\,,
\qquad\qquad\qquad
\Delta_S[\psi_0^{\mbox{\tiny{$+-$}}}]=\sigma(s)
\left[
\psi_0^{\mbox{\tiny{$+-$}}}(s) - \stackrel{{ }_{\mbox{\scriptsize{$o$}}}}{\psi} \stackrel{{\mbox{\tiny{$+-$}}}}{_{\!\!\!\! 0}} \! \! (s)
\right]\,,\nonumber
\\
\Delta_S[\varphi_0^{x}]&=&-
\frac{\sigma(s)}{2} \left[ \varphi_0^{x}(s) - \stackrel{{ }_{\mbox{\scriptsize{$o$}}}}{\varphi} \stackrel{x}{_{0}} (s) \right]\,,\qquad\qquad\qquad
\Delta_S[\psi_0^{x}]= -
\frac{\sigma(s)}{2} \left[ \psi_0^{x}(s) - \stackrel{{ }_{\mbox{\scriptsize{$o$}}}}{\psi} \stackrel{x}{_{0}} (s) \right]\,,
\nonumber\\
\Delta_S[\sigma\varphi_0^{x}]&=&\frac{1}{2} \left[ \sigma (s) - (1 + 2 \sqrt{3} \epsilon_2) \sigma_0 (s) \right]
\varphi_0^{x}(s)\,,
\quad
\Delta_S[\sigma\psi_0^{x}]=
\frac{1}{2} \left[ \sigma (s) - (1 + 2 \sqrt{3} \epsilon_2) \sigma_0 (s) \right]
\psi_0^{x} (s) \,, \qquad \nonumber
\\
\Delta_S[ {\mbox{\large{f}}}]&=&
\frac{1}{2}
\sigma_0 (s) \varphi_0^{x}(s) 
\left[ (1 + 2 \sqrt{3} \epsilon_2) {\mbox{\large{f}}}_0^{\mbox{\tiny{$+-$}}}(s,s_\ell)  - {\mbox{\large{f}}}_0^{00}(s,s_\ell) 
\right]=\sigma_0 (s) \varphi_0^{x}(s) \Delta {\mbox{\large{f}}}_0(s,s_\ell)\,.
\end{eqnarray}

\begin{table}[t]
\begin{center}\begin{tabular}{c|c|c|c|c|c|c|c|c||c||c}
$M_{\pi\pi}$ (MeV) & $\Delta_S[\varphi_0^{\mbox{\tiny{$+-$}}}]$ &$\Delta_S[\psi_0^{\mbox{\tiny{$+-$}}}]$&
$\Delta_S[\varphi_0^{x}]$&$\Delta_S[\psi_0^{x}]$ & $\Delta_S[\sigma\varphi_0^{x}]$ & $\Delta_S[\sigma\psi_0^{x}]$
& $\Delta_S[ {\mbox{\large{f}}}]$
&
$\delta_s(s,0)-\delta_0(s)$ & $\delta_p(s)-\delta_1(s)$ & $\Delta_{\mbox{\scriptsize{IB}}}$\\
\hline
286.06 & 1.95 & 0.00 & 0.56 & 0.00 & 12.07 & 0.07 & -2.00 & 12.65 &
 0.01 & 12.64\\
295.95 & 3.02 & -0.01 & 0.88 & 0.01 & 9.77 & 0.13 & -2.69 & 11.11 &
 0.03 & 11.08
\\
304.88 & 3.72 & -0.03 & 1.10 & 0.01 & 9.03 & 0.17& -3.28 & 10.72 & 
 0.05 & 10.67
\\
313.48 & 4.29 & -0.05 & 1.27 & 0.02 & 8.72 & 0.21 & -3.85 & 10.61 &
0.08 & 10.53
\\
322.02 & 4.78 & -0.07 & 1.43 & 0.02 & 8.60 & 0.24 & -4.42 & 10.59 & 
0.10 & 10.48
\\
330.80 & 5.24 & -0.10 & 1.58 & 0.03 & 8.61 & 0.27 & -5.02 & 10.61 & 
0.14 & 10.47
\\
340.17 & 5.69 & -0.14& 1.74 & 0.03 & 8.70 & 0.30 & -5.68 & 10.64 & 
0.17 & 10.47\\
350.94 & 6.18 & -0.19 & 1.90 & 0.03 & 8.89 & 0.33 & -6.47 & 10.67 &
0.21 & 10.46\\
364.57 & 6.75 & -0.27 & 2.10 & 0.04 & 9.21 & 0.36 & -7.53 & 10.67 &
0.26  & 10.41\\
389.95 & 7.74 & -0.48 & 2.46 & 0.03 & 9.98 & 0.41 & -9.68 & 10.47  & 
0.36 &  10.11\\
\end{tabular}\end{center}
\caption{Contributions to $\delta_s(s,s_\ell)-\delta_0(s)$ at $s_\ell=0$, $a_0^0=0.22$, $a_0^2=-0.045$,
as a function of the dipion invariant mass $M_{\pi\pi}=\sqrt{s}$.
 All the intermediate columns add up to the ninth one. The two remaining contributions correspond to isospin breaking in $P$-wave phase shifts and the total isospin-breaking correction $\Delta_{\mbox{\scriptsize{IB}}}(s,0)$. All phase shifts are given in milliradians.}
\label{tab:DeltaIBfull}
\end{table}

The result of this splitting for $s_\ell = 0$ and for the same values of $a_0^0$ and $a_0^2$ as before
is given in Tab.~\ref{tab:DeltaIBfull}, which also displays
the $P$-wave contribution and the total isospin-breaking correction $\Delta_{\rm IB} (s , 0)$.
The main contribution can be seen as coming, on the one hand, from pure phase-space effects in $\Delta_S[\sigma\varphi_0^{x}]$, 
which dominates in the low-energy region, 
and on the other hand, from the significant (especially at higher energies) universal contribution 
$\Delta_S[\varphi_0^{\mbox{\tiny{$+-$}}}]$   
and the form-factor dependent one $\Delta_S[ {\mbox{\large{f}}}]$, with opposite signs.
As in the case of the scalar and vector pion form factors discussed in Ref.~\cite{DescotesGenon:2012gv},
we see that the form-factor dependent part tends to decrease the size of the correction, and a significant cancellation 
takes place between the universal and non-universal contributions to isospin breaking in the two-loop phase shifts. Even though the pattern is similar, one should also notice that the individual contributions are more significant in the case of 
the $K_{\ell 4}$ form factor. The contributions to $\Delta_{\rm IB} (s , s_\ell)$ from the $P$-wave term, which are completely 
universal, are very small, in agreement with the results in Ref.~\cite{DescotesGenon:2012gv}. Going away from
$s_\ell = 0$ does not change the above picture, since only the factor $\Delta {\mbox{\large{f}}}_0 (s,s_\ell)$ 
in $\Delta_S[ {\mbox{\large{f}}}]$ introduces an $s_\ell$-dependence which, as we have already seen,
remains well below the noise created by the uncertainties on the other inputs.

\begin{figure}[t]
\center\epsfig{figure=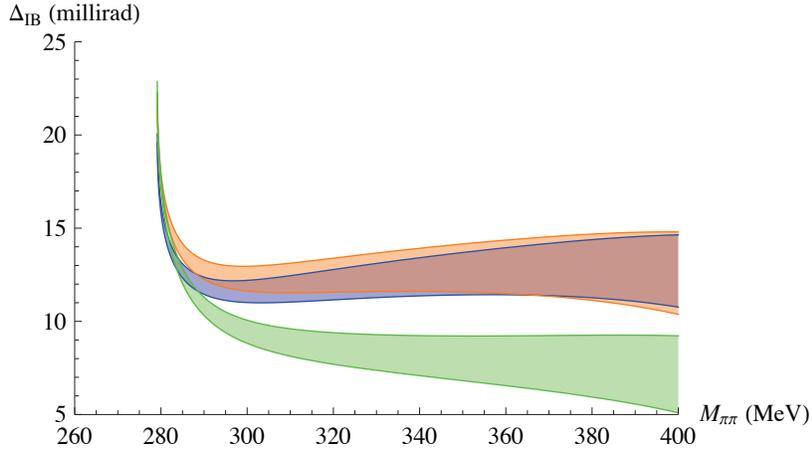,height=6.0cm}
\caption{Isospin breaking in the phase of the two-loop form factors, $\Delta_{\mbox{\scriptsize{IB}}}(s,s_\ell)$
as a function of the dipion invariant mass $M_{\pi\pi}=\sqrt{s}$,
for $s_\ell=0$.
The middle (light-blue) band corresponds to the $(a_0^0,a_0^2) = (0.182,-0.052)$,
whereas the other two cases shown correspond to $(a_0^0,a_0^2) = (0.205, -0.055)$ (upper orange band)
and to $(a_0^0,a_0^2) = (0.24, -0.035)$ (lower green band). The widths of these bands result from the
uncertainty on the various inputs needed at two loops.}
\label{fig_IB_in_phase_NLO_comb}
\end{figure}

Finally, we can discuss how $\Delta_{\mbox{\scriptsize{IB}}}$ varies in the $(a_0^0,a_0^2)$ plane, and how large the uncertainty on this correction is. For different values of $(a_0^0,a_0^2)$ within the so-called Universal Band, we have computed the isospin-breaking difference $\Delta_{\mbox{\scriptsize{IB}}}$, varying the various low-energy constants within their range to combine the resulting uncertainties in quadrature. The corresponding results are quoted in the left-hand part of Tab.~\ref{tab:DeltaIBa00a02plane}. The uncertainty quoted for  $\Delta_{\mbox{\scriptsize{IB}}}$ results mainly from the variation of the three $N_f=3$ strong low-energy constants $L_1,L_2,L_3$ and the two $N_f=2$ electromagnetic constants ${\widehat k}_1$ and ${\widehat k}_3$. The uncertainty on $R$ plays only a minor role. Moreover, the uncertainty on our results is only very weakly dependent on the poorly known value of the decay constant in the three-flavour chiral limit. 
We observe that at large $s$, the correction is reduced compared to the leading-order results illustrated in 
Tab.~\ref{tab:DeltaLO}. This is also illustrated in Fig.~\ref{fig_IB_in_phase_NLO_comb}, which can be compared to the lower plot of Fig.~\ref{fig_IB_in_phase_LO}. An approximate numerical expression for both the central value and the uncertainty of as $\Delta_{\mbox{\scriptsize{IB}}}$ functions of $a_0^0$, $a_0^2$, $s$ and $s_\ell$ can be found in App.~\ref{app:approximate}.

Another systematic effect should also be considered. In the above discussion, we have used the 
one-loop relations Eqs.~(\ref{IB_in_a_+-})-(\ref{IB_in_b_x})
between the scattering parameters $a_{\mbox{\tiny{$+-$}}}$, $a_x$, $b_{\mbox{\tiny{$+-$}}}$, $b_x$, 
and the two scattering lengths $a_0^0$ and $a_0^2$ in the isospin limit, 
and we have implemented the corresponding numerical values in all the places where they occur 
in $\Delta_{\rm IB} (s , s_\ell)$. 
This means that we have included effects belonging to higher
orders in our chiral counting than those considered here. Indeed, at the order that we have been considering, we could have  truncated the isospin-breaking corrections for the scattering parameters at tree level
in all the contributions to $\Delta_{\rm IB}  (s , s_\ell)$ starting at next-to-leading order ($\Delta_S[\psi_0^{\mbox{\tiny{$+-$}}}]$,
$\Delta_S[\psi_0^{x}]$,
$\Delta_S[\sigma\psi_0^{x}]$,
$\Delta_S[ {\mbox{\large{f}}}]$, and the obviously defined 
$\Delta_P[\psi_1^{\mbox{\tiny{$+-$}}}]$), 
keeping the one-loop expressions in the contributions starting at 
leading order ($\Delta_S[\varphi_0^{\mbox{\tiny{$+-$}}}]$,
$\Delta_S[\varphi_0^{x}]$,
$\Delta_S[\sigma\varphi_0^{x}]$, and the obviously defined
$\Delta_P[\varphi_1^{\mbox{\tiny{$+-$}}}]$).
We quote the corresponding central values and uncertainties in the right-hand part of Tab.~\ref{tab:DeltaIBa00a02plane}. 
This truncation of the isospin-breaking contributions to the scattering parameters 
$a_{\mbox{\tiny{$+-$}}}$, $a_x$, $b_{\mbox{\tiny{$+-$}}}$, and $b_x$  has a non-negligible impact on the isospin-breaking 
correction $\Delta_{\rm IB} (s , s_\ell)$ at large $s$. Let us notice that the size of this effect is covered by 
the uncertainties generated in any case by the variation of the inputs. This difference 
is part of higher orders than those considered here. As far as the
differences shown by the two sides of Tab.~\ref{tab:DeltaIBa00a02plane} are representative
for the size of these higher order effects, we could conclude that the next-to-leading approximation
considered here is quite appropriate for the treatment of the data currently available.

Following a similar line, a comparison between Fig.~\ref{fig_IB_in_phase_NLO_comb} and the lower half of Fig.~\ref{fig_IB_in_phase_LO}
provides an indication of the corrections when going from leading to next-to-leading order. As can be seen,
the main impact is a flattening and a diminution of the corrections for large values of $M_{\pi\pi}$.

\begin{table}[t]
\begin{center}
{\footnotesize \begin{tabular}{c||c|c|c|c||c|c|c|c}
& \multicolumn{4}{c||}{One-loop case for $(a_0^0,a_0^2)=$} 
& \multicolumn{4}{c}{Truncated case for $(a_0^0,a_0^2)=$}\\
$M_{\pi\pi}$ (MeV) &  $(0.22,-0.045)$ &
$(0.26,-0.03)$ &
$(0.22,-0.03)$ &
$(0.18,-0.045)$ &
$(0.22,-0.045)$ &
$(0.26,-0.03)$ &
$(0.22,-0.03)$ &
$(0.18,-0.045)$\\
\hline
286.06 & 12.64$\pm$0.43& 11.88$\pm$0.51 & 10.88$\pm$0.37 & 11.56$\pm$0.37
 & 13.22$\pm$0.44 & 12.51$\pm$0.51 & 11.32$\pm$0.38 & 11.97$\pm$0.39\\
295.95 & 11.08$\pm$0.58 & 9.32$\pm$0.68 & 8.92$\pm$0.50 & 10.58$\pm$0.50
 & 11.82$\pm$0.60 & 10.09$\pm$0.66 & 9.48$\pm$0.51 & 11.10$\pm$0.53\\
304.88 & 10.67$\pm$0.70 & 8.33$\pm$0.81 & 8.24$\pm$0.59 & 10.46$\pm$0.60
 & 11.52$\pm$0.72 & 9.20$\pm$0.82 & 8.87$\pm$0.60 & 11.07$\pm$0.64\\
313.48 & 10.53$\pm$0.79 & 7.78$\pm$0.93 & 7.89$\pm$0.67 & 10.52$\pm$0.68
 &11.48$\pm$0.83 & 8.71$\pm$0.94  & 8.58$\pm$0.69 & 11.21$\pm$0.73\\
322.02 & 10.48$\pm$0.89 & 7.41$\pm$1.04 & 7.68$\pm$0.75 & 10.64$\pm$0.76
 & 11.51$\pm$0.93 & 8.38$\pm$1.06 & 8.41$\pm$0.77 & 11.41$\pm$0.81\\
330.80 & 10.47$\pm$0.99 & 7.12$\pm$1.16 & 7.53$\pm$0.84 & 10.78$\pm$0.84
 & 11.59$\pm$1.03 & 8.13$\pm$1.18 & 8.31$\pm$0.86 & 11.62$\pm$0.89\\
340.17 & 10.47$\pm$1.10 & 6.88$\pm$1.30 & 7.42$\pm$0.93 & 10.93$\pm$0.94
 & 11.67$\pm$1.14 & 7.91$\pm$1.32 & 8.24$\pm$0.95 & 11.83$\pm$0.99\\
350.94 & 10.46$\pm$1.24 & 6.64$\pm$1.46 & 7.32$\pm$1.04 & 11.07$\pm$1.05
 & 11.74$\pm$1.28 & 7.68$\pm$1.49 & 8.17$\pm$1.07 & 12.06$\pm$1.10\\
364.57 & 10.41$\pm$1.44 & 6.35$\pm$1.70 & 7.20$\pm$1.21 & 11.19$\pm$1.21
 & 11.79$\pm$1.48 & 7.39$\pm$1.73 & 8.08$\pm$1.24 & 12.29$\pm$1.26\\
389.95 & 10.11$\pm$1.89 & 5.77$\pm$2.24& 6.92$\pm$1.59 & 11.23$\pm$1.58
 & 11.70$\pm$1.94 & 6.77$\pm$2.28 & 7.84$\pm$1.64 & 12.56$\pm$1.64
\end{tabular}}
\caption{Isospin-breaking correction $\Delta_{\mbox{\scriptsize{IB}}}(s,0)$ 
as a function of the dipion invariant mass $M_{\pi\pi}=\sqrt{s}$,
for different values of the $S$-wave scattering lengths, with the uncertainty coming from the variations of our inputs. On the left-hand side, the scattering parameters $a_{\mbox{\tiny{$+-$}}}$, $a_x$, $b_{\mbox{\tiny{$+-$}}}$, $b_x$ and $a_0^0$, $a_0^2$ are computed at one loop, while on the right-hand side, they are computed at one loop for the contributions to $\Delta_{\mbox{\scriptsize{IB}}}(s,0)$ starting at leading order, but at tree level for the contributions starting at next-to-leading order. All phase shifts are given in milliradians.}
\label{tab:DeltaIBa00a02plane}
\end{center}
\end{table}

\section{Re-analysis of NA48/2 results}\label{sec:fits}
\setcounter{equation}{0}

We now come to a particularly interesting application of our present work. 
We will use our computation of the isospin-breaking correction $\Delta_{\rm IB} (s , s_\ell)$ 
as a function of the two scattering lengths $a_0^0$ and $a_0^2$ to perform an
analysis of the available phase shifts from {\bf $K_{e4}^\pm$} decays, as provided by the old Geneva-Saclay
experiment \cite{Rosselet:1976pu}, the BNL-E865 experiment ~\cite{Pislak:2001bf}, 
and finally the quite recent NA48/2 experiment~\cite{Batley:2007zz,Batley:2010zza} at the CERN SPS.
Actually, the high accuracy of the latter analysis dominates completely the discussion, and we will only consider the data coming  from NA48/2 in the following.

We proceed along the lines of Ref.~\cite{DescotesGenon:2001tn}, using the same solutions of the Roy equations in the isospin limit. As discussed extensively
in Ref.~\cite{Ananthanarayan:2000ht}, the Roy equations rely on dispersive representations and data at higher energies to interpolate the phase shifts from threshold
up to a matching point carefully chosen (taken at $s_0=0.8$ GeV) to ensure the unicity of the solution. The solution can be written as
\begin{equation} \label{eq:deltaIl}
\tan \delta^I_l(s)
 =\sqrt{1-\frac{4M_\pi^2}{s}} q^{2l}
 \left\{A^I_l+B^I_l q^2+C^I_l q^4+D^I_l q^6\right\} 
 \left(\frac{4M_\pi^2-s^I_l}{s-s^I_l}\right)\,.
\end{equation}
where the Schenk parametrisation~\cite{Schenk:1991xe} involves coefficients $A^I_l ,B^I_l ,C^I_l ,D^I_l ,s^I_l $ with
a known dependence on the two scattering lengths $a_0^0, a_0^2$ (chosen
as subtraction parameters for the Roy equations) and the phases at the matching point ($\delta_0^0(s_0)\equiv\theta_0, \delta_1^1(s_0)\equiv\theta_1$), which are constrained experimentally.

These expressions were obtained in the isospin limit, but we are now able to supplement them with the expression for isospin-breaking corrections derived in the previous sections.
We can then fit the NA48/2 data on the $S$-$P$ interference through a $\chi^2$ function including  isospin-breaking corrections
\begin{eqnarray}
\chi^2_{S-P}(a_0^0,a_0^2,\theta_0,\theta_1) &=& \sum_i 
\frac{([\delta_0^0-\delta_1^1]^{\rm Roy}(s_i^{\rm exp})
     -[\delta_{\mbox{\scriptsize{s}}}-\delta_{\mbox{\scriptsize{p}}}]_i^{\rm exp}+\Delta_{\mbox{\scriptsize{IB}}}(s_i^{\rm exp}))^2}{(\sigma_i^{\rm exp})^2+\delta\Delta_{\mbox{\scriptsize{IB}}}^2(s_i^{\rm exp})}\nonumber\\
&&\qquad+\left(\frac{\theta_0-82.3^\circ}{3.4^\circ}\right)^2
+\left(\frac{\theta_1-108.9^\circ}{2^\circ}\right)^2\,,
\end{eqnarray}
where we use the same input for the phases at the matching point $\theta_0,\theta_1$ as in Refs.~\cite{Ananthanarayan:2000ht,DescotesGenon:2001tn}.
In view of the discussion of the previous section, it is enough to
evaluate both the central value of the isospin-breaking correction $\Delta_{\mbox{\scriptsize{IB}}}$ and its uncertainty $\delta \Delta_{\mbox{\scriptsize{IB}}}$ at $s_\ell=0$.

Actually, the $S$-$P$ interference from the $K^\pm_{e4}$ angular analysis  provides a strong correlation between $a_0^0$ and $a_0^2$, but a weaker constraint on each of them separately. 
We can circumvent this problem by performing the extended fit described in Ref.~\cite{DescotesGenon:2001tn}, where we supplement the NA48/2 data set with information from the $I=2$ $S$ wave~\footnote{The isospin-breaking corrections attached to the $I=2$ channel cannot be estimated in our framework and are certainly subleading compared to the large uncertainties for this set of data.} in order to constrain each of the two scattering lengths more tightly. The corresponding $\chi^2$ reads
\begin{equation}
\chi^2_{\rm extended}(a_0^0,a_0^2,\theta_0,\theta_1)=\chi^2_{S-P}(a_0^0,a_0^2,\theta_0,\theta_1)+
\sum_j \left(
\frac{(\delta_0^2)^{\rm Roy}(s_j^{\rm exp})
     -(\delta_0^2)_j^{\rm exp}}{\sigma_j^{\rm exp}}\right)^2\,.
\end{equation}
For completeness, we also perform another fit proposed in the literature, using a theoretical constraint from $N_f=2$ $\chi$PT between $a_0^0$ and $a_0^2$, which stems from the scalar radius of the pion~\cite{Colangelo:2000jc}:
\begin{equation}
\chi^2_{\rm scalar}(a_0^0,a_0^2) =
\left(\frac{a_0^2-G(a_0^0)}{.0008}\right)^2
 + \sum_i 
\frac{([\delta_0^0-\delta_1^1]^{\rm Roy}(s_i^{\rm exp})
     -[\delta_{\mbox{\scriptsize{s}}}-\delta_{\mbox{\scriptsize{p}}}]_i^{\rm exp}+\Delta_{\mbox{\scriptsize{IB}}}(s_i^{\rm exp}))^2}{(\sigma_i^{\rm exp})^2+\delta\Delta_{\mbox{\scriptsize{IB}}}^2(s_i^{\rm exp})}\,,
\end{equation}
with $\theta_0$ and $\theta_1$ set at their central values, and with
the constraint described by
\begin{equation}\label{eq:corrscal}
 a_0^2=G(a_0^0)\pm   .0008,\qquad
G(a_0^0)=-.0444 +.236 (a_0^0-.22) -.61 (a_0^0-.22)^2  -9.9(a_0^0-.22)^3\,.
\end{equation}

\begin{figure}[t]
\center\epsfig{figure=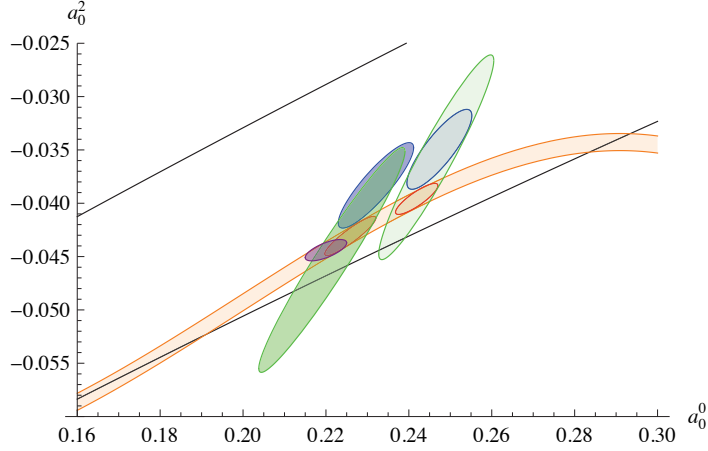,height=6.0cm}
\caption{Results of the fits to the NA48/2 data in the $(a_0^0,a_0^2)$ plane. 
The two black solid lines indicate the universal band where the two $S$-wave scattering lengths 
comply with dispersive constraints (Roy equations) and high-energy data on $\pi\pi$ scattering.
The orange band is the constraint  coming from the scalar radius of the pion given in Eq.~(\ref{eq:corrscal}).
The small dark (purple) ellipse represents 
the prediction based on $N_f=2$ chiral perturbation theory described in Ref.~\cite{Colangelo:2000jc}.
The three other ellipses on the left represent, in order of increasing sizes,
the 1-$\sigma$ ellipses corresponding to the scalar (orange ellipse), $S$-$P$ (blue ellipse) and extended (green ellipse),
respectively, when isospin-breaking corrections are included.
The light-shaded ellipses on the right represent the same outputs, but obtained without including 
isospin-breaking corrections.}
\label{fig:fitres}
\end{figure}

\begin{table}[b]
{\small 
\begin{center}
\begin{tabular}{|c||c|c|c||c|c|c|}
\hline & \multicolumn{3}{|c||}{With isospin-breaking corrections} &
\multicolumn{3}{|c|}{Without isospin-breaking corrections} \\
\hline
& {$S$-$P$}
& {Extended}
& {Scalar} 
& {$S$-$P$}
& {Extended}
& {Scalar}
\\
\hline
$a_0^0$ & $0.221\pm 0.018$ 
              & $0.232\pm 0.009$ 
              & $0.226 \pm 0.007$
              & $0.247\pm 0.014$
              & $0.247\pm 0.008$
               & $0.242\pm$ 0.006\\
$a_0^2$  & $-0.0453\pm 0.0106$ 
              & $-0.0383\pm 0.0040$ 
              & $-0.0431\pm 0.0019$
              & $-0.0357\pm 0.0096$
              & $-0.0349\pm 0.0038$
               & $-0.0396\pm 0.0015$ \\
$\rho_{a_0^0,a_0^2}$ & 0.964
            & 0.881
            & 0.914
            & 0.945
            & 0.842
            & 0.855\\
$\theta_0$ &    $(82.3\pm 3.4)^\circ$
&     $(82.3\pm 3.4)^\circ$
&    $82.3^\circ$
&     $(82.3\pm 3.4)^\circ$
&     $(82.3\pm 3.4)^\circ$
&   $82.3^\circ$
\\
$\theta_1$ &    $(108.9\pm 2)^\circ$
&    $(108.9\pm 2)^\circ$
&    $108.9^\circ$
&    $(108.9\pm 2)^\circ$
&    $(108.9\pm 2)^\circ$
&    $108.9^\circ$
\\
$\chi^2/N$ & 7.6/6
   & 16.6/16
   & 7.8/8
   & 7.2/6
   & 15.7/16
   & 7.3/8
   \\
\hline
 $\alpha$  & $1.043\pm 0.548$ 
              & $1.340\pm 0.231$ 
              & $1.179\pm 0.123$
              & $1.637\pm 0.472$
              & $1.672\pm 0.208$
               & $1.458\pm 0.098$ \\
 $\beta$     & $1.124\pm 0.053$ 
              & $1.088\pm 0.020$ 
              & $1.116\pm 0.007$
              & $1.103\pm 0.055$
              & $1.098\pm 0.021$
               & $1.128\pm 0.008$ \\
 $\rho_{\alpha\beta}$ & 0.47
        & 0.31
        & 0.02
        & 0.47
        & 0.32
        & 0.00\\
 $\lambda_1 \cdot 10^3$  
              & $-3.56\pm 0.68$
              & $-3.80\pm 0.58$ 
              & $-3.89\pm 0.10$ 
              & $-3.79\pm 0.68$
              & $-3.78\pm 0.57$
               & $-3.74\pm 0.11$
               \\
 $\lambda_2 \cdot 10^3$
               & $9.08\pm 0.28$
              & $8.94\pm 0.10$ 
              & $9.14\pm 0.04$ 
              & $9.02\pm 0.23$
              & $9.02\pm 0.11$
               & $9.21\pm 0.42$ \\

 $\lambda_3 \cdot 10^4$ 
              & $2.38\pm 0.18$ 
              & $2.30\pm 0.14$ 
              & $2.32\pm 0.04$
              & $2.34\pm 0.18$
              & $2.34\pm 0.14$
               & $2.41\pm 3.67$ \\
 $\lambda_4 \cdot 10^4 $
              & $-1.46\pm 0.10$ 
              & $-1.39\pm 0.04$ 
              & $-1.45\pm 0.02$
              & $-1.41\pm 0.10$
              & $-1.40\pm 0.04$
               & $-1.46\pm 0.02$ \\
\hline
$\bar\ell_3$
& $3.15\pm 9.9$ 
              & $-10.2\pm 5.7$ 
              & $-2.7\pm 6.6$
              & $-39.9\pm 20.3$
              & $-43.5\pm 19.1$
               & $-19.6\pm 7.8$ \\
$\bar\ell_4$ 
& 5.3$\pm 0.8$ 
              & $4.4\pm 0.6$ 
              & $5.1\pm 0.3$
              & $5.2\pm 0.8$
              & $5.2\pm 0.7$
               & $6.0\pm 0.4$ \\
$X(2)$        
              & $0.88\pm 0.05$ 
              & $0.80\pm 0.06$
              & $0.82\pm 0.02$
              & $0.72\pm 0.05$
              & $0.71\pm 0.05$ 
               & $0.75\pm 0.03$ \\

$Z(2)$         & $0.87\pm 0.03$ 
              & $0.89\pm 0.02$ 
              & $0.86\pm 0.01$
              & $0.87\pm 0.02$
              & $0.87\pm 0.02$
               & $0.85\pm 0.01$ \\
\hline
\end{tabular}
\caption{Scattering lengths, subthreshold parameters and chiral low-energy constants for the different fits considered, with and without the isospin-breaking correction $\Delta_{\mbox{\scriptsize{IB}}}$.}\label{tab:fitresults}
\end{center}
}
\end{table}

The results of these analyses are shown in Fig.~\ref{fig:fitres}, and summarised in Tab.~\ref{tab:fitresults}. We perform the analysis both in presence and in absence of the isospin-breaking correction terms. We obtain results that are in good agreement with the ones from the NA48/2 collaboration for the $S$-$P$ fit (so-called Model B
in Ref.~\cite{Batley:2010zza}: $a_0^0=0.222\pm0.013$ and $a_0^2=-0.043\pm 0.009$) but with slightly larger errors once isospin-breaking corrections are included. This is not surprising since our isospin-breaking correction varies with $a_0^0$ and $a_0^2$. In addition, we notice that the outcome of our fit provides values of $\lambda_1$ and $\lambda_2$ which are compatible with our inputs, see Eq.~(\ref{eq:lambda12}) -- in agreement with the fact that the determination of these two subthreshold parameters has remained very stable over time~\cite{Knecht:1995ai,Colangelo:2001df,DescotesGenon:2001tn}.  We see that in absence of isospin breaking, larger values of $a_0^0$ are preferred, which brings in better agreement the extended fit (including data in $I=2$ channel) with the scalar fit. Once isospin-breaking corrections are included, one recovers the mild discrepancy previously observed between these two kinds of fits~\cite{DescotesGenon:2001tn}, whereas the larger uncertainty of the $S$-$P$ fit covers both solutions.

\begin{table}[t]
{\small 
\begin{center}
\begin{tabular}{|c||c|c|c|c|c|c|}
\hline
& \multicolumn{3}{c|}{Leading order}
& \multicolumn{3}{c|}{Truncated version}
\\
\hline
& {$S$-$P$}
& {Extended}
& {Scalar} 
& {$S$-$P$}
& {Extended}
& {Scalar} 
\\
\hline
$a_0^0$ & $0.221\pm 0.015$
              & $0.227\pm 0.008$
               & $0.220\pm 0.006$
               & $0.219\pm 0.018$
               & $0.230\pm 0.009$
               & $0.224\pm 0.006$ \\
$a_0^2$  & $-0.0435\pm 0.0097$ 
              & $-0.0390\pm 0.0038$ 
               & $-0.0444\pm 0.0017$
               & $-0.0456\pm 0.0104$
               & $-0.0388\pm 0.0040$
               & $-0.0435 \pm  0.0019$ \\
$\rho_{a_0^0,a_0^2}$ & 0.957
    & 0.865
    & 0.903
    & 0.965
    & 0.883
    & 0.918
    \\
$\chi^2/N$ & 8.0/6
   & 16.6/16
   & 8.0/8
   & 7.7/6
   & 16.6/16
   & 7.8/8
   \\
   \hline
\end{tabular}
\caption{Scattering lengths for the different fits considered, using
the leading-order isospin-breaking correction Eq.~(\ref{Delta_IB_LO_expl}), with $R=36.6\pm 3.8$ and $F=86.2\pm 0.5$ MeV, or the truncation discussed at the end of Sec.~\ref{sec:IBtwoloops}. }
\label{tab:LOfits}
\end{center}}
\end{table}

The numbers presented in Tab.~\ref{tab:fitresults} do not include a systematic error due the impact of higher orders. One can add a further systematic error to the fit due to higher orders by exploiting
our discussion of truncation around Tab.~\ref{tab:DeltaIBa00a02plane}. An estimate of higher orders is thus
provided by taking the difference of central values between the full or truncated expressions. If we combine this uncertainty in quadrature with the other uncertainties, it amounts numerically to multiplying $\delta \Delta_{\mbox{\scriptsize{IB}}}$ by a factor of $\simeq\sqrt{2}$, which changes barely the outcome of the above fits, as this error is much smaller than the experimental uncertainties involved.

Another way of assessing the systematics due to higher orders consists in using the one-loop expression of Eq.~(\ref{Delta_IB_LO_expl}) (and the value $R=36.6\pm 3.8$) or performing the truncation  discussed
at the end of Sec.~\ref{sec:IBtwoloops}, leading to the results in Tab.~\ref{tab:LOfits}. We notice that the various fits have slightly different outcomes from the previous results, mainly due to the fact that the isospin-breaking correction at large $s$ is different at one and two loops.
If we focus on the $S$-$P$ fit (equivalent to the Model B fit performed in Ref.~\cite{Batley:2010zza}), the central values for $a_0^0$ and $a_0^2$ shift  by 0.002 and 0.0018 respectively when moving from leading-order to next-to-leading-order  isospin-breaking corrections. This estimate 
can be compared to the impact of higher-order (actually next-to-leading-order) 
contributions to isospin-breaking corrections, quoted in Tab.~5 in Ref.~\cite{Batley:2010zza} as 
0.0035 and 0.0005 for $a_0^0$ and $a_0^2$, respectively. These estimates could be obtained following the $\chi$PT analysis of Ref.~\cite{Colangelo:2008sm}, which proposed  two very different estimations, either by varying the inputs entering the leading-order $\chi$PT estimate of isospin-breaking corrections  (mainly the decay constant, similarly to Fig.~\ref{fig_IB_in_phase_LO_Fpi}, varying $F$ only between 86.2 and 92.2 MeV), or by considering the size of higher-order corrections for the scalar form factor of the pion (and assuming that the same correction applied to $K^+_{e 4}$ form factors). From our own results in Tabs.~\ref{tab:fitresults} and \ref{tab:LOfits}, we agree with the estimate provided in Ref.~\cite{Batley:2010zza} for the systematic uncertainty attached to $a_0^0$ (0.0035), but we consider that the systematics for $a_0^2$ (0.0005) has been underestimated (by a factor 3). Fortunately, the uncertainty on the results of the fit stemming from statistical (experimental) uncertainty is large enough to cover this underestimated systematic uncertainty.
We can pursue this assessment of systematics one order higher by comparing the outcome of the $S$-$P$ fits using the full and truncated versions of the isospin-breaking corrections at next-to-leading order. The scattering lengths $a_0^0$ and $a_0^2$ show a spread narrower than 0.001 and 0.0006 respectively, providing an indication of the systematics attached to higher orders (starting from next-to-next-to-leading order) and suggesting that these effects will affect our results for the scattering lengths in a very marginal way.

Once the scattering lengths in the isospin limit have been determined,
we can test $N_f=2$ $\chi$PT by comparing the dispersive and chiral descriptions of the low-energy $\pi\pi$ amplitude in the isospin limit. First, the solutions of the Roy equations are used to reconstruct the $\pi\pi$ amplitude in the unphysical (subthreshold) region where $\chi$PT should converge particularly well. 
 As explained in Refs.~\cite{Stern:1993rg,Knecht:1995tr} and recalled in Ref.~\cite{DescotesGenon:2001tn}, in the isospin limit, one can describe the $\pi\pi$ amplitude in  terms of only six parameters ($\alpha,\beta,\lambda_1,\lambda_2,\lambda_3,\lambda_4$)
up to and including terms of order $(E/\Lambda_H)^6$ in the low-energy
expansion. These subthreshold parameters yield the $N_f=2$ chiral low-energy constants 
$\bar{\ell}_3,\bar{\ell}_4$, or equivalently the two-flavour 
quark condensate and pion decay constant measured in physical units
\begin{equation}
X(2)=\frac{2m\Sigma(2)}{F_\pi^2M_\pi^2}\,, \qquad Z(2)=\frac{F^2(2)}{F_\pi^2}\,,
\qquad  \Sigma(2)=-\lim_{m_u,m_d\to 0} \langle 0|\bar{u}u|0\rangle\,, \qquad F(2)=\lim_{m_u,m_d\to 0} F_\pi\,,
\end{equation}
by matching the chiral expansions to the subthreshold parameters $\alpha,\beta$. These expansions
are expected to exhibit a good convergence since they involve the $\pi\pi$ scattering amplitude far from  singularities. The corresponding values 
of the subthreshold parameters and of the chiral low-energy constants are gathered in Tab.~\ref{tab:fitresults}.
For comparison, we also show the results obtained
without including the isospin corrections.
One should emphasize that the minor difference in $a_0^2$ between the three fits once isospin-breaking corrections are included is sufficient to yield significant differences in the estimate of the $N_f=2$ chiral order parameters and low-energy constants.

Following the discussion of Ref.~\cite{DescotesGenon:2001tn}, we take as our final results for the reanalysis of the NA48/2 data the first two columns of Tab.~\ref{tab:fitresults}, i.e. the $S$-$P$ fit and the extended fit including isospin-breaking corrections. These two analyses indeed are mostly driven by experiment and do not involve additional theoretical assumptions (contrary to the scalar fit). The extended fit includes $I=2$ data in order to constrain the two $S$-wave scattering lengths more precisely, which leads to a smaller value of $a_0^2$, a negative $\bar\ell_3$ and a smaller two-flavour quark condensate compared to the $S$-$P$ fit. In the two fits, both $X(2)$ and $Z(2)$, i.e. the two-flavour quark condensate and decay constant in physical units, are close to 1 indicating a fast convergence of $N_f=2$ chiral expansions.

\section{Summary and conclusions}\label{sec:conclusion}
\setcounter{equation}{0}

$K_{\ell 4}$ decays provide a very accurate probe of low-energy $\pi\pi$ scattering, and thus constitute a crucial test
of chiral symmetry breaking in the $N_f=2$ chiral limit ($m_u,m_d\to 0$). Powerful methods have been
devised to extract information on the pattern of chiral symmetry breaking from the interference between $S$ and $P$ waves in this 
decay. The Roy equations, based on dispersion relations, allow one to describe $S$- and $P$-wave $\pi\pi$ phase shifts in terms of 
the two $S$-wave scattering lengths $a_0^0$ and $a_0^2$. Solutions to these equations can be constructed using
data at higher energies, for a range of values of the scattering lengths inside the Universal Band.
However, these solutions have been obtained in the isospin limit, while
the level of accuracy reached by recent $K_{\ell 4}$ data requires  isospin-breaking 
corrections to be applied before the above methods can be implemented.

The effects of real and virtual photons are taken into account at the level of the data analyses, 
but the isospin violations induced by the difference between charged and neutral pion masses have to be assessed theoretically. 
This has been done in Chiral Perturbation Theory, with the $S$- and $P$-wave phases accounted for at lowest order \cite{Colangelo:2008sm}.
In principle the result obtained this way can be used only for values of the $S$-wave scattering 
lengths very close to their values predicted by $\chi$PT, which
is precisely to be put under test by the data in question. In order to address the issue 
of isospin-breaking corrections when $a_0^0$ and $a_0^2$ can take values outside this very narrow range, 
one needs a framework where the usual expansion in powers of energy divided by a typical hadronic scale $\Lambda_H \sim 1$ GeV
can be implemented, while keeping at the same time the values of the scattering lengths as
arbitrary parameters, requiring only that they behave dominantly as quantities of order ${\cal O} (E^2/\Lambda_H^2)$
in the low-energy expansion. Such a framework has been presented here. It starts with a dispersive representation
of the form factors with an adequate (from the point of view of the low-energy expansion) number of subtractions.
Implementing the general properties of analyticity, unitarity and crossing, we have shown that this 
dispersive representation allows for an iterative construction of the relevant $K_{\ell 4}$ form factors
up to and including next-to-next-to-leading order in the low-energy expansion. Isospin-breaking effects due to the pion-mass
differences arise naturally, and the resulting expressions depend on a limited set of subtraction
constants and of scattering data corresponding to the various two-meson intermediate states
that can contribute to the unitarity sum in the energy range covered by the $K_{\ell 4}$ decay.

Along with the form factors themselves, our approach also requires the similar iterative construction 
of the $\pi\pi$ scattering amplitudes in various channels, in the presence of isospin breaking. This
aspect had been dealt with in Ref.~\cite{DescotesGenon:2012gv}
[where the same method as here has also been applied to the pion vector and scalar form factors for illustration].
We have gone through the first step of this iteration explicitly, and we have shown that the 
resulting expressions of the $K_{\ell 4}$ form factors $F$ and $G$ can be brought into a form that displays the same
structure, dictated by unitarity and chiral counting, as the calculations performed directly in $\chi$PT at one loop, 
but without the restriction on the scattering lengths that restrains the scope of the latter. We have
briefly indicated how one could proceed from this result, through a second iteration, towards a 
representation of the form factors valid at two loops and accounting for isospin breaking properly.
We have not pursued the matter further in the present work, focusing on the phases of the two-loop form factors. 

First, we have obtained the expression for the phases at leading order for general values of $a_0^0$ and $a_0^2$. It reduces to the results obtained 
previously in $\chi$PT \cite{Colangelo:2008sm} once we identify the two scattering lengths with their tree-level
$\chi$PT expressions. We have analysed the effect of varying the $S$-wave scattering lengths within a reasonable range 
of values allowed by the solutions of the Roy equations, and we have found that it could be significant, even when the
uncertainties generated by the various inputs are taken into account. 

Going beyond leading order, we have seen that 
a discussion of the phases at next-to-leading order requires only the 
one-loop form factors and $\pi\pi$ amplitudes, as well as the imaginary parts of the two-loop form factors. The latter follow from the former through the unitarity condition.
We have analysed the resulting expressions for isospin breaking in the experimentally measured difference of phase shifts $\Delta_{\rm IB}$, which contain both universal 
contributions (from $\pi\pi$ scattering) and process-dependent ones (stemming from form factors). 
This is the first place where
a dependence with respect to $s_\ell$, the square of the dilepton invariant mass, becomes 
possible. We have shown that this dependence remains actually very weak, well below the 
level of sensitivity allowed by the uncertainties introduced by the various numerical inputs
in our theoretical treatment, or, for that matter, by the experimental error bars
on the data themselves. Working at $s_\ell = 0$, we have shown that the next-to-leading
corrections have an impact on the determination of the scattering lengths $a_0^0$ and $a_0^2$,
but the values obtained remain compatible, within the error bars, with those that result from 
an analysis at leading-order only.

At two loops, the results are 
affected by somewhat larger uncertainties, due to a larger number of inputs contributing in a non negligible way 
to the result. One observes a significant cancellation between process-dependent and 
universal contributions, leaving most of isospin breaking to differences in phase spaces and form factor normalisations. $\Delta_{\rm IB}$ shows 
significant variations when $a_0^0$ and $a_0^2$ are away from the chiral prediction. We have studied truncation effects in the matching between 
the dispersive subtraction parameters and their chiral expansion, and found that the effect was within the uncertainty stemming from 
the variation of the inputs. This leads us to expect that our results are stable with respect to effects of higher orders
than those considered here.
We have finally reanalysed the recent NA48/2 data using our correction for isospin breaking, leading to results in agreement with the NA48/2 
analyses based on the estimate of isospin-breaking corrections from Ref.~\cite{Colangelo:2008sm}.

In summary, we have extended the previous work of Ref.~\cite{Colangelo:2008sm} on isospin breaking
in $K_{\ell 4}$ decays in two respects. First, we have devised a formalism which allows us
to keep the scattering lengths as free parameters in the corrections to the phases of the
form factors arising from isospin breaking. Second, we have determined the corrections
to the phases in this general framework at next-to-leading order.

A natural extension of our work would consist in working out not only the phases, but also the real parts 
of the $K_{\ell 4}$ form factors, in order 
to compute isospin breaking in these quantities which are experimentally available.
In view of Ref.~\cite{DescotesGenon:2012gv} where the pion vector and scalar form factors have been expressed up to two loops, 
and contrary to the case of the phases, a full analytical treatment seems out of reach. 
But the outcome would involve a limited number
of one-dimensional dispersive integrals amenable to a numerical treatment. 
For instance, it would  provide a theoretical framework suitable to 
analyse the cusp recently observed by the NA48/2 experiment in $K^\pm \to\pi^0\pi^0 e^\pm \nu_e$~\cite{BBD-kaon2013}. 
The same approach could also be used to address the effects of mass differences in other low-energy 
observables or to provide phenomenologically useful representations for them, 
see e.g. Refs.~\cite{P_to_3pi_1,P_to_3pi_2,P_to_3pi_3,eta_to_3pi} for some examples.

\section*{Acknowledgments}

We thank B. Bloch-Devaux, G. Colangelo, J. Gasser, B. Kubis, L. Lellouch and A. Rusetsky for discussions and comments. Work supported in part by the EU Integrated Infrastructure Initiative HadronPhysics3.

\appendix

\renewcommand{\theequation}{\Alph{section}.\arabic{equation}}

\section{Mesonic scattering amplitudes} \label{app:loampl}
\setcounter{equation}{0}

In Sec.~\ref{sec:1-loop_ff}, we have given expressions of one-loop form factors
describing the amplitudes for the decay channels $K^+ \to \pi^+ \pi^- \ell^+ \nu_\ell$
and $K^+ \to \pi^0 \pi^0 \ell^+ \nu_\ell$. These expressions involve a certain number
of mesonic scattering amplitudes, corresponding to the various intermediate states that can
contribute to the unitarity conditions. Apart from the $\pi\pi$ amplitudes, which
contribute up to next-to-leading order in the isospin-breaking corrections to the 
phase-shifts in Sec.~\ref{sec:IB_in_phases}, the amplitudes
involving also the other pseudoscalar mesons (kaons and eta) occur only at lowest
order and we begin the discussion with the latter. For the $\pi\pi$ amplitudes, 
we adopt a parametrisation in terms of threshold parameters (scattering lengths, effective ranges), 
but describe the remaining amplitudes  with a parametrisation in terms of subthreshold parameters.
They could easily be converted into expressions involving the scattering lengths if necessary. 
We rely on the definitions given in Sec.~\ref{subsec:1-loop_ff_general}.
Since no confusion can arise, in this appendix we denote the relevant kinematical
variables by $s,t,u$ and $\theta$, instead of the notation $s,{\hat t}, {\hat u}$,
and ${\hat \theta}$ used there.

\subsection{$\pi K$ and other scattering amplitudes at lowest order}

Typically, the amplitudes of interest are of the general form 
\begin{equation}
\left.
A^{a b ; a^\prime b^\prime} (s,t,u)\right\vert_{LO} \,=\,
\frac{\beta^{a b ; a^\prime b^\prime}}{F_\pi^2} \left( s - \frac{1}{3} \Sigma \right)
+ \frac{\alpha^{a b ; a^\prime b^\prime}}{3} \, \frac{ \Sigma}{F_\pi^2}
+ 3 \gamma^{a b ; a^\prime b^\prime} \frac{t - u}{F_\pi^2}
,
\end{equation}
where $\Sigma \equiv M_a^2 + M_b^2 + M_{a^\prime}^2 + M_{b^\prime}^2$.
This form is preserved under crossing. For instance
\begin{eqnarray}
\alpha^{{\bar a}^\prime b ; {\bar a} b^\prime} = \lambda_a \lambda_{a^\prime} \alpha^{a b ; a^\prime b^\prime}
,
\ \beta^{{\bar a}^\prime b ; {\bar a} b^\prime} = - \frac{1}{2} \, \lambda_a \lambda_{a^\prime} 
\left( \beta^{a b ; a^\prime b^\prime} + 9 \gamma^{a b ; a^\prime b^\prime} \right)
,
\ \gamma^{{\bar a}^\prime b ; {\bar a} b^\prime} = \frac{1}{6} \, \lambda_a \lambda_{a^\prime}
\left( 3 \gamma^{a b ; a^\prime b^\prime} - \beta^{a b ; a^\prime b^\prime} \right)
\!,
\quad{ }
\end{eqnarray}
or
\begin{eqnarray}
\alpha^{{\bar b}^\prime b ; a^\prime {\bar a}} = \lambda_a \lambda_{b^\prime} \alpha^{a b ; a^\prime b^\prime}
,
\ \beta^{{\bar b}^\prime b ; a^\prime {\bar a}} = - \frac{1}{2} \, \lambda_a \lambda_{b^\prime} 
\left( \beta^{a b ; a^\prime b^\prime} - 9 \gamma^{a b ; a^\prime b^\prime} \right)
,
\ \gamma^{{\bar b}^\prime b ; a^\prime {\bar a}} = \frac{1}{6} \, \lambda_a \lambda_{b^\prime}
\left( 3 \gamma^{a b ; a^\prime b^\prime} + \beta^{a b ; a^\prime b^\prime} \right)
\!.
\quad{ }
\end{eqnarray}
Equivalently, one may write $\left. A^{a b ; a^\prime b^\prime} (s,t,u)\right\vert_{LO} =
16 \pi [ \varphi_0^{a b ; a^\prime b^\prime} (s) + 3 \varphi_1^{a b ; a^\prime b^\prime} (s) \cos \theta_{ab}]$, with
\begin{eqnarray}
16 \pi \varphi_0^{a b ; a^\prime b^\prime} (s) &=&
\frac{\beta^{a b ; a^\prime b^\prime}}{F_\pi^2} \, s 
+ \frac{1}{3} \left( \alpha^{a b ; a^\prime b^\prime} - \beta^{a b ; a^\prime b^\prime} \right)
\frac{\Sigma}{F_\pi^2}
- 3 \frac{\gamma^{a b ; a^\prime b^\prime}}{F_\pi^2} \, \frac{\Delta_{ab} \Delta_{a^\prime b^\prime}}{s }
,
\nonumber\\
16 \pi \varphi_1^{a b ; a^\prime b^\prime} (s) &=&
 \frac{\gamma^{a b ; a^\prime b^\prime}}{F_\pi^2} \, \frac{\lambda^{1/2}_{ab} (s) \lambda^{1/2}_{a^\prime b^\prime} (s) }{s }
 ,
\end{eqnarray}
and $\Delta_{ab} \equiv M_a^2 - M_b^2$, $\Delta_{a^\prime b^\prime} \equiv M_{a^\prime}^2 - M_{b^\prime}^2$.

The subthreshold parameters $\beta^{a b ; a^\prime b^\prime}$ and $\gamma^{a b ; a^\prime b^\prime}$ 
in these expressions are the same as those defined in Eq.~(\ref{beta_and_gamma_def}), as
can be seen upon forming the combinations $\varphi_P^{a b ; a^\prime b^\prime}$ and $\varphi_S^{a b ; a^\prime b^\prime} (s)$,
as defined in Eqs.~(\ref{varphi_P}) and (\ref{varphi_S}), respectively. Some of these coefficients 
are given, at lowest order, in Tab.~\ref{tab:beta_gamma}. Those not shown in this table can be
obtained through the crossing relations given above. These expressions involve the two mixing angles 
$\epsilon_1$ and $\epsilon_2$, defined in terms of the isospin-breaking matrix elements
of the flavour-diagonal octet axial currents \cite{Gasser:1984gg}. At linear order in isospin breaking,
we have
\begin{eqnarray}
\langle \Omega\,\vert\, A_\mu^8 (0) \,\vert\, \pi^0(p) \rangle \ =\ i p_\mu \,\epsilon_1 F_\pi\ ,
\quad
\langle \Omega\,\vert\, A_\mu^3 (0) \,\vert\, \eta(p) \rangle \ =\ -i p_\mu \,\epsilon_2 F_\eta
.
\end{eqnarray}
which at lowest order read
\bea
\epsilon_1 \ =\ \epsilon_2 \ =\ \frac{\sqrt{3}}{4R}\,, \qquad R=\frac{m_d - m_u}{m_s - {\widehat m}}\,.
\eea

Finally, one has
\begin{equation}
16 \pi \varphi_S^{a b ; a^\prime b^\prime} (0) =
\frac{1}{3} \left( \alpha^{a b ; a^\prime b^\prime} - \beta^{a b ; a^\prime b^\prime} \right)
\frac{\Sigma}{F_\pi^2}
.
\end{equation}
with the crossing relations 
\begin{eqnarray}
16 \pi \varphi_S^{{\bar a}^\prime b ; {\bar a} b^\prime} (0) &=& \lambda_a \lambda_{a^\prime} 
\left[
16 \pi \varphi_S^{a b ; a^\prime b^\prime} (0) 
+ \frac{1}{2} \left( \beta^{a b ; a^\prime b^\prime} + 3 \gamma^{a b ; a^\prime b^\prime} \right) \frac{\Sigma}{F_\pi^2}
\right]\,,
\nonumber\\
16 \pi \varphi_S^{{\bar b}^\prime b ; a^\prime {\bar a}} (0) &=& \lambda_a \lambda_{b^\prime} 
\left[
16 \pi \varphi_S^{a b ; a^\prime b^\prime} (0) 
+ \frac{1}{2} \left( \beta^{a b ; a^\prime b^\prime} - 3 \gamma^{a b ; a^\prime b^\prime} \right) \frac{\Sigma}{F_\pi^2}
\right]
.
\end{eqnarray}

\begin{table}[t]
\begin{center}
\begin{tabular}{|c|c||c|c|}
\hline
  $a b$    &  $a^\prime b^\prime$  & $(F_0^2/F_\pi^2) \cdot \beta^{a b ; a^\prime b^\prime}$ & 
$(F_0^2/F_\pi^2) \cdot \gamma^{a b ; a^\prime b^\prime}$
\\ \hline\hline
 $\pi^+ \pi^-$ & $K^+ K^-$ &  ${\frac{1}{4}} $  &  $\frac{1}{12} $
\\ \hline
 $\pi^0 \pi^- $ & $K^0 K^- $ & $-\frac{1}{2} \sqrt{\frac{3}{2}} \epsilon_2$ & $\frac{1}{6\sqrt{2}}$
\\ \hline
 $\pi^0 \pi^0$ & $K^+ K^-$  & $- \frac{1}{4} \left( 1 + 2\sqrt{3} \epsilon_2 \right)$ &  $0$
\\ \hline
 $\pi^0 \pi^0$ & $K^0 {\bar K}^0$  & $ \frac{1}{4} \left( 1 - 2\sqrt{3} \epsilon_2 \right)$ & $0$
\\ \hline
 $\pi^+ \pi^-$ & $K^0 {\bar K}^0$  & $ - \frac{1}{4}$   & $ \frac{1}{12}$
\\ \hline
  $\pi^+ K^-$ & $\eta {\bar K}^0$ & $-\frac{1}{4}\sqrt{\frac{3}{2}}\left( 1 - \sqrt{3}{\epsilon_1} \right)$  & 
$\frac{1}{12}\sqrt{\frac{3}{2}}\left( 1 + \frac{\epsilon_1}{\sqrt{3}} \right)$
\\ \hline
  $\pi^0 K^-$ & $\eta K^-$ & $ - \frac{\sqrt{3}}{8}\left( 1 - \frac{\epsilon_1}{\sqrt{3}} + \sqrt{3}{\epsilon_2} \right)$  &  $\frac{1}{8\sqrt{3}}\left( 1 - \frac{\epsilon_1}{\sqrt{3}} + \sqrt{3}{\epsilon_2} \right) $
\\ \hline\hline
  $\pi^+ \pi^-$ & $\eta\eta$ & $0$ & $0$
\\ \hline
  $\pi^+ \pi^-$ & $\pi^0 \eta$ & $\epsilon_1$ & $0$
\\ \hline
  $\pi^0 \pi^0$ & $\eta\eta$ & $0 $ & $0 $
\\ \hline
  $\pi^0 \pi^0$ & $\pi^0 \eta$ & $0 $ & $0 $
\\ \hline
\end{tabular}
\caption{The parameters $\beta^{a b ; a^\prime b^\prime}$ and $\gamma^{a b ; a^\prime b^\prime}$
corresponding to various lowest-order amplitudes needed for the one-loop expressions of the
form factors $F^{ab}(s,t,u)$ and $G^{ab}(s,t,u)$ discussed in Sec.~\ref{sec:1-loop_ff}.}\label{tab:beta_gamma}
\end{center}
\end{table}

\subsection{$\pi\pi$ scattering amplitudes}

The $\pi\pi$ scattering amplitudes have already been discussed in quite some detail in \cite{DescotesGenon:2012gv},
see in particular App.~F therein.
We need the following lowest-order partial-wave projections
\begin{eqnarray}
&&
\varphi^{\mbox{\tiny{$+- ; +-$}}}_0(s) \,=\, a_{\mbox{\tiny{$+-$}}} + b_{\mbox{\tiny{$+-$}}}\,\frac{s - 4 M_{\pi^\pm}^2}{F_\pi^2}
\,,\quad
\varphi^{\mbox{\tiny{$+- ; +-$}}}_1(s) \,=\, \frac{c_{\mbox{\tiny{$+-$}}}}{3} \, \frac{s - 4 M_{\pi^\pm}^2}{F_\pi^2}
,
\nonumber\\
&&
\varphi^{\mbox{\tiny{$+- ; 00$}}}_0(s) \,=\, a_{x} + b_{x}\,\frac{s - 4 M_{\pi^\pm}^2}{F_\pi^2} 
\,,\quad
\varphi^{\mbox{\tiny{$+- ; 00$}}}_1(s) \,=\, 0
,
\nonumber\\
&&
\varphi^{\mbox{\tiny{$00 ; 00$}}}_0(s) \,=\, a_{00}  
\,,\quad
\varphi^{\mbox{\tiny{$00 ; 00$}}}_1(s) \,=\, 0
.
\label{pion_PW_LO}
\end{eqnarray}
The parameters 
entering these amplitudes in terms of  $a^0_0$ and $a^0_2$ (the $S$-wave scattering lengths in the isospin limit,
in the isospin channels $I=0$ and $I=2$, respectively) can be obtained at the lowest order upon combining
the formulas (3.3), (3.4), and (3.5) of Ref.~\cite{DescotesGenon:2012gv},
which gives [$\Delta_{\pi}=M_{\pi}^2 - M_{\pi^0}^2$]
\begin{eqnarray}
&&
a_{\mbox{\tiny{$+-$}}}\,=\,\frac{2}{3}\,a^0_0\,+\,\frac{1}{3}\,a^2_0\,-\,2 a^2_0\,\frac{\Delta_\pi}{M_{\pi}^2}\ ,
\quad
b_{\mbox{\tiny{$+-$}}}\,=\,c_{\mbox{\tiny{$+-$}}} \,=\, \frac{1}{24} \frac{F_\pi^2}{M_\pi^2} \left( 2 a^2_0 - 5 a^2_0 \right)\,,
\nonumber\\
&&
a_x \,=\, - \frac{2}{3}\,a^0_0\,+\,\frac{2}{3}\,a^2_0\,+\, a^2_0\,\frac{\Delta_\pi}{M_{\pi}^2}\ ,
\quad
b_x\,=\,- \frac{1}{12} \frac{F_\pi^2}{M_\pi^2} \left( 2 a^2_0 - 5 a^2_0 \right)\,,
\nonumber\\
&&
a_{00}\,=\,\frac{2}{3}\,a^0_0\,+\,\frac{4}{3}\,a^2_0\,-\,\frac{2}{3}\left(a^0_0 + 2 a^2_0\right)\frac{\Delta_\pi}{M_{\pi}^2}
.
\label{IB_in_a_and_b_LO}
\end{eqnarray}

In order to proceed with the analysis of isospin-breaking in the $K^+_{e4}$ data,
we need to go beyond lowest order, and work out the contributions 
$\psi^{\mbox{\tiny{$+- ; +-$}}}_{0,1}(s)$, $\psi^{\mbox{\tiny{$+- ; 00$}}}_{0}(s)$
to the partial-wave projections at next-to-leading order.
These can be obtained from Eqs.~(4.6), (4.10), (4.15) of Ref.~\cite{DescotesGenon:2012gv},
combined with the formulas (F.8)-(F.12) of that same reference.
Likewise, the formulas for $a_{\mbox{\tiny{$+-$}}}$, $a_x$, 
$b_{\mbox{\tiny{$+-$}}}$, and $b_x$
need also to be determined at next-to-leading
order. In the case of the scattering lengths, this can be achieved upon using the expressions (F.13) 
and the formulas given in Appendix E of \cite{DescotesGenon:2012gv}~\footnote{
We take this opportunity to correct three typos in Apps.~E and F in this reference. In Eq.~(E.3), the term proportional to $L_\pi$ in $\beta_x$ should read $(10\beta-7\alpha)\beta$ instead of $(13\alpha-10\beta)\beta$.
In Eq.~(F.4), the expression given for $P^x(s,t,u)$ is actually that for $-P^x$. In Eq.~(F.13), the term proportional
to $\bar{J}_0(4M_\pi^2)$ should read $(\beta_x(10M_\pi^2-2 M_{\pi^0}^2)+\alpha_x M_{\pi^0}^2)^2$ 
instead of $(8\beta_x M_\pi^2+\alpha_x M_{\pi^0}^2)^2$.}. 
The result reads
\begin{eqnarray}
16 \pi \left[ a_{\mbox{\tiny{$+-$}}} - \frac{1}{3} \left( 2 a_0^0 + a_0^2 \right) \right] &=&
-32 \pi a_0^2 \, \frac{\Delta_\pi}{M_\pi^2} 
- 16 \pi \frac{1}{6} \left( 2 a_0^0 + 7 a_0^2 \right) \frac{\Delta_\pi}{M_\pi^2} 
\nonumber\\
&&
+ 16 \left( \lambda_1 + 2 \lambda_2 \right) \frac{\Delta_\pi}{F_\pi^2} \frac{M_\pi^2}{F_\pi^2} 
\nonumber\\
&&
-
\frac{8}{27} \frac{\Delta_\pi}{M_\pi^2} \left[
37 \left( a_0^0 \right)^2  -
\frac{1499}{4} \left( a_0^2 \right)^2 
+ 241 a_0^0 a_0^2 
\right]
\nonumber\\
&&
+
\frac{8}{27} L_\pi \!\left[
53 \left( a_0^0 \right)^2  -
\frac{1153}{4} \left( a_0^2 \right)^2 
+ 26 a_0^0 a_0^2 
\right]
\nonumber\\
&&
+
16 \pi \frac{1}{18} \left( 2 a_0^0 - 5 a_0^2 \right) 
\frac{e^2}{32 \pi^2} \left[ 3 {\widehat{\cal K}}^{\mbox{\tiny{$+-$}}}_1 +  {\widehat{\cal K}}^{\mbox{\tiny{$+-$}}}_2 
- 3 \left( {\widehat{\cal K}}^{00}_1 + {\widehat{\cal K}}^{00}_2 \right)  \right]
\nonumber\\
&&
- 32 \pi a_0^2 \frac{e^2}{32 \pi^2} {\widehat{\cal K}}^{\mbox{\tiny{$+-$}}}_3 
+ {\cal O} \left( \frac{\Delta_\pi^2}{M_\pi^4} \right)
,
\label{IB_in_a_+-}
\\
\nonumber\\
16 \pi \left[ a_x + \frac{2}{3} \left( a_0^0 - a_0^2 \right) \right] &=&
16 \pi a_0^2 \, \frac{\Delta_\pi}{M_\pi^2} 
+ 16 \pi \frac{1}{12} \left( 2 a_0^0 + 7 a_0^2 \right) \frac{\Delta_\pi}{M_\pi^2} 
- 4 \frac{\Delta_\pi}{F_\pi^2} \frac{M_{\pi}^2}{F_\pi^2} \left( 3 \lambda_1 + 4 \lambda_2 \right)
\nonumber\\
&&
+
\frac{1}{81} \frac{\Delta_\pi}{M_\pi^2} \left[
644 \left( a_0^0 \right)^2  -
3895 \left( a_0^2 \right)^2 
+
6212 a_0^0 a_0^2 
\right]
\nonumber\\
&&
+ \frac{16}{27} \left[ \frac{M_{\pi^0}^2}{\Delta_\pi} L_\pi - 1 \right] \left[
4 \left( a_0^0 \right)^2 
- 11 \left( a_0^2 \right)^2 
+
16 a_0^0 a_0^2 
\right]
\nonumber\\
&&
- \frac{1}{9} \,  L_\pi
\left[
12 \left( a_0^0 \right)^2  +
133 \left( a_0^2 \right)^2 +
32 a_0^0 a_0^2 
\right]
\nonumber\\
&&
-
16 \pi \frac{1}{36} \left( 2 a_0^0 - 5 a_0^2 \right) 
\frac{e^2}{32 \pi^2} \left[ 10 {\widehat{\cal K}}^x_1 
+  {\widehat{\cal K}}^x_2 
- 3 \left( {\widehat{\cal K}}^{00}_1 + {\widehat{\cal K}}^{00}_2 \right)  \right]
\nonumber\\
&&
+ 16 \pi a_0^2 \frac{e^2}{32 \pi^2} {\widehat{\cal K}}^x_3 
+ {\cal O} \left( \frac{\Delta_\pi^2}{M_\pi^4} \right)
\label{IB_in_a_x}
,\end{eqnarray}
When we consider isospin-breaking corrections to the phase shifts at next-to-leading order, we also need $a_{\mbox{\tiny{$+$}}0}$,
$a_{\mbox{\tiny{$++$}}}$ and $a_{00}$ from the universal contributions due to $\pi\pi$ (re)scattering. These scattering parameters enter only next-to-leading order corrections, and are thus needed at leading order. However, we will also estimate the impact of higher-order effects in the  isospin-breaking corrections to the phase shifts  by comparing the numerical results obtained using either leading- or next-to-leading-order expressions for these scattering parameters. The latter read
\begin{eqnarray}
16 \pi  \left[a_{\mbox{\tiny{$+$}}0} - a_0^2\right] = &&-16 \pi a_0^2 \frac{\Delta_\pi}{M_\pi^2} -\frac{4 \pi}{3} (2 a_0^0 +7 a_0^2)\frac{\Delta_\pi}{M_\pi^2}
+4 \frac{\Delta_\pi}{F_\pi^2}\frac{M_\pi^2}{F_\pi^2}(\lambda_1  + 2 \lambda_2)
\nonumber \\
&& +\frac{1}{9} \frac{\Delta_\pi}{M_\pi^2}\left[4 (a_0^0)^2  +345 (a_0^2)^2  -76 a_0^0  a_0^2 \right] \nonumber \\
&&+ \frac{1}{27} \left[\frac {M_{\pi^0}^2}{\Delta_\pi} L_\pi -1\right]\left[-64 (a_0^0)^2 - 544 (a_0^2)^2+ 32 a_0^0 a_0^2 \right] \nonumber \\
&&+ \frac{1}{27} L_\pi \!\left[ 180 (a_0^0)^2  - 1221 (a_0^2)^2 + 384 a_0^0 a_0^2 \right] \nonumber\\
&& + 16 \pi \frac{1}{36} \frac{e^2}{32 \pi^2} (2 a_0^0-5 a_0^2)\left[-2 {\widehat{\cal K}}^x_1
+ {\widehat {\cal K}}_2^x -3 ({\widehat {\cal K}}_1^{00} 
+ {\widehat {\cal K}}_2^{00}) \right] \nonumber \\
&&-16 \pi a_0^2  \frac{e^2}{32 \pi^2} 
{\widehat {\cal K}}_3^x + {\cal O} \left( \frac{\Delta_\pi^2}{M_\pi^4} \right)\label{IB_in_a_+0},\\
16 \pi  \left[a_{\mbox{\tiny{$++$}}} -2 a_0^2\right] = &&-32 \pi a_0^2 \frac{\Delta_\pi}{M_\pi^2} -\frac{8 \pi}{3} (2 a_0^0 +7 a_0^2)\frac{\Delta_\pi}{M_\pi^2}
+16 \frac{\Delta_\pi}{F_\pi^2}\frac{M_\pi^2}{F_\pi^2}(\lambda_1  + 2 \lambda_2)
\nonumber \\
&& -\frac{2}{27} \frac{\Delta_\pi}{M_\pi^2}\left[52 (a_0^0)^2  -1955 (a_0^2)^2 + 436 a_0^0  a_0^2\right] \nonumber \\
&&+ \frac{2}{27} L_\pi \!\left[ 308 (a_0^0)^2  - 1993 (a_0^2)^2 + 200 a_0^0 a_0^2 \right] \nonumber\\
&& + 16 \pi \frac{1}{18} \frac{e^2}{32 \pi^2} (2 a_0^0-5 a_0^2)\left[ {\widehat{\cal K}}_2^{+-}
-3( {\widehat {\cal K}}_1^{+-}  +{\widehat {\cal K}}_1^{00}
+ {\widehat {\cal K}}_2^{00}) \right] \nonumber \\
&&-32 \pi a_0^2  \frac{e^2}{32 \pi^2}
{\widehat {\cal K}}_3^{+-} + {\cal O} \left( \frac{\Delta_\pi^2}{M_\pi^4} \right)\label{IB_in_a_++}
,\\
16 \pi  \left[a_{00} -\frac{2}{3} (a_0^0 +2 a_0^2)\right] = &&
-16 \pi \frac{2}{3} (a_0^0 +2 a_0^2)\frac {\Delta_\pi}{M_\pi^2}-16 \pi \frac{1}{4} (2 a_0^0+7 a_0^2)\frac{\Delta_\pi}{M_\pi^2}
\nonumber \\
&& -\frac{1}{9} \frac{\Delta_\pi}{M_\pi^2}\left[5 (2a_0^0+a_0^2)^2 -128( a_0^0  -a_0^2)^2 \right] \nonumber \\
&&+ \frac{1}{9} L_\pi \!\left[ 100 (a_0^0)^2  +475 (a_0^2)^2 + 424 a_0^0 a_0^2 \right] \nonumber\\
&& + 16 \pi \frac{1}{12} \frac{e^2}{32 \pi^2} (2 a_0^0-5 a_0^2)\left[{\widehat{\cal K}}^{00}_1+{\widehat {\cal K}}_2^{00} \right] 
-16 \pi \frac{2}{3}(a_0^0+2 a_0^2)  \frac{e^2}{32 \pi^2}
{\widehat {\cal K}}_2^{00} \nonumber \\
&&-
16\pi\frac{4}{9} \frac{\sqrt{\Delta_\pi}}{ M_\pi}(a_0^0-a_0^2)^2 
+ {\cal O} \left( \frac{\Delta_\pi^{3/2}}{M_\pi^3} \right)\label{IB_in_a_00}
,
\end{eqnarray}
where the last term in Eq.~(\ref{IB_in_a_00}) comes from the expansion of $\bar{J}(4 M_{\pi^0}^2)$
in powers of $\Delta_\pi$ (see the expression of $a_{00}$ in Eq.~(F.13) in Ref.~\cite{DescotesGenon:2012gv}). We have also
\begin{eqnarray}
16 \pi b_{\mbox{\tiny{$+-$}}} &=& \frac{2 \pi}{3} \frac{F_\pi^2}{M_\pi^2} \left( 2 a_0^0 - 5 a_0^2 \right) 
- \frac{4}{3} \frac{F_\pi^2}{M_\pi^2} \left[ 2 \left( a_0^0 \right)^2 - 5 \left( a_0^2 \right)^2 \right]
\nonumber\\
&&
+ \frac{10}{3} \frac{\Delta_\pi}{M_\pi^2} \frac{F_\pi^2}{M_\pi^2} \, a_0^2 \left( 2 a_0^0 - 5 a_0^2 \right)
- \frac{2}{9} \frac{F_\pi^2}{M_\pi^2} \left( 2 a_0^0 - 5 a_0^2 \right)
\left( 2 a_0^0 + 7 a_0^2  \right) L_\pi
\nonumber\\
&&
+
16 \pi \frac{F_\pi^2}{M_{\pi}^2} \frac{1}{24} \left( 2 a_0^0 - 5 a_0^2 \right) \frac{e^2}{32 \pi^2} {\widehat{\cal K}}^{\mbox{\tiny{$+-$}}}_1
+ {\cal O} \left( \frac{\Delta_\pi^2}{M_\pi^4} \right)
,
\label{IB_in_b_+-}
\\
\nonumber\\
16 \pi b_x &=& - \frac{4 \pi}{3} \frac{F_\pi^2}{M_\pi^2} \left( 2 a_0^0 - 5 a_0^2 \right) 
+ \frac{8}{3} \frac{F_\pi^2}{M_\pi^2} \left[ 2 \left( a_0^0 \right)^2 - 5 \left( a_0^2 \right)^2 \right]
- 2 \lambda_1 \frac{\Delta_\pi}{F_\pi^2}
\nonumber\\
&&
- \frac{1}{18} \frac{\Delta_\pi}{M_\pi^2} \frac{F_\pi^2}{M_\pi^2}
\left[
28 \left( a_0^0 \right)^2 - 425 \left( a_0^2 \right)^2 + 100 a_0^0 a_0^2 
\right]
+ \frac{1}{3} \frac{F_\pi^2}{M_\pi^2} 
\left( 2 a_0^0 - 5 a_0^2 \right) \left( 2 a_0^0 + 9 a_0^2 \right) L_\pi
\nonumber\\
&&
+ \frac{4}{3}  \frac{F_\pi^2}{M_\pi^2}
\left[ \frac{M_{\pi^0}^2}{\Delta_\pi} L_\pi - 1 \right]
a_0^2 \left( 2 a_0^0 - 5 a_0^2 \right)
-
16 \pi \frac{F_\pi^2}{M_{\pi}^2}\frac{1}{12} \left( 2 a_0^0 - 5 a_0^2 \right) 
\frac{e^2}{32 \pi^2}  {\widehat{\cal K}}^x_1
+ {\cal O} \left( \frac{\Delta_\pi^2}{M_\pi^4} \right)
.
\label{IB_in_b_x}
\end{eqnarray}
whereas $b_{+0}=b_x/2$.
In each case, the first term on the right-hand side of the equation corresponds to the leading-order contribution while further contributions are chirally suppressed. 
The combinations ${\widehat{\cal K}}^{ab}_i$ of electromagnetic low-energy constants appearing in these expressions 
are given in Eq.~(E.4) of \cite{DescotesGenon:2012gv}. In Eq.~(E.3) of the same reference, 
one also finds the connection between the subthreshold 
parameters $\lambda^{(1,2)}_{ab}$ in presence of isospin and those in the isospin limit, $\lambda_{1,2}$.
In both cases, the $\pi\pi$ subthreshold parameter $\beta$ defined in the isospin limit is involved.
For our purposes, it is only needed at lowest order, where it is related to the two $S$-wave scattering lengths by
\begin{equation}
\beta = \frac{4 \pi}{3} \frac{F_\pi^2}{M_\pi^2} \left( 2 a^2_0 - 5 a^2_0 \right)
.
\end{equation}

\section{Subtraction polynomials}\label{app:poly}
\setcounter{equation}{0}

The expressions of the one-loop $K_{\ell 4}$ form factors obtained in Sec.~\ref{sec:1-loop_ff}
involve subtraction polynomials $P^{ab}_F(s,t,u)$ and $P^{ab}_G(s,t,u)$, which themselves
depend on a certain number of coefficients $\pi^{ab}_{i, F/G}$, that will be discussed in this appendix.

In the effective theory framework, the subtraction coefficients are given in terms
of the low-energy constants and chiral logarithms. It is therefore possible to extract
the expressions of the polynomials $P^{ab}_F(s,t,u)$ and $P^{ab}_G(s,t,u)$ from a
comparison of the expressions obtained in Secs.~\ref{subsec:1-loop_ff_+-}
and \ref{subsec:1-loop_ff_00} with a one-loop computation of the relevant form factors.
Although such computations, including isospin breaking, are available, it is 
only necessary to compute the contributions from the tree diagrams and 
from the tadpoles, since we have been careful in expressing the unitarity parts in terms 
of the appropriate loop functions. All we need to take into account from the
unitarity parts are the terms generated by the differences between ${\bar J}_{ab}$ and $J^r_{ab}$, 
or between $M_{ab}$ and $M^r_{ab}$. As far as the tree-level contributions are concerned,
they include those arising from the strong $N_f=3$ low-energy constants $L_i$ \cite{Gasser:1984gg}, and those coming from
the counterterms ${\widehat K}_i$ describing isospin-breaking corrections \cite{Urech:1994hd}.
The tadpole contributions are most easily computed upon using the corresponding
expression of the one-loop generating functional \cite{Gasser:1984gg}. 
We have actually redone this calculation, taking into account the contributions from the 
mixing angles $\epsilon_1$ and $\epsilon_2$.

We want to stress that only tadpoles generated by the 
kinematic term with two covariant derivatives at lowest order
contribute to the $K_{e4}$ form factors at one loop. Therefore, their dependence
on the meson masses comes from the invariant products of the meson momenta 
involved, and are not reconstructed from the $B_0$ term in the leading-order $N_f=3$ chiral Lagrangian~\cite{Gasser:1984gg}.
According to the chiral counting, these tadpoles contribute only to $\pi_{0,F}^{\mbox{\tiny{$+-$}}}$,
$\pi_{0,G}^{\mbox{\tiny{$+-$}}}$, and $\pi_{0,F}^{\mbox{\tiny{$00$}}}$ at the one-loop order. 
These three quantities will thus require a separate discussion once we have dealt with
the remaining subtraction constants.

In order to illustrate the procedure just outlined, let us consider in some detail the case
of $\pi_{3,F}^{\mbox{\tiny{$+-$}}}$, cf. Eq.~(\ref{P_F_and_P_G_+-}). Here and in what follows, we
drop contributions that are of second order in isospin
breaking whenever convenient (e.g. quadratic terms in the mixing angles 
$\epsilon_{1,2}$). Our expressions and statements will always be exact up to 
isospin-violating corrections of higher order. Concerning $\pi_{3,F}^{\mbox{\tiny{$+-$}}}$,
we thus find
\begin{eqnarray}
\pi_{3,F}^{\mbox{\tiny{$+-$}}} &=& - 2 L_3 
\nonumber\\
&&
+
\frac{1}{2} \,\frac{F_0^2}{F_\pi^2} \,
 ( \beta^{\mbox{\tiny{$+ K^-$}};\mbox{\tiny{$+ K^-$}}} + 2 \beta^{\mbox{\tiny{$K^- -$}};\mbox{\tiny{$K^- -$}}}
- 3 \gamma^{\mbox{\tiny{$+ K^-$}};\mbox{\tiny{$+ K^-$}}} ) 
k_{K \pi} (\mu)
\nonumber\\
&&
- \frac{3}{2} \frac{1}{\sqrt{2}} \frac{F_0^2}{F_\pi^2}
\left[ \left( 1 + \frac{\epsilon_2}{\sqrt{3}} \right) \beta^{\mbox{\tiny{$+ K^-$}};\mbox{\tiny{$0 {\bar K}^0$}}} -
( 1 - \sqrt{3} \epsilon_2 ) \gamma^{\mbox{\tiny{$+ K^-$}};\mbox{\tiny{$0 {\bar K}^0$}}}
\right] 
k_{K^0 \pi^0} (\mu)
\nonumber\\
&&
- \frac{1}{2}  \sqrt{\frac{3}{2}} \frac{F_0^2}{F_\pi^2}
\left[
( 1 - \sqrt{3} \epsilon_1 ) \beta^{\mbox{\tiny{$+ K^-$}};\mbox{\tiny{$\eta {\bar K}^0$}}} +
3 \left( 1 + \frac{\epsilon_1}{\sqrt{3}} \right) \gamma^{\mbox{\tiny{$+ K^-$}};\mbox{\tiny{$\eta {\bar K}^0$}}}
\right]
k_{\eta K^0} (\mu)
,
\end{eqnarray}
which clearly exhibits the two types of contributions, $L_3$ at  the tree level (tadpoles are absent in this case) and 
the terms proportional to the $k_{ab} (\mu)$ coming from the unitarity loops. $\pi_{3,F}^{\mbox{\tiny{$+-$}}}$ should not depend on the subtraction scale $\mu$ scale, as can be checked 
using the information concerning
the values of the coefficients $\beta^{ a b ; a^\prime b^\prime }$ and
$\gamma^{ a b ; a^\prime b^\prime }$ given in App.~\ref{app:loampl}
since $L_3$ is not 
renormalized and
\begin{equation}
\mu \frac{d}{d \mu} k_{ab} (\mu) = - \frac{1}{16 \pi^2}
.
\end{equation}
Equivalently, we can rewrite the expression for $\pi_{3,F}^{\mbox{\tiny{$+-$}}}$
in a manner where scale independence is manifest, upon eliminating
the last term such as to make appear the differences $k_{K \pi} (\mu) - k_{\eta K^0} (\mu)$
and $k_{K^0 \pi^0} (\mu) - k_{\eta K^0} (\mu)$ only (the result of this operation
is shown below).
Concerning the remaining coefficients in the channel with two charged pions,
we proceed along similar lines. 
Besides $L_3$, their expressions also involve the renormalized low-energy
constants $L_1^r (\mu)$, $L_2^r (\mu)$, and $L_9^r (\mu)$. The scale 
dependence of the latter, which has been worked out in Ref.~\cite{Gasser:1984gg}, can be found
in Eqs.~(\ref{scale_dep_1}) and (\ref{scale_dep_2}) below. Together with the information on the 
coefficients $\beta^{a b ; a^\prime b^\prime}$ and $\gamma^{a b ; a^\prime b^\prime}$ in App.~\ref{app:loampl}, we have checked that these expressions 
indeed do not depend on the renormalization scale $\mu$. Making use of this property
and the crossing relations between the coefficients
$\beta^{ a b ; a^\prime b^\prime }$ and $\gamma^{ a b ; a^\prime b^\prime }$
to simplify these expressions further,
we obtain the following manifestly scale-independent formulas
\begin{eqnarray}
\pi_{1,F}^{\mbox{\tiny{$+-$}}} &=& 10 L_3  + 16 \left[ 2  L_1^r (\mu) - L_2^r (\mu)  \right] 
- 6 \, \frac{F_0^2}{F_\pi^2} \gamma^{\mbox{\tiny{$+ -$}};\mbox{\tiny{$K^+ K^-$}}} 
\left[ k_{K \pi} (\mu) - \frac{16}{3} L_2^r (\mu) \right]
\nonumber\\
&&
+ \frac{1}{\sqrt{2}} \, \frac{F_0^2}{F_\pi^2} \left[ \beta^{\mbox{\tiny{$0 -$}};\mbox{\tiny{$K^0 K^-$}}} -
6 \left( 1 + \frac{\sqrt{3}}{2} \epsilon_2 \right) \gamma^{\mbox{\tiny{$0 -$}};\mbox{\tiny{$K^0 K^-$}}} ) \right]
\left[ k_{K^0 \pi^0} (\mu) - \frac{16}{3} L_2^r (\mu) \right]
\nonumber\\
&&
+ \frac{1}{2} \sqrt{\frac{3}{2}} \, \frac{F_0^2}{F_\pi^2} \left[
( 1 - \sqrt{3} \epsilon_1 ) \beta^{\mbox{\tiny{$+ K^-$}};\mbox{\tiny{$\eta {\bar K}^0$}}} +
3 \left( 1 + \frac{\epsilon_1}{\sqrt{3}} \right) \gamma^{\mbox{\tiny{$+ K^-$}};\mbox{\tiny{$\eta {\bar K}^0$}}}
\right]
\left[ k_{\eta K^0} (\mu) - \frac{16}{3} L_2^r (\mu) \right]
\nonumber\\
&&
- \frac{F_0^2}{F_\pi^2} \left\{
 2 \beta^{\mbox{\tiny{$+ -$}};\mbox{\tiny{$+ -$}}} \left[ k_{\pi\pi} (\mu) - \frac{16}{3} L_2^r (\mu) \right] 
- (1 + 2 \sqrt{3} \epsilon_2 ) \beta^{\mbox{\tiny{$+ -$}};\mbox{\tiny{$0 0$}}} \left[ k_{\pi^0 \pi^0} (\mu) - \frac{16}{3} L_2^r (\mu) \right]
\right.
\nonumber\\
&&
\left.
+ 4 \beta^{\mbox{\tiny{$+ -$}};\mbox{\tiny{$K^+ K^-$}}} \left[ k_{KK} (\mu) - \frac{16}{3} L_2^r (\mu) \right]
- 2 \beta^{\mbox{\tiny{$+ -$}};\mbox{\tiny{$K^0 {\bar K}^0$}}} \left[ k_{K^0 {\bar K}^0} (\mu) - \frac{16}{3} L_2^r (\mu) \right] 
\right.
\nonumber\\
&&
\left.
- 3 \left( 1 - \frac{2}{\sqrt{3}} \, \epsilon_1 \right) \beta^{\mbox{\tiny{$+ -$}};\mbox{\tiny{$\eta\eta$}}} \!
\left[ k_{\eta\eta} (\mu) - \frac{16}{3} L_2^r (\mu) \right]
- 2 \sqrt{3} \left( 1 - \frac{\epsilon_1}{\sqrt{3}} + \sqrt{3} \epsilon_2 \right)
\beta^{\mbox{\tiny{$+ -$}};\mbox{\tiny{$\pi^0 \eta$}}} \!
\left[ k_{\eta\pi^0} (\mu) - \frac{16}{3} L_2^r (\mu) \right]
\right\}
\! ,
\nonumber\\
\pi_{2,F}^{\mbox{\tiny{$+-$}}} &=& - 2 L_3  + 2 \left[ L_9^r (\mu)  - \frac{4}{3} L_2^r (\mu) \right] 
+ 6 \, \frac{F_0^2}{F_\pi^2} \gamma^{\mbox{\tiny{$+ -$}};\mbox{\tiny{$K^+ K^-$}}} 
\left[ k_{K \pi} (\mu) - \frac{16}{3} L_2^r (\mu) \right]
\nonumber\\
&&
- \frac{1}{\sqrt{2}} \, \frac{F_0^2}{F_\pi^2} \left[ \beta^{\mbox{\tiny{$0 -$}};\mbox{\tiny{$K^0 K^-$}}} -
6 \left( 1 + \frac{\sqrt{3}}{2} \epsilon_2 \right) \gamma^{\mbox{\tiny{$0 -$}};\mbox{\tiny{$K^0 K^-$}}} ) \right]
\left[ k_{K^0 \pi^0} (\mu) - \frac{16}{3} L_2^r (\mu) \right]
\nonumber\\
&&
- \frac{1}{2}  \sqrt{\frac{3}{2}} \frac{F_0^2}{F_\pi^2}
\left[
( 1 - \sqrt{3} \epsilon_1 ) \beta^{\mbox{\tiny{$+ K^-$}};\mbox{\tiny{$\eta {\bar K}^0$}}} +
3 \left( 1 + \frac{\epsilon_1}{\sqrt{3}} \right) \gamma^{\mbox{\tiny{$+ K^-$}};\mbox{\tiny{$\eta {\bar K}^0$}}}
\right]
\left[ k_{\eta \pi^0} (\mu) - \frac{16}{3} L_2^r (\mu) \right]
,
\nonumber\\
\pi_{3,F}^{\mbox{\tiny{$+-$}}} &=& - 2 L_3 
\nonumber\\
&&
-
\frac{F_0^2}{F_\pi^2} \,
 ( \beta^{\mbox{\tiny{$+ -$}};\mbox{\tiny{$K^+ K^-$}}} + 3 \gamma^{\mbox{\tiny{$+ -$}};\mbox{\tiny{$K^+ K^-$}}} ) 
\left[ k_{K \pi} (\mu) - k_{\eta K^0} (\mu) \right]
\nonumber\\
&&
- \frac{1}{\sqrt{2}} \frac{F_0^2}{F_\pi^2}
\left[ \beta^{\mbox{\tiny{$0 -$}};\mbox{\tiny{$K^0 K^-$}}} -
6 \left( 1 + \frac{\sqrt{3}}{2} \epsilon_2 \right) \gamma^{\mbox{\tiny{$0 -$}};\mbox{\tiny{$K^0 K^-$}}} ) 
\right] 
\left[ k_{K^0 \pi^0} (\mu) - k_{\eta K^0} (\mu) \right]
,
\end{eqnarray}
and
\begin{eqnarray}
\pi_{1,G}^{\mbox{\tiny{$+-$}}} &=& - 2 L_3 
\nonumber\\
&&
- \frac{2}{3} \, \frac{F_0^2}{F_\pi^2} 
( \beta^{\mbox{\tiny{$+ -$}};\mbox{\tiny{$K^+ K^-$}}} + 3 \gamma^{\mbox{\tiny{$+ -$}};\mbox{\tiny{$K^+ K^-$}}} ) 
\left[ k_{K \pi} (\mu) - k_{\eta K^0} (\mu) \right]
\nonumber\\
&&
- \frac{1}{3\sqrt{2}}  \frac{F_0^2}{F_\pi^2} \left[
2 \left( 1 + \frac{\sqrt{3}}{2} \epsilon_2 \right) \beta^{\mbox{\tiny{$0 -$}};\mbox{\tiny{$K^0 K^-$}}} -
3 ( 7 + 2 \sqrt{3} \epsilon_2 ) \gamma^{\mbox{\tiny{$0 -$}};\mbox{\tiny{$K^0 K^-$}}}
\right]
\left[ k_{K^0 \pi^0} (\mu) - k_{\eta K^0} (\mu) \right]
\nonumber\\
&&
- 2 \frac{F_0^2}{F_\pi^2} \left\{
 \gamma^{\mbox{\tiny{$+ -$}};\mbox{\tiny{$+ -$}}} \left[ k_{\pi\pi} (\mu) - k_{\eta K^0} (\mu) \right] 
+ 2 \gamma^{\mbox{\tiny{$+ -$}};\mbox{\tiny{$K^+ K^-$}}} \left[ k_{KK} (\mu) - k_{\eta K^0} (\mu) \right]
\right.
\nonumber\\
&&
\left.
- \gamma^{\mbox{\tiny{$+ -$}};\mbox{\tiny{$K^0 {\bar K}^0$}}} \left[ k_{K^0 {\bar K}^0} (\mu) - k_{\eta K^0} (\mu) \right]  
\right\}
,
\nonumber\\
\pi_{2,G}^{\mbox{\tiny{$+-$}}} &=& + 2 L_3 
\nonumber\\
&&
+ \frac{2}{3} \, \frac{F_0^2}{F_\pi^2} 
( \beta^{\mbox{\tiny{$+ -$}};\mbox{\tiny{$K^+ K^-$}}} + 3 \gamma^{\mbox{\tiny{$+ -$}};\mbox{\tiny{$K^+ K^-$}}} ) 
\left[ k_{K \pi} (\mu) - 4  L_9^r (\mu) \right] 
\nonumber\\
&&
+ \frac{1}{3\sqrt{2}}  \frac{F_0^2}{F_\pi^2} \left[
2 \left( 1 + \frac{\sqrt{3}}{2} \epsilon_2 \right) \beta^{\mbox{\tiny{$+ -$}};\mbox{\tiny{$K^+ K^-$}}} -
3 ( 7 + 2 \sqrt{3} \epsilon_2 ) \gamma^{\mbox{\tiny{$+ -$}};\mbox{\tiny{$K^+ K^-$}}}
\right]
\left[ k_{K^0 \pi^0} (\mu) - 4  L_9^r (\mu) \right] 
\nonumber\\
&&
+ \frac{1}{2} \sqrt{\frac{3}{2}} \left[
( 1 - \sqrt{3} \epsilon_1 ) \beta^{\mbox{\tiny{$+ K^-$}};\mbox{\tiny{$\eta {\bar K}^0$}}} -
 \left( 1 + \frac{\epsilon_1}{\sqrt{3}} \right) \gamma^{\mbox{\tiny{$+ K^-$}};\mbox{\tiny{$\eta {\bar K}^0$}}}
\right]
\left[ k_{\eta K^0} (\mu) - 4  L_9^r (\mu) \right] 
,
\nonumber\\
\pi_{3,G}^{\mbox{\tiny{$+-$}}} &=& + 2 L_3 
\nonumber\\
&&
- \frac{1}{3} \, \frac{F_0^2}{F_\pi^2} ( \beta^{\mbox{\tiny{$+ -$}};\mbox{\tiny{$K^+ K^-$}}} 
+ 21 \gamma^{\mbox{\tiny{$+ -$}};\mbox{\tiny{$K^+ K^-$}}} ) 
\left[ k_{K \pi} (\mu) - \frac{16}{3} L_2^r (\mu) \right]
\nonumber\\
&&
+ \frac{1}{3\sqrt{2}}  \frac{F_0^2}{F_\pi^2} \left[
2 \left( 1 + \frac{\sqrt{3}}{2} \epsilon_2 \right) \beta^{\mbox{\tiny{$+ -$}};\mbox{\tiny{$K^+ K^-$}}} -
3 ( 7 + 2 \sqrt{3} \epsilon_2 ) \gamma^{\mbox{\tiny{$+ -$}};\mbox{\tiny{$K^+ K^-$}}}
\right]
\left[ k_{K^0 \pi^0} (\mu) - \frac{16}{3} L_2^r (\mu) \right]
\nonumber\\
&&
+ \frac{1}{2} \sqrt{\frac{3}{2}} \frac{F_0^2}{F_\pi^2} \left[
( 1 - \sqrt{3} \epsilon_1 ) \beta^{\mbox{\tiny{$+ K^-$}};\mbox{\tiny{$\eta {\bar K}^0$}}} -
 \left( 1 + \frac{\epsilon_1}{\sqrt{3}} \right) \gamma^{\mbox{\tiny{$+ K^-$}};\mbox{\tiny{$\eta {\bar K}^0$}}}
\right]
\left[ k_{\eta K^0} (\mu) - \frac{16}{3} L_2^r (\mu) \right]
.
\end{eqnarray}
In the channel with two neutral pions, some coefficients vanish due to
Bose symmetry, see Eq.~(\ref{P_F_and_P_G_00}). For the remaining ones, we obtain
manifestly scale-independent expressions
\begin{eqnarray}
\pi_{1,F}^{\mbox{\tiny{$00$}}} &=& + 2 ( 5 + 2 \sqrt{3} \epsilon_2) L_3  + 16 \left[ 2 L_1^r (\mu) - L_2^r (\mu) \right] 
\nonumber\\
&&
-
\frac{F_0^2}{F_\pi^2} \left\{
3 (1 + 2 \sqrt{3} \epsilon_2 ) \gamma^{\mbox{\tiny{$0 0 $;$K^+ K^-$ }}} ) 
\left[ k_{K\pi^0} (\mu) - \frac{16}{3} L_2^r (\mu) \right]
\right.
\nonumber\\
&&
\left.
+ {\sqrt{2}} \left[
\beta^{\mbox{\tiny{$0 -$;$K^0 K^-$}}} +
6 \left(1 - \frac{\sqrt{3}}{2} \epsilon_2 \right) \gamma^{\mbox{\tiny{$0 -$;$K^0 K^-$}}}
\right] 
\left[ k_{K^0 \pi} (\mu) - \frac{16}{3} L_2^r (\mu) \right]
\right.
\nonumber\\
&&
\left.
- \frac{\sqrt{3}}{2} \left( 1 - \frac{\epsilon_1}{\sqrt{3}} + \sqrt{3} \epsilon_2 \right)
( \beta^{\mbox{\tiny{$0 K^-$;$\eta K^-$ }}} + 3 \gamma^{\mbox{\tiny{$0 K^-$;$\eta K^-$ }}} ) 
\left[ k_{\eta K} (\mu) - \frac{16}{3} L_2^r (\mu) \right]
\right.
\nonumber\\
&&
\left.
- 2 \beta^{\mbox{\tiny{$00$;$+-$}}} \left[ k_{\pi\pi} (\mu) - \frac{16}{3} L_2^r (\mu) \right]
- 4 \beta^{\mbox{\tiny{$00$;$K^+ K^-$}}} \left[ k_{KK} (\mu) - \frac{16}{3} L_2^r (\mu) \right]
\right.
\nonumber\\
&&
\left.
+ 2 \beta^{\mbox{\tiny{$00$;$K^0 {\bar K}^0$}}} \left[ k_{K^0 {\bar K}^0} (\mu) - \frac{16}{3} L_2^r (\mu) \right]
\right\}
,      
\nonumber\\
\pi_{2,F}^{\mbox{\tiny{$00$}}} &=& - 2 \left[ 4 L_2^r (\mu) - 3 L_9^r (\mu) \right] 
- 2 \left( 1 - \frac{2}{\sqrt{3}} \epsilon_2 \right) L_3
\nonumber\\
&&
+ 
\frac{F_0^2}{F_\pi^2} \left\{
3 (1 + 2 \sqrt{3} \epsilon_2 ) \gamma^{\mbox{\tiny{$0 0 $;$K^+ K^-$ }}} ) 
\left[ k_{K\pi^0} (\mu) - 4  L_9^r (\mu) \right] 
\right.
\nonumber\\
&&
\left.
+ {\sqrt{2}} \left[
\beta^{\mbox{\tiny{$0 -$;$K^0 K^-$}}} +
6 \left(1 - \frac{\sqrt{3}}{2} \epsilon_2 \right) \gamma^{\mbox{\tiny{$0 -$;$K^0 K^-$}}}
\right] 
\left[ k_{K^0 \pi} (\mu) - 4  L_9^r (\mu) \right] 
\right.
\nonumber\\
&&
\left.
- \frac{\sqrt{3}}{2} \left( 1 - \frac{\epsilon_1}{\sqrt{3}} + \sqrt{3} \epsilon_2 \right)
( \beta^{\mbox{\tiny{$0 K^-$;$\eta K^-$ }}} + 3 \gamma^{\mbox{\tiny{$0 K^-$;$\eta K^-$ }}} ) 
\left[ k_{\eta K} (\mu) - 4  L_9^r (\mu) \right] 
\right\}
,
\nonumber\\
\pi_{3,G}^{\mbox{\tiny{$00$}}} &=& + 2 \left( 1 - \frac{2}{\sqrt{3}} \epsilon_2 \right) L_3 
\nonumber\\
&&
+
\frac{F_0^2}{F_\pi^2} \left\{ 
\frac{1}{3} (1 + 2 \sqrt{3} \epsilon_2 ) ( \beta^{\mbox{\tiny{$0 0$;$K^+ K^-$ }}} - 6 \gamma^{\mbox{\tiny{$0 0$;$K^+ K^-$ }}} ) 
\left[ k_{K\pi^0} (\mu) - \frac{16}{3} L_2^r (\mu) \right]
\right.
\nonumber\\
&&
\left.
- {\sqrt{2}} \left[
\frac{2}{3} \left( 1 - \frac{\sqrt{3} \epsilon_2}{2} \right) \beta^{\mbox{\tiny{$0 +$;$K^+ {\bar K}^0$ }}} +
(7 - 2 \sqrt{3} \epsilon_2 ) \gamma^{\mbox{\tiny{$0 +$;$K^+ {\bar K}^0$ }}}
\right] 
\left[ k_{K^0 \pi} (\mu) - \frac{16}{3} L_2^r (\mu) \right]
\right.
\nonumber\\
&&
\left.
+ \frac{\sqrt{3}}{2} \left( 1 - \frac{\epsilon_1}{\sqrt{3}} + \sqrt{3} \epsilon_2 \right)
( \beta^{\mbox{\tiny{$0 K^-$;$\eta K^-$ }}} - \gamma^{\mbox{\tiny{$0 K^-$;$\eta K^-$ }}} ) 
\left[ k_{\eta K} (\mu) - \frac{16}{3} L_2^r (\mu) \right]
\right\}
.
\end{eqnarray}

Let us now consider the coefficients $\pi^{ab}_{0,F/G}$ left aside so far.
Their expressions exhibit the same decomposition
\begin{equation}
\pi^{ab}_{0,F/G} = \big( \pi^{ab}_{0,F/G} \big)_T + \big( \pi^{ab}_{0,F/G} \big)_L
 , \ ab = +-, 00
,
\label{pi_0_T+L}
\end{equation} 
into a contribution from tree and tadpole terms (chiral logarithms from tadpoles are now present), and
a contribution from the unitarity loops, as describe before. Starting with the former,
we find
\begin{eqnarray}
\big( \pi_{0,F}^{\mbox{\tiny{$+-$}}} \big)_{T} (\mu) &=&- 64 M_{\pi}^2 L_1^r (\mu) + 8 M_K^2 L_2^r (\mu)+
2 ( M_K^2 - 8 M_{\pi}^2 ) L_3 + 8 (6 {\widehat m} - m_s ) B_0 L_4^r (\mu)
\nonumber\\
&&
 + 
e^2 F_0^2 \left[ - \frac{4}{3} \left( {\widehat K}_1^r (\mu) - 11 {\widehat K}_2^r (\mu)  \right) + 
\frac{4}{9} \left( 2 {\widehat K}_5^r (\mu) + 11 {\widehat K}_6^r (\mu)  \right) + 2 {\widehat K}_{12}^r (\mu)  
\right]
\nonumber\\
&&
 + \frac{1}{256 \pi^2}
\left[
10 M_{\pi}^2 \ln \frac{M_{\pi}^2}{\mu^2} + \left( 11 - 2 \sqrt{3} \epsilon_2 \right) M_{\pi^0}^2 \ln \frac{M_{\pi^0}^2}{\mu^2} +
2 M_{K}^2 \ln \frac{M_{K}^2}{\mu^2} + 4 M_{K^0}^2 \ln \frac{M_{K^0}^2}{\mu^2}
\right.
\nonumber\\
&&
\left.
- 3 \left( 1 - \frac{2}{\sqrt{3}} \epsilon_1 \right) M_{\eta}^2  \ln \frac{M_{\eta}^2}{\mu^2}
\right]
,
\end{eqnarray}
\begin{eqnarray}
\big( \pi_{0,G}^{\mbox{\tiny{$+-$}}} \big)_{T} (\mu) &=& - 2 M_K^2 L_3 - 8 (2 {\widehat m} + m_s ) B_0 L_4^r (\mu)
\nonumber\\
&&
 + 
e^2 F_0^2 \left[ - \frac{4}{3} \left( {\widehat K}_1^r (\mu) + {\widehat K}_2^r (\mu)  \right) + 
4 \left( 2 {\widehat K}_3^r (\mu) + {\widehat K}_4^r (\mu)  \right) - 
\frac{4}{9} \left( 4 {\widehat K}_5^r (\mu) - 5 {\widehat K}_6^r (\mu)  \right) + 2 {\widehat K}_{12}^r (\mu)  
\right]
\nonumber\\
&&
 + \frac{1}{256 \pi^2}
\left[
6 M_{\pi}^2 \ln \frac{M_{\pi}^2}{\mu^2} + \left( 5 + 2 \sqrt{3} \epsilon_2 \right) M_{\pi^0}^2 \ln \frac{M_{\pi^0}^2}{\mu^2} +
6 M_{K}^2 \ln \frac{M_{K}^2}{\mu^2} + 4 M_{K^0}^2 \ln \frac{M_{K^0}^2}{\mu^2}
\right.
\nonumber\\
&&
\left.
+ 3 \left( 1 - \frac{2}{\sqrt{3}} \epsilon_1 \right) M_{\eta}^2  \ln \frac{M_{\eta}^2}{\mu^2}
\right]
,
\end{eqnarray}
and
\begin{eqnarray}
\big( \pi_{0,F}^{\mbox{\tiny{$00$}}} \big)_T &=& - 64 M_{\pi^0}^2 L_1^r (\mu) + 8 M_K^2 L_2^r (\mu) +
2 \left[ \left( 1 - \frac{2}{\sqrt{3}} \epsilon_2 \right) M_K^2 - 8 \left( 1 + \frac{2}{\sqrt{3}} \epsilon_2 \right) M_{\pi^0}^2  \right] L_3 
\nonumber\\
&&
+ 
8 \left[ 6 {\widehat m} - m_s -2 \sqrt{3} \epsilon_2 ( 2 {\widehat m} + m_s) \right] B_0 L_4^r (\mu) -
8 \left( \frac{1}{R} - 4 \frac{\epsilon_2}{ \sqrt{3}}  \right) ( m_s - {\widehat m}) B_0 L_5^r (\mu) 
\nonumber\\
&&
 -
e^2 F_0^2 \left[ \frac{4}{3} \left( {\widehat K}_1^r (\mu) + {\widehat K}_2^r (\mu)  \right) 
+ \frac{4}{9} \left( {\widehat K}_5^r (\mu) + {\widehat K}_6^r (\mu)  \right) - 
2   {\widehat K}_{12}^r (\mu)  
\right]
\nonumber\\
&&
 + \frac{1}{256 \pi^2}
\left[
2 ( 11 + 26 \sqrt{3} \epsilon_2 ) M_{\pi}^2 \ln \frac{M_{\pi}^2}{\mu^2} - 
\left( 1 + 4 \sqrt{3} \epsilon_2 \right) M_{\pi^0}^2 \ln \frac{M_{\pi^0}^2}{\mu^2}
\right.
\nonumber\\
&&
\left. 
- 4  ( 1 - 2 \sqrt{3} \epsilon_2 ) M_{K}^2 \ln \frac{M_{K}^2}{\mu^2} + 
2  ( 5 - 2 \sqrt{3} \epsilon_2 ) M_{K^0}^2 \ln \frac{M_{K^0}^2}{\mu^2}
\right.
\nonumber\\
&&
\left.
- 3 \left( 1 - 2 \frac{ \epsilon_1}{\sqrt{3}} + 2 \sqrt{3} \epsilon_2 \right) M_{\eta}^2  \ln \frac{M_{\eta}^2}{\mu^2}
\right]
.
\end{eqnarray}
A few remarks are in order at this stage. First, one notices that in the isospin limit these expressions differ from the 
corresponding ones in Refs.~\cite{Bijnens94,Bijnens:1994me} by the contribution proportional to $L_4$ and by
the absence of a contribution proportional to $L_5$ [a
contribution proportional to $L_5$ appears in $\big( \pi_{0,F}^{\mbox{\tiny{$00$}}} \big)_T$, but it vanishes in the
isospin limit, and is identically zero at lowest order, where $\epsilon_{1,2} = \sqrt{3}/(4R)$]. 
These differences are entirely due to the choice of the overall normalization
in Eqs.~(\ref{FG_+-_1loop}) and (\ref{FG_00_1loop}), with $F_0$ instead
of $F_\pi$ in the denominator. This explains
also the differences in the tadpole contribution. Secondly, we also note the appearance of the low-energy
constants ${\widehat K}_i^r (\mu)$, already present in equations (\ref{IB_in_a_+-}), (\ref{IB_in_a_x}) and (\ref{IB_in_a_+0}).
Finally, once the normalization issue is accounted for, the contributions
proportional to $L_4$ are written in terms of ${\widehat m}B_0$ and $m_s B_0$, which is their actual form when
computed from the chiral Lagrangian. Usually, these contributions are directly expressed in terms of the meson masses, but we refrain from following this practice: as we will show shortly, and for reasons similar to those already discussed in the introduction [cf.  Eqs.~(\ref{example_1})-(\ref{example_3})], the expression of these terms is again dictated 
by the contributions coming from the unitarity loops, 
$\big( \pi_{0,F}^{\mbox{\tiny{$+-$}}} \big)_{L}$, $\big( \pi_{0,G}^{\mbox{\tiny{$+-$}}} \big)_{L}$,
and $\big( \pi_{0,F}^{\mbox{\tiny{$00$}}} \big)_{L}$. 
Turning to the latter, we obtain the somewhat lengthy expressions
\begin{eqnarray}
\big( \pi_{0,F}^{\mbox{\tiny{$+-$}}} \big)_{L} (\mu) &=& \frac{F_0^2}{F_\pi^2}
\left[
\frac{1}{2} \, ( \beta^{\mbox{\tiny{$+ K^-$}};\mbox{\tiny{$+ K^-$}}} - 2 \beta^{\mbox{\tiny{$K^- -$}};\mbox{\tiny{$K^- -$}}}
- 3 \gamma^{\mbox{\tiny{$+ K^-$}};\mbox{\tiny{$+ K^-$}}} ) (M_K^2 + 2 M_\pi^2)
\right. 
\nonumber\\
&&
\left. 
 +
( - \beta^{\mbox{\tiny{$+ K^-$}};\mbox{\tiny{$+ K^-$}}} + 6 \gamma^{\mbox{\tiny{$K^- -$}};\mbox{\tiny{$K^- -$}}}
+ 4 \gamma^{\mbox{\tiny{$+ K^-$}};\mbox{\tiny{$+ K^-$}}} ) (M_K^2 - M_\pi^2)
\right. 
\nonumber\\
&&
\left. 
+ 16 \pi F_\pi^2 \varphi_S^{\mbox{\tiny{$+ K^-$}};\mbox{\tiny{$+ K^-$}}} (0)
- 32 \pi F_\pi^2 \varphi_S^{\mbox{\tiny{$K^- -$}};\mbox{\tiny{$K^- -$}}} (0)
\right] k_{K\pi} (\mu)
\nonumber\\
&&
+ \frac{F_0^2}{F_\pi^2} \frac{1}{\sqrt{2}} \left\{
 \frac{3}{2} 
\left[ - \left( 1 + \frac{\epsilon_2}{\sqrt{3}} \right) \beta^{\mbox{\tiny{$+ K^-$}};\mbox{\tiny{$0 {\bar K}^0$}}} +
( 1 - \sqrt{3} \epsilon_2 ) \gamma^{\mbox{\tiny{$+ K^-$}};\mbox{\tiny{$0 {\bar K}^0$}}}
\right]  (M_K^2 + 2 M_\pi^2)
\right. 
\nonumber\\
&&
\left. 
 -
\left[ 9 \left( 1 + \frac{\epsilon_2}{\sqrt{3}} \right) \gamma^{\mbox{\tiny{$+ K^-$}};\mbox{\tiny{$0 {\bar K}^0$}}} -
( 1 - \sqrt{3} \epsilon_2 ) \beta^{\mbox{\tiny{$+ K^-$}};\mbox{\tiny{$0 {\bar K}^0$}}} \right] (M_{K^0}^2 - M_{\pi^0}^2)
\right. 
\nonumber\\
&&
\left. 
- ( 1 - \sqrt{3} \epsilon_2 ) \gamma^{\mbox{\tiny{$+ K^-$}};\mbox{\tiny{$0 {\bar K}^0$}}} (M_{K}^2 - M_{\pi}^2)
- 3 \left( 1 + \frac{\epsilon_2}{\sqrt{3}} \right) 16 \pi F_\pi^2 \varphi_S^{\mbox{\tiny{$+ K^-$}};\mbox{\tiny{$0 {\bar K}^0$}}} (0)
\right \}
k_{K^0 \pi^0} (\mu)
\nonumber\\
&&
+ \frac{F_0^2}{F_\pi^2} \sqrt{\frac{3}{2}} \left\{
- \frac{1}{2} \left[
( 1 - \sqrt{3} \epsilon_1 ) \beta^{\mbox{\tiny{$+ K^-$}};\mbox{\tiny{$\eta {\bar K}^0$}}} +
3 \left( 1 + \frac{\epsilon_1}{\sqrt{3}} \right) \gamma^{\mbox{\tiny{$+ K^-$}};\mbox{\tiny{$\eta {\bar K}^0$}}}
\right] (M_K^2 + 2 M_\pi^2)
\right. 
\nonumber\\
&&
\left. 
+ \left[ 3 ( 1 - \sqrt{3} \epsilon_1 ) \gamma^{\mbox{\tiny{$+ K^-$}};\mbox{\tiny{$\eta {\bar K}^0$}}}
+ \left( 1 + \frac{\epsilon_1}{\sqrt{3}} \right) \beta^{\mbox{\tiny{$+ K^-$}};\mbox{\tiny{$\eta {\bar K}^0$}}}
\right] (M_\eta^2 - M_{K^0}^2 )
\right. 
\nonumber\\
&&
\left. 
+ \left( 1 + \frac{\epsilon_1}{\sqrt{3}} \right) \gamma^{\mbox{\tiny{$+ K^-$}};\mbox{\tiny{$\eta {\bar K}^0$}}} ( M_K^2 - M_\pi^2 )
- ( 1 - \sqrt{3} \epsilon_1 ) 16 \pi F_\pi^2 \varphi_S^{\mbox{\tiny{$+ K^-$}};\mbox{\tiny{$\eta {\bar K}^0$}}} (0)
\right\}
k_{\eta K^0} (\mu)
\nonumber\\
&&
+ 16 \pi F_0^2 \left[
(1 + 2 \sqrt{3} \epsilon_2 ) \varphi_S^{\mbox{\tiny{$+-$;$00$}}} (0) k_{\pi^0 \pi^0} (\mu)
- 2 \varphi_S^{\mbox{\tiny{$+-$;$+-$}}} (0) k_{\pi \pi} (\mu)
\right.
\nonumber\\
&&
\left.
- 4 \varphi_S^{\mbox{\tiny{$+-$;$K^+ K^-$}}} (0) k_{KK} (\mu)
+ 2 \varphi_S^{\mbox{\tiny{$+-$;$K^0 {\bar K}^0$}}} (0) k_{K^0 {\bar K}^0} (\mu)
\right.
\nonumber\\
&&
\left.
+ 3 \left( 1 - \frac{2}{\sqrt{3}} \epsilon_1 \right) \varphi_S^{\mbox{\tiny{$+-$;$\eta\eta$}}} (0) k_{\eta \eta} (\mu)
+ 2 \sqrt{3} \left( 1 - \frac{\epsilon_1}{\sqrt{3}} + \sqrt{3}\epsilon_2 \right)
\varphi_S^{\mbox{\tiny{$+-$;$0\eta$}}} (0) k_{\eta \pi^0} (\mu)
\right]
,
\label{pi_0_F_+-_L}
\end{eqnarray}
\begin{eqnarray}
\big( \pi_{0,G}^{\mbox{\tiny{$+-$}}} \big)_{L} (\mu) &=&\frac{F_0^2}{F_\pi^2} 
\left[
- \frac{1}{2} \, ( \beta^{\mbox{\tiny{$+ K^-$}};\mbox{\tiny{$+ K^-$}}} + 2 \beta^{\mbox{\tiny{$K^- -$}};\mbox{\tiny{$K^- -$}}}
+ \gamma^{\mbox{\tiny{$+ K^-$}};\mbox{\tiny{$+ K^-$}}} ) (M_K^2 + 2 M_\pi^2)
\right. 
\nonumber\\
&&
\left. 
 +
( \beta^{\mbox{\tiny{$+ K^-$}};\mbox{\tiny{$+ K^-$}}} + 6 \gamma^{\mbox{\tiny{$K^- -$}};\mbox{\tiny{$K^- -$}}}
- 4 \gamma^{\mbox{\tiny{$+ K^-$}};\mbox{\tiny{$+ K^-$}}} ) (M_K^2 - M_\pi^2)
\right. 
\nonumber\\
&&
\left. 
- 16 \pi F_\pi^2 \varphi_S^{\mbox{\tiny{$+ K^-$}};\mbox{\tiny{$+ K^-$}}} (0)
- 32 \pi F_\pi^2 \varphi_S^{\mbox{\tiny{$K^- -$}};\mbox{\tiny{$K^- -$}}} (0)
\right]
k_{K \pi} (\mu)
\nonumber\\
&&
+ \frac{F_0^2}{F_\pi^2} \frac{1}{\sqrt{2}} \left\{
 \frac{3}{2} 
\left[ \left( 1 + \frac{\epsilon_2}{\sqrt{3}} \right) \beta^{\mbox{\tiny{$+ K^-$}};\mbox{\tiny{$0 {\bar K}^0$}}} +
\frac{1}{3} ( 1 - \sqrt{3} \epsilon_2 ) \gamma^{\mbox{\tiny{$+ K^-$}};\mbox{\tiny{$0 {\bar K}^0$}}}
\right]  (M_K^2 + 2 M_\pi^2)
\right. 
\nonumber\\
&&
\left. 
 + 
\left[ 9 \left( 1 + \frac{\epsilon_2}{\sqrt{3}} \right) \gamma^{\mbox{\tiny{$+ K^-$}};\mbox{\tiny{$0 {\bar K}^0$}}} -
( 1 - \sqrt{3} \epsilon_2 ) \beta^{\mbox{\tiny{$+ K^-$}};\mbox{\tiny{$0 {\bar K}^0$}}} \right] (M_{K^0}^2 - M_{\pi^0}^2)
\right. 
\nonumber\\
&&
\left. 
+ ( 1 - \sqrt{3} \epsilon_2 ) \gamma^{\mbox{\tiny{$+ K^-$}};\mbox{\tiny{$0 {\bar K}^0$}}} (M_{K}^2 - M_{\pi}^2)
+ 3 \left( 1 + \frac{\epsilon_2}{\sqrt{3}} \right) 16 \pi F_\pi^2 \varphi_S^{\mbox{\tiny{$+ K^-$}};\mbox{\tiny{$0 {\bar K}^0$}}} (0)
\right \}
k_{K^0 \pi^0} (\mu)
\nonumber\\
&&
+ \frac{F_0^2}{F_\pi^2} \sqrt{\frac{3}{2}} \left\{
\frac{1}{2} \left[
( 1 - \sqrt{3} \epsilon_1 ) \beta^{\mbox{\tiny{$+ K^-$}};\mbox{\tiny{$\eta {\bar K}^0$}}} -
 \left( 1 + \frac{\epsilon_1}{\sqrt{3}} \right) \gamma^{\mbox{\tiny{$+ K^-$}};\mbox{\tiny{$\eta {\bar K}^0$}}}
\right] (M_K^2 + 2 M_\pi^2)
\right. 
\nonumber\\
&&
\left. 
- \left[ 3 ( 1 - \sqrt{3} \epsilon_1 ) \gamma^{\mbox{\tiny{$+ K^-$}};\mbox{\tiny{$\eta {\bar K}^0$}}}
+ \left( 1 + \frac{\epsilon_1}{\sqrt{3}} \right) \beta^{\mbox{\tiny{$+ K^-$}};\mbox{\tiny{$\eta {\bar K}^0$}}}
\right] (M_\eta^2 - M_{K^0}^2 )
\right. 
\nonumber\\
&&
\left. 
- \left( 1 + \frac{\epsilon_1}{\sqrt{3}} \right) \gamma^{\mbox{\tiny{$+ K^-$}};\mbox{\tiny{$\eta {\bar K}^0$}}} ( M_K^2 - M_\pi^2 )
+ ( 1 - \sqrt{3} \epsilon_1 ) 16 \pi F_\pi^2 \varphi_S^{\mbox{\tiny{$+ K^-$}};\mbox{\tiny{$\eta {\bar K}^0$}}} (0)
\right\}
k_{\eta K^0} (\mu)
,
\label{pi_0_G_+-_L}
\end{eqnarray}
and
\begin{eqnarray}
\big( \pi_{0,F}^{\mbox{\tiny{$00$}}} \big)_L &=& - \frac{F_0^2}{F_\pi^2} (1 + 2 \sqrt{3} \epsilon_2 ) \left[
\frac{1}{2} 
( \beta^{\mbox{\tiny{$0 K^-$;$0 K^-$ }}} + 3 \gamma^{\mbox{\tiny{$0 K^-$;$0 K^-$ }}} ) (M_K^2 + 2 M_{\pi^0}^2 )
\right.
\nonumber\\
&&
\left.
+ ( \beta^{\mbox{\tiny{$0 K^-$;$0 K^-$ }}} + 2 \gamma^{\mbox{\tiny{$0 K^-$;$0 K^-$ }}} ) (M_K^2 - M_{\pi^0}^2 )
+ 16 \pi F_\pi^2 \varphi_S^{\mbox{\tiny{$0 K^-$;$0 K^-$ }}} (0)
\right] k_{K\pi^0} (\mu)
\nonumber\\
&&
+ {\sqrt{2}} \frac{F_0^2}{F_\pi^2} \left\{
- \frac{3}{2} \left[
\left( 1 - \frac{\epsilon_2}{\sqrt{3}} \right) \beta^{\mbox{\tiny{$0 K^-$;$- {\bar K}^0$ }}} -
 (1 + \sqrt{3} \epsilon_2 ) \gamma^{\mbox{\tiny{$0 K^-$;$- {\bar K}^0$ }}}
\right]  (M_K^2 + 2 M_{\pi^0}^2 )
\right.
\nonumber\\
&&
\left.
+
\left[
(1 + \sqrt{3} \epsilon_2 ) \beta^{\mbox{\tiny{$0 K^-$;$- {\bar K}^0$ }}} - 9 \left( 1 - \frac{\epsilon_2}{\sqrt{3}} \right) \gamma^{\mbox{\tiny{$0 K^-$;$- {\bar K}^0$ }}}
\right] (M_{K^0}^2 - M_\pi^2)
\right.
\nonumber\\
&&
\left.
- (1 + \sqrt{3} \epsilon_2 ) \gamma^{\mbox{\tiny{$0 K^-$;$- {\bar K}^0$ }}} (M_K^2 - M_{\pi^0}^2 )
- 3 \left( 1 - \frac{\epsilon_2}{\sqrt{3}} \right) 16 \pi \varphi_S^{\mbox{\tiny{$0 K^-$;$- {\bar K}^0$ }}} (0)
\right\} k_{K^0 \pi} (\mu)
\nonumber\\
&&
+ \sqrt{3} \left( 1 - \frac{\epsilon_1}{\sqrt{3}} + \sqrt{3} \epsilon_2 \right) \frac{F_0^2}{F_\pi^2}
\left[
\frac{1}{2} ( \beta^{\mbox{\tiny{$0 K^-$;$\eta K^-$ }}} + 3 \gamma^{\mbox{\tiny{$0 K^-$;$\eta K^-$ }}} ) (2 M_\eta^2 - 3 M_K^2 - 2 M_{\pi^0}^2 )
\right.
\nonumber\\
&&
\left.
+ \gamma^{\mbox{\tiny{$0 K^-$;$\eta K^-$ }}} (M_K^2 - M_{\pi^0}^2 )
- 16 \pi \varphi_S^{\mbox{\tiny{$0 K^-$;$\eta K^-$ }}} (0)
\right] k_{\eta K} (\mu)
\nonumber\\
&&
+ \frac{F_0^2}{F_\pi^2} \left[
- (1 + 2 \sqrt{3} \epsilon_2 ) 16 \pi F_\pi^2 \varphi_S^{\mbox{\tiny{$00$;$00$}}} (0) k_{\pi^0 \pi^0} (\mu)
+ 32 \pi F_\pi^2 \varphi_S^{\mbox{\tiny{$00$;$+-$}}} (0) k_{\pi \pi} (\mu)
\right.
\nonumber\\
&&
\left.
+ 64 \pi F_\pi^2 \varphi_S^{\mbox{\tiny{$00$;$K^+ K^-$}}} (0) k_{KK} (\mu)
- 32 \pi F_\pi^2 \varphi_S^{\mbox{\tiny{$00$;$K^0 {\bar K}^0$}}} (0) k_{K^0 {\bar K}^0} (\mu)
\right.
\nonumber\\
&&
\left.
- 3 \left( 1 - \frac{2}{\sqrt{3}} \epsilon_1 \right) 16 \pi F_\pi^2 \varphi_S^{\mbox{\tiny{$00$;$\eta\eta$}}} (0) k_{\eta \eta} (\mu)
- 2\sqrt{3} \left( 1 - \frac{\epsilon_1}{\sqrt{3}} + \sqrt{3} \epsilon_2 \right) 
16 \pi F_\pi^2 \varphi_S^{\mbox{\tiny{$00$;$0\eta$}}} (0) k_{\pi^0 \eta} (\mu)
\right]
\!.
\qquad{ }
\label{pi_0_F_00_L}
\end{eqnarray}
As mentioned above, the sums in (\ref{pi_0_T+L}) should no longer
depend on the renormalization scale $\mu$. The scale dependences
of $\big( \pi_{0,F}^{\mbox{\tiny $+-$}} \big)_T (\mu)$,
$\big( \pi_{0,G}^{\mbox{\tiny $+-$}} \big)_T (\mu)$, and
$\big( \pi_{0,F}^{\mbox{\tiny $00$}} \big)_T (\mu)$ can be worked out through
\begin{equation}
\mu \frac{d}{d \mu} L_i^r (\mu) = - \frac{1}{16 \pi^2} \Gamma_i ,
\quad
\mu \frac{d}{d \mu} e^2 {\widehat K}_i^r ( \mu ) = - \frac{1}{16 \pi^2} \frac{\Delta_\pi}{F_0^2} {\widehat \Sigma}_i
,
\label{scale_dep_1}
\end{equation}
where the coefficients $\Gamma_i$ were worked out in \cite{Gasser:1984gg},
and the coefficients ${\widehat \Sigma}_i$ can be obtained from the coefficients
$\Sigma_i$ of Ref.~\cite{Urech:1994hd}, as explained in Ref.~\cite{DescotesGenon:2012gv}:
\begin{eqnarray}
&&
\Gamma_1 = \frac{3}{32} , \ \Gamma_2 = \frac{3}{16} , \ \Gamma_3 = 0 , \ \Gamma_4 = \frac{1}{8} ,
\ \Gamma_5 = \frac{3}{8} , \ \Gamma_9 = \frac{1}{4} ,
\nonumber\\
&&
{\widehat \Sigma}_1 = 0 , \ {\widehat \Sigma}_2 = \frac{1}{2} , \ {\widehat \Sigma}_3 = 0 , \ {\widehat \Sigma}_4 = 1 ,
\ {\widehat \Sigma}_5 = 0 , \ {\widehat \Sigma}_6 = \frac{3}{4} ,
\nonumber\\
&&
{\widehat \Sigma}_7 = 0 , \ {\widehat \Sigma}_8 = \frac{1}{2} , \ {\widehat \Sigma}_9 = 0 , \ {\widehat \Sigma}_{10} = \frac{3}{4}
, \ {\widehat \Sigma}_{11} = 0 , \ {\widehat \Sigma}_{12} = 0 ,
\label{scale_dep_2}
\end{eqnarray}
The corresponding expressions for $\mu (d / d \mu) \big( \pi_{0,F}^{\mbox{\tiny $+-$}} \big)_L (\mu)$,
$\big( \pi_{0,G}^{\mbox{\tiny $+-$}} \big)_L (\mu)$ and
$\big( \pi_{0,F}^{\mbox{\tiny $00$}} \big)_L (\mu)$ can be directly read off from Eqs.~(\ref{pi_0_F_+-_L}),
(\ref{pi_0_G_+-_L}), and (\ref{pi_0_F_00_L}), respectively, upon replacing $k_{ab} (\mu)$ by $-1/(16\pi^2)$. 
In order to complete this exercise, the lowest-order expressions of the various
quantities $\varphi_S^{ab ; a^\prime b^\prime} (0)$ are requested, in addition to the information
already provided by Tab.~\ref{tab:beta_gamma}. These expressions can be found in Tab.~\ref{tab:varphi(0)}. 
The lowest-order expressions of the masses and mixing angles are also needed, see Refs.~\cite{Gasser:1984gg} and \cite{Urech:1994hd}.
Scale invariance is achieved through identities which, e.g. in the case of $\pi_{0,F}^{\mbox{\tiny $+-$}}$ and at lowest order, relates  the combination $8 (2 {\widehat m} + m_s ) B_0 $  to a certain
combination of masses and  parameters describing the lowest-order scattering amplitudes, which multiply the chiral logarithms arising from the tadpole and loop contributions. 
Instead, one may use this identity in order to rewrite $\pi_{0,F}^{\mbox{\tiny $+-$}}$ in an equivalent form,
which is explicitly scale independent but with no explicit reference to $B_0$ any longer
\begin{table}[t]
\begin{center}
\begin{tabular}{|c|c||c|}
\hline
  $a b$    &  $a^\prime b^\prime$  & $ 16 \pi F_0^2 \varphi_S^{a b ; a^\prime b^\prime} (0)$ 
\\ \hline\hline
 $\pi^+ \pi^-$ & $ K^+ K^-$ &  
$ \frac{1}{6} \left[ ( 3 {\widehat m} + m_s ) B_0 - \frac{m_s - {\widehat m} }{2 R} B_0 - M_\pi^2 - M_K^2 + 8 \Delta_\pi \right] $  
\\ \hline
 $\pi^0 \pi^- $ & $K^0 K^- $ & 
$ \frac{1}{6 \sqrt{2}} \left[ \sqrt{3} \epsilon_2 \left( M_\pi^2 + M_{\pi^0}^2 + M_K^2 + M_{K^0}^2\right)  
+ ( m_s - {\widehat m} ) B_0 \left( \frac{1}{R} + 2 \frac{\epsilon_2}{\sqrt{3}}\right) + 3 \Delta_\pi \right]$
\\ \hline
 $\pi^0 \pi^0$ & $ K^+ K^-$ &  
$ - \frac{1}{6} \left[ ( 3 {\widehat m} + m_s ) B_0 - ( m_s - {\widehat m} ) B_0 \left( \frac{3}{2 R} + 2 \frac{\epsilon_2}{\sqrt{3}}\right) 
- ( 1 + 2 \sqrt{3} \epsilon_2 ) ( M_{\pi^0}^2 + M_K^2 - \Delta_\pi ) \right] $  
\\ \hline
 $\pi^0 \pi^0$ & $ K^0 {\bar K}^0$ &  
$ \frac{1}{6} \left[ ( 3 {\widehat m} + m_s ) B_0 + ( m_s - {\widehat m} ) B_0 \left( \frac{3}{2 R} + 2 \frac{\epsilon_2}{\sqrt{3}}\right) 
- ( 1 - 2 \sqrt{3} \epsilon_2 ) ( M_{\pi^0}^2 + M_{K^0}^2 ) \right] $
\\ \hline
 $\pi^+ \pi^-$ & $ K^0 {\bar K}^0$ & 
$ - \frac{1}{6} \left[ ( 3 {\widehat m} + m_s ) B_0 + \frac{ m_s - {\widehat m} }{2 R} B_0 - M_\pi^2 - M_{K^0}^2 + \Delta_\pi \right] $  
\\ \hline\hline
  $\pi^+ K^-$ & $\eta {\bar K}^0$ & 
$ \frac{1}{12}\sqrt{\frac{3}{2}}\left[ ( 1 - \sqrt{3}{\epsilon_1} ) (M_\eta^2 +M_\pi^2 + M_K^2 + M_{K^0}^2 ) - 
\frac{4}{3} (m_s - {\widehat m} ) B_0 - 2 \Delta_\pi \right]$ 
\\ \hline
  $\pi^0 K^-$ & $\eta K^-$ & 
$
\frac{\sqrt{3}}{24}\left[ \left( 1 - \frac{\epsilon_1}{\sqrt{3}} + \sqrt{3}{\epsilon_2} \right) (M_\eta^2 + 2 M_K^2 + M_{\pi^0}^2)
- \frac{4}{3} (m_s - {\widehat m}) B_0 \right.
$
\\
 & &
$ \left.
 - \frac{4}{3} \frac{ m_s - {\widehat m} }{2 R} B_0 
- \frac{4}{3} ( m_s + 3{\widehat m} ) B_0 \sqrt{3} \epsilon_1 + \frac{4}{3} ( 3 m_s + {\widehat m} ) B_0 \sqrt{3} \epsilon_2
+ 4 \Delta_\pi \right]
$
\\ \hline\hline
  $\pi^+ \pi^-$ & $\eta\eta$ & $- \frac{2}{3} {\widehat m} B_0$ 
\\ \hline
  $\pi^+ \pi^-$ & $\pi^0 \eta$ & 
$ - \frac{1}{3} \left[  \epsilon_1 ( M_\eta^2 + 2 M_\pi^2 + M_{\pi^0}^2) - \frac{1}{\sqrt{3}} \frac{ m_s - {\widehat m} }{R} B_0 \right]$ 
\\ \hline
  $\pi^0 \pi^0$ & $\eta\eta$ & $\frac{2}{3} {\widehat m} B_0 $ 
\\ \hline
  $\pi^0 \pi^0$ & $\pi^0 \eta$ & $ - \frac{1}{\sqrt{3}} \frac{ m_s - {\widehat m} }{R} B_0 $ 
\\ \hline
\end{tabular}
\caption{The quantities $\varphi_S^{a b ; a^\prime b^\prime} (0)$
corresponding to various lowest-order amplitudes appearing in the one-loop expressions of the
form factors $F^{ab}(s,t,u)$ and $G^{ab}(s,t,u)$ discussed in Sec.~\ref{sec:1-loop_ff}.}\label{tab:varphi(0)}
\end{center}
\end{table}
\begin{eqnarray}
\pi_{0,F}^{\mbox{\tiny{$+-$}}} &=& - 64 M_{\pi}^2 \left[ L_1^r (\mu) - \frac{3}{4}  L_4^r (\mu) \right] 
+ 8 M_K^2 \left[ L_2^r (\mu) - \frac{3}{2}  L_4^r (\mu) \right] +
2 ( M_K^2 - 8 M_{\pi}^2 ) L_3 
\nonumber\\
&&
 - 
  \left[ \frac{4}{3}e^2F_0^2 \left( {\widehat K}_1^r (\mu) - 11 {\widehat K}_2^r (\mu)  \right) - 
\frac{4}{9}e^2F_0^2 \left( 2 {\widehat K}_5^r (\mu) + 11 {\widehat K}_6^r (\mu)  \right) - 2 e^2F_0^2 {\widehat K}_{12}^r (\mu) 
+  88\Delta_\pi L_4^r (\mu) 
\right]
\nonumber\\
&&
 + 
10 M_{\pi}^2 \left( \frac{1}{256 \pi^2} \ln \frac{M_{\pi}^2}{\mu^2} - L_4^r (\mu) \right)
+ \left( 11 - 2 \sqrt{3} \epsilon_2 \right) M_{\pi^0}^2 \left( \frac{1}{256 \pi^2} \ln \frac{M_{\pi^0}^2}{\mu^2} - L_4^r (\mu) \right)
\nonumber\\
&&
+ 2 M_{K}^2 \left( \frac{1}{256 \pi^2} \ln \frac{M_{K}^2}{\mu^2} - L_4^r (\mu) \right) 
+ 4 M_{K^0}^2 \left( \frac{1}{256 \pi^2} \ln \frac{M_{K^0}^2}{\mu^2} - L_4^r (\mu) \right)
\nonumber\\
&&
- 3 \left( 1 - \frac{2}{\sqrt{3}} \epsilon_1 \right) M_{\eta}^2 \left( \frac{1}{256 \pi^2}\ln \frac{M_{\eta}^2}{\mu^2} - L_4^r (\mu) \right)
\nonumber\\
&&
+ \frac{F_0^2}{F_\pi^2}
\left[
\frac{1}{6} \beta^{\mbox{\tiny{$+ -$}};\mbox{\tiny{$K^+ K^-$}}} ( 7 M_K^2 - 19 M_\pi^2 )
- \frac{17}{2} \gamma^{\mbox{\tiny{$+ -$}};\mbox{\tiny{$K^+ K^-$}}} ( M_K^2 - M_\pi^2 )
\right. 
\nonumber\\
&&
\left. 
- 16 \pi F_\pi^2 \varphi_S^{\mbox{\tiny{$+ -$}};\mbox{\tiny{$K^+ K^-$}}} (0)
\right] \left[ k_{K\pi} (\mu) - 8 L_4^r (\mu) \right]
\nonumber\\
&&
+ \frac{F_0^2}{F_\pi^2} \frac{1}{\sqrt{2}} \left\{
\left[ -  \beta^{\mbox{\tiny{$0 -$}};\mbox{\tiny{$K^0 K^-$}}} +
6 \left( 1 + \frac{\sqrt{3}}{2} \epsilon_2 \right) \gamma^{\mbox{\tiny{$0 -$}};\mbox{\tiny{$K^0 K^-$}}}
\right]  (M_K^2 + 2 M_\pi^2)
\right. 
\nonumber\\
&&
\left. 
+ \frac{1}{6} ( 1 - \sqrt{3} \epsilon_2 )
\left[ \beta^{\mbox{\tiny{$0 -$}};\mbox{\tiny{$K^0 K^-$}}} + 3 \gamma^{\mbox{\tiny{$0 -$}};\mbox{\tiny{$K^0 K^-$}}}
\right]  (M_K^2 - M_\pi^2)
 + 2
\left[ 
\beta^{\mbox{\tiny{$0 -$}};\mbox{\tiny{$K^0 K^-$}}} + 3 \sqrt{3} \epsilon_2 \gamma^{\mbox{\tiny{$0 -$}};\mbox{\tiny{$K^0 K^-$}}} 
\right] (M_{K^0}^2 - M_{\pi^0}^2)
\right. 
\nonumber\\
&&
\left. 
+ \frac{3}{2} \left( 1 + \frac{\epsilon_2}{\sqrt{3}} \right)
\left[ \beta^{\mbox{\tiny{$0 -$}};\mbox{\tiny{$K^0 K^-$}}} - 3 \gamma^{\mbox{\tiny{$0 -$}};\mbox{\tiny{$K^0 K^-$}}}
\right] (M_K^2 + M_{K^0}^2 + M_\pi^2 + M_{\pi^0}^2 )
\right. 
\nonumber\\
&&
\left. 
+ 3 \left( 1 + \frac{\epsilon_2}{\sqrt{3}} \right) 16 \pi F_\pi^2 \varphi_S^{\mbox{\tiny{$0 -$}};\mbox{\tiny{$K^0 K^-$}}} (0)
\right \}
\left[ k_{K^0 \pi^0} (\mu) - 8 L_4^r (\mu) \right]
\nonumber\\
&&
+ \frac{F_0^2}{F_\pi^2} \sqrt{\frac{3}{2}} \left\{
- \frac{1}{2} \left[
( 1 - \sqrt{3} \epsilon_1 ) \beta^{\mbox{\tiny{$+ K^-$}};\mbox{\tiny{$\eta {\bar K}^0$}}} +
3 \left( 1 + \frac{\epsilon_1}{\sqrt{3}} \right) \gamma^{\mbox{\tiny{$+ K^-$}};\mbox{\tiny{$\eta {\bar K}^0$}}}
\right] (M_K^2 + 2 M_\pi^2)
\right. 
\nonumber\\
&&
\left. 
+ \left[ 3 ( 1 - \sqrt{3} \epsilon_1 ) \gamma^{\mbox{\tiny{$+ K^-$}};\mbox{\tiny{$\eta {\bar K}^0$}}}
+ \left( 1 + \frac{\epsilon_1}{\sqrt{3}} \right) \beta^{\mbox{\tiny{$+ K^-$}};\mbox{\tiny{$\eta {\bar K}^0$}}}
\right] (M_\eta^2 - M_{K^0}^2 )
\right. 
\nonumber\\
&&
\left. 
+ \left( 1 + \frac{\epsilon_1}{\sqrt{3}} \right) \gamma^{\mbox{\tiny{$+ K^-$}};\mbox{\tiny{$\eta {\bar K}^0$}}} ( M_K^2 - M_\pi^2 )
- ( 1 - \sqrt{3} \epsilon_1 ) 16 \pi F_\pi^2 \varphi_S^{\mbox{\tiny{$+ K^-$}};\mbox{\tiny{$\eta {\bar K}^0$}}} (0)
\right\}
\left[ k_{\eta K^0} (\mu) - 8 L_4^r (\mu) \right]
\nonumber\\
&&
+ \frac{F_0^2}{F_\pi^2} \left\{
(1 + 2 \sqrt{3} \epsilon_2 ) 16 \pi F_\pi^2 \varphi_S^{\mbox{\tiny{$+-$;$00$}}} (0) \left[ k_{\pi^0 \pi^0} (\mu) - 8 L_4^r (\mu) \right]
- 32 \pi F_\pi^2 \varphi_S^{\mbox{\tiny{$+-$;$+-$}}} (0) \left[ k_{\pi \pi} (\mu) - 8 L_4^r (\mu) \right]
\right.
\nonumber\\
&&
\left.
- 64 \pi F_\pi^2 \varphi_S^{\mbox{\tiny{$+-$;$K^+ K^-$}}} (0) \left[ k_{KK} (\mu) - 8 L_4^r (\mu) \right]
+ 32 \pi F_\pi^2 \varphi_S^{\mbox{\tiny{$+-$;$K^0 {\bar K}^0$}}} (0) 
\left[ k_{K^0 {\bar K}^0} (\mu) - 8 L_4^r (\mu) \right]
\right.
\nonumber\\
&&
\left.
+ 48 \pi F_\pi^2 \left( 1 - \frac{2}{\sqrt{3}} \epsilon_1 \right) \varphi_S^{\mbox{\tiny{$+-$;$\eta\eta$}}} (0) 
\left[ k_{\eta \eta} (\mu) - 8 L_4^r (\mu) \right]
\right.
\nonumber\\
&&
\left.
+ 32 \pi \sqrt{3} \left( 1 - \frac{\epsilon_1}{\sqrt{3}} + \sqrt{3}\epsilon_2 \right)
\varphi_S^{\mbox{\tiny{$+-$;$0\eta$}}} (0) 
\left[ k_{\eta \pi^0} (\mu) - 8 L_4^r (\mu) \right]
\right\}
.
\end{eqnarray}
Proceeding in the same way with $\pi_{0,G}^{\mbox{\tiny{$+-$}}}$ and $\pi_{0,F}^{\mbox{\tiny{$00$}}}$, we obtain
\begin{eqnarray}
\pi_{0,G}^{\mbox{\tiny{$+-$}}} &=& - 2 M_K^2 L_3 
 - 
 \left[ \frac{4}{3}e^2 F_0^2 \left( {\widehat K}_1^r (\mu) + {\widehat K}_2^r (\mu)  \right) - 
4 e^2 F_0^2\left( 2 {\widehat K}_3^r (\mu) + {\widehat K}_4^r (\mu)  \right) + \frac{4}{9}e^2 F_0^2 \left( 4 {\widehat K}_5^r (\mu) - 5 {\widehat K}_6^r (\mu)  \right)
\right.
\nonumber\\
&&
\left.
- 2e^2 F_0^2 {\widehat K}_{12}^r (\mu)  
+  40\Delta_\pi L_4^r (\mu) 
\right]
\nonumber\\
&&
 + 
6 M_{\pi}^2 \left( \frac{1}{256 \pi^2} \ln \frac{M_{\pi}^2}{\mu^2} - L_4^r (\mu) \right) 
+ \left( 5 + 2 \sqrt{3} \epsilon_2 \right) M_{\pi^0}^2 \left( \frac{1}{256 \pi^2} \ln \frac{M_{\pi^0}^2}{\mu^2} - L_4^r (\mu) \right) 
\nonumber\\
&&
+ 6 M_{K}^2 \left( \frac{1}{256 \pi^2} \ln \frac{M_{K}^2}{\mu^2} - L_4^r (\mu) \right) 
+ 4 M_{K^0}^2 \left( \frac{1}{256 \pi^2} \ln \frac{M_{K^0}^2}{\mu^2} - L_4^r (\mu) \right)
\nonumber\\
&&
+ 3 \left( 1 - \frac{2}{\sqrt{3}} \epsilon_1 \right) M_{\eta}^2  \left( \frac{1}{256 \pi^2} \ln \frac{M_{\eta}^2}{\mu^2} - L_4^r (\mu) \right)
\nonumber\\
&&
- \frac{F_0^2}{F_\pi^2} 
\left[
\frac{1}{2} \left( 5 \beta^{\mbox{\tiny{$+ -$}};\mbox{\tiny{$K^+ K^-$}}} + 3 \gamma^{\mbox{\tiny{$+ -$}};\mbox{\tiny{$K^+ K^-$}}} \right) M_K^2
+ \frac{3}{2} \left( \beta^{\mbox{\tiny{$+ -$}};\mbox{\tiny{$K^+ K^-$}}} - \gamma^{\mbox{\tiny{$+ -$}};\mbox{\tiny{$K^+ K^-$}}} \right) M_\pi^2
\right. 
\nonumber\\
&&
\left. 
+ 3 \cdot 16 \pi F_\pi^2 \varphi^{\mbox{\tiny{$+ -$}};\mbox{\tiny{$K^+ K^-$}}}_S (0)
\right]
\left[ k_{K \pi} (\mu) - 8 L_4^r (\mu) \right]
\nonumber\\
&&
+ \frac{F_0^2}{F_\pi^2} \frac{1}{\sqrt{2}} \left\{
 \frac{1}{3} 
\left[ 2 \left( 1 + \frac{\sqrt{3}}{2} \epsilon_2 \right) \beta^{\mbox{\tiny{$0 -$}};\mbox{\tiny{$K^0 K^-$}}} -
3 ( 7 + 2 \sqrt{3} \epsilon_2 ) \gamma^{\mbox{\tiny{$0 -$}};\mbox{\tiny{$K^0 K^-$}}}
\right]  (M_K^2 + 2 M_\pi^2)
\right. 
\nonumber\\
&&
\left. 
 - 
2 \left( \beta^{\mbox{\tiny{$0 -$}};\mbox{\tiny{$K^0 K^-$}}} +
3 \gamma^{\mbox{\tiny{$0 -$}};\mbox{\tiny{$K^0 K^-$}}} \sqrt{3} \epsilon_2 \right) (M_{K^0}^2 - M_{\pi^0}^2) 
- \frac{1}{6} ( 1 - \sqrt{3} \epsilon_2 )
( \beta^{\mbox{\tiny{$0 -$}};\mbox{\tiny{$K^0 K^-$}}} + 3 \gamma^{\mbox{\tiny{$0 -$}};\mbox{\tiny{$K^0 K^-$}}} )  (M_{K}^2 - M_{\pi}^2)
\right. 
\nonumber\\
&&
\left.
- \frac{3}{2} \left( 1 + \frac{\epsilon_2}{\sqrt{3}} \right)
( \beta^{\mbox{\tiny{$0 -$}};\mbox{\tiny{$K^0 K^-$}}} - 3 \gamma^{\mbox{\tiny{$0 -$}};\mbox{\tiny{$K^0 K^-$}}} ) 
( M_{K}^2 + M_{K^0}^2 + M_{\pi}^2 + M_{\pi^0}^2 )
\right. 
\nonumber\\
&&
\left. 
- 3 \left( 1 + \frac{\epsilon_2}{\sqrt{3}} \right) 16 \pi F_\pi^2 \varphi_S^{\mbox{\tiny{$0 -$}};\mbox{\tiny{$K^0 K^-$}}} (0)
\right \}
\left[ k_{K^0 \pi^0} (\mu) - 8 L_4^r (\mu) \right]
\nonumber\\
&&
+ \frac{F_0^2}{F_\pi^2} \sqrt{\frac{3}{2}} \left\{
\frac{1}{2} \left[
( 1 - \sqrt{3} \epsilon_1 ) \beta^{\mbox{\tiny{$+ K^-$}};\mbox{\tiny{$\eta {\bar K}^0$}}} -
 \left( 1 + \frac{\epsilon_1}{\sqrt{3}} \right) \gamma^{\mbox{\tiny{$+ K^-$}};\mbox{\tiny{$\eta {\bar K}^0$}}}
\right] (M_K^2 + 2 M_\pi^2)
\right. 
\nonumber\\
&&
\left. 
- \left[ 3 ( 1 - \sqrt{3} \epsilon_1 ) \gamma^{\mbox{\tiny{$+ K^-$}};\mbox{\tiny{$\eta {\bar K}^0$}}}
+ \left( 1 + \frac{\epsilon_1}{\sqrt{3}} \right) \beta^{\mbox{\tiny{$+ K^-$}};\mbox{\tiny{$\eta {\bar K}^0$}}}
\right] (M_\eta^2 - M_{K^0}^2 )
\right. 
\nonumber\\
&&
\left. 
- \left( 1 + \frac{\epsilon_1}{\sqrt{3}} \right) \gamma^{\mbox{\tiny{$+ K^-$}};\mbox{\tiny{$\eta {\bar K}^0$}}} ( M_K^2 - M_\pi^2 )
+ ( 1 - \sqrt{3} \epsilon_1 ) 16 \pi F_\pi^2 \varphi_S^{\mbox{\tiny{$+ K^-$}};\mbox{\tiny{$\eta {\bar K}^0$}}} (0)
\right\}
\left[ k_{\eta K^0} (\mu) - 8 L_4^r (\mu) \right]
\! ,
\quad { }
\end{eqnarray}
and
\begin{eqnarray}
\pi_{0,F}^{\mbox{\tiny{$00$}}} &=& - 64 M_{\pi^0}^2 \left[ L_1^r (\mu) - \frac{3}{4}  L_4^r (\mu) \right]  
+ 8 M_K^2 \left[ L_2^r (\mu) - \frac{3}{2}  L_4^r (\mu) \right]  
\nonumber\\
&&
+ 2 \left[ \left( 1 - \frac{2}{\sqrt{3}} \epsilon_2 \right) M_K^2 - 8 \left( 1 + \frac{2}{\sqrt{3}} \epsilon_2 \right) M_{\pi^0}^2  \right] L_3 
\nonumber\\
&&
 - 
 \left[ \frac{4}{3}e^2 F_0^2  \left( {\widehat K}_1^r (\mu) + {\widehat K}_2^r (\mu)  \right) + 
\frac{4}{9} e^2 F_0^2 \left( {\widehat K}_5^r (\mu) + {\widehat K}_6^r (\mu)  \right) - 
2e^2 F_0^2  {\widehat K}_{12}^r (\mu)  - 8\Delta_\pi  L_4^r (\mu)
\right]
\nonumber\\
&&
 + 2 ( 11 + 26 \sqrt{3} \epsilon_2 ) M_{\pi}^2 \left( \frac{1}{256 \pi^2} \ln \frac{M_{\pi}^2}{\mu^2} - L_4^r (\mu) \right) - 
\left( 1 + 4 \sqrt{3} \epsilon_2 \right) M_{\pi^0}^2 \left( \frac{1}{256 \pi^2} \ln \frac{M_{\pi^0}^2}{\mu^2} - L_4^r (\mu) \right)
\nonumber\\
&&
- 4  ( 1 - 2 \sqrt{3} \epsilon_2 ) M_{K}^2 \left( \frac{1}{256 \pi^2} \ln \frac{M_{K}^2}{\mu^2} - L_4^r (\mu) \right) + 
2  ( 5 - 2 \sqrt{3} \epsilon_2 ) M_{K^0}^2 \left( \frac{1}{256 \pi^2} \ln \frac{M_{K^0}^2}{\mu^2} - L_4^r (\mu) \right)
\nonumber\\
&&
- 3 \left( 1 - 2 \frac{ \epsilon_1}{\sqrt{3}} + 2 \sqrt{3} \epsilon_2 \right) M_{\eta}^2  \left( \frac{1}{256 \pi^2} \ln \frac{M_{\eta}^2}{\mu^2} - L_4^r (\mu) \right)
\nonumber\\
&&
- \frac{F_0^2}{F_\pi^2} (1 + 2 \sqrt{3} \epsilon_2 ) \left[
 \frac{1}{6} \beta^{\mbox{\tiny{$0 0$;$K^+ K^-$ }}}  (M_K^2 - M_{\pi^0}^2 )
+ 16 \pi F_\pi^2 \varphi_S^{\mbox{\tiny{$0 K^-$;$0 K^-$ }}} (0)
\right] \left[ k_{K\pi^0} (\mu) - 8 L_4^r (\mu) \right]
\nonumber\\
&&
+ {\sqrt{2}} \frac{F_0^2}{F_\pi^2} \left\{
\left[
\beta^{\mbox{\tiny{$0 -$;$K^0 K^-$ }}} + 6 \left(1 - \frac{\sqrt{3}}{2} \epsilon_2 \right) \gamma^{\mbox{\tiny{$0 -$;$K^0 K^-$ }}}
\right]  (M_K^2 + 2 M_{\pi^0}^2 )
\right.
\nonumber\\
&&
\left.
- 2 \left[
\beta^{\mbox{\tiny{$0 -$;$K^0 K^-$ }}} + 3 \sqrt{3} \epsilon_2 \gamma^{\mbox{\tiny{$0 -$;$K^0 K^-$ }}}
\right] (M_{K^0}^2 - M_\pi^2)
\right.
\nonumber\\
&&
\left.
- \frac{1}{6} (1 + \sqrt{3} \epsilon_2 ) \left[
\beta^{\mbox{\tiny{$0 -$;$K^0 K^-$ }}} - 3  \gamma^{\mbox{\tiny{$0 -$;$K^0 K^-$ }}}
\right] (M_K^2 - M_{\pi^0}^2 )
\right.
\nonumber\\
&&
\left.
- \frac{3}{2} \left( 1 - \frac{\epsilon_2}{\sqrt{3}} \right)  \left[
\beta^{\mbox{\tiny{$0 -$;$K^0 K^-$ }}} + 3   \gamma^{\mbox{\tiny{$0 -$;$K^0 K^-$ }}}
\right] (M_K^2 +M_{K^0}^2 + M_\pi^2 +  M_{\pi^0}^2 )
\right.
\nonumber\\
&&
\left.
- 3 \left( 1 - \frac{\epsilon_2}{\sqrt{3}} \right) 16 \pi F_\pi^2 \varphi_S^{\mbox{\tiny{$0 -$;$K^0 K^-$ }}} (0)
\right\} \left[ k_{K^0 \pi} (\mu) - 8 L_4^r (\mu) \right]
\nonumber\\
&&
+ \sqrt{3} \left( 1 - \frac{\epsilon_1}{\sqrt{3}} + \sqrt{3} \epsilon_2 \right) \frac{F_0^2}{F_\pi^2}
\left[
\frac{1}{2} ( \beta^{\mbox{\tiny{$0 K^-$;$\eta K^-$ }}} + 3 \gamma^{\mbox{\tiny{$0 K^-$;$\eta K^-$ }}} ) (2 M_\eta^2 - 3 M_K^2 - 2 M_{\pi^0}^2 )
\right.
\nonumber\\
&&
\left.
+ \gamma^{\mbox{\tiny{$0 K^-$;$\eta K^-$ }}} (M_K^2 - M_{\pi^0}^2 )
- 16 \pi \varphi_S^{\mbox{\tiny{$0 K^-$;$\eta K^-$ }}} (0)
\right] \left[ k_{\eta K} (\mu) - 8 L_4^r (\mu) \right]
\nonumber\\
&&
+ \frac{F_0^2}{F_\pi^2} \left\{
- (1 + 2 \sqrt{3} \epsilon_2 ) 16 \pi F_\pi^2 \varphi_S^{\mbox{\tiny{$00$;$00$}}} (0) \left[ k_{\pi^0 \pi^0} (\mu) - 8 L_4^r (\mu) \right]
+ 32 \pi F_\pi^2 \varphi_S^{\mbox{\tiny{$00$;$+-$}}} (0) \left[ k_{\pi \pi} (\mu) - 8 L_4^r (\mu) \right]
\right.
\nonumber\\
&&
\left.
+ 64 \pi F_\pi^2 \varphi_S^{\mbox{\tiny{$00$;$K^+ K^-$}}} (0) \left[ k_{KK} (\mu) - 8 L_4^r (\mu) \right]
- 32 \pi F_\pi^2 \varphi_S^{\mbox{\tiny{$00$;$K^0 {\bar K}^0$}}} (0) \left[ k_{K^0 {\bar K}^0} (\mu) - 8 L_4^r (\mu) \right]
\right.
\nonumber\\
&&
\left.
- 3 \left( 1 - \frac{2}{\sqrt{3}} \epsilon_1 \right) 16 \pi F_\pi^2 \varphi_S^{\mbox{\tiny{$00$;$\eta\eta$}}} (0) \left[ k_{\eta \eta} (\mu) - 8 L_4^r (\mu) \right]
\right.
\nonumber\\
&&
\left.
- 2\sqrt{3} \left( 1 - \frac{\epsilon_1}{\sqrt{3}} + \sqrt{3} \epsilon_2 \right) 
16 \pi F_\pi^2 \varphi_S^{\mbox{\tiny{$00$;$0\eta$}}} (0) \left[ k_{\pi^0 \eta} (\mu) - 8 L_4^r (\mu) \right]
\right\}
.
\end{eqnarray}

\section{Indefinite integrals of loop functions}\label{app:indefinite}
\setcounter{equation}{0}

In this appendix, we give integrals of
the loop function ${\bar J}_{ab}(t)$ that are useful for the computation
of the one-loop partial-wave projections in Sec.~\ref{sec:IB_in_phases}.
We need the following (indefinite) integrals
\begin{equation}
\int dt \left\{ t^n {\bar J}_{ab} (t) \, ; {\bar J}_{ab} (t) \, ; K_{ab} (t) \, ; M_{ab} (t) \, ; L_{ab} (t) - t M_{ab} (t) \right\}
\equiv 
\left\{ {\cal J}_{ab}^{(n)} (t) \,; {\cal J}_{ab} (t) \,; {\cal K}_{ab} (t) \,; {\cal M}_{ab} (t) \,; {\cal L}_{ab} (t) \right\}
,
\end{equation}
with $n = 1, 2$. From their definitions, it follows that the combination 
\begin{equation}
 6 {\cal L}_{ab} (t) - \Sigma_{ab} {\cal J}_{ab} (t) + \Delta_{ab} {\cal K}_{ab} (t)
+ \frac{1}{2} {\cal J}_{ab}^{(1)} (t) - 2 \Delta_{ab}^2 {\bar J}_{ab}^\prime (0) t + \frac{t^2}{96 \pi^2}
\end{equation}
of these functions is a constant.
In the kinematical range of interest for us, $t < (M_a + M_b)^2$,
so that the loop function ${\bar J}_{ab}(t)$ is real and can be expressed as
\begin{equation}
{\bar J}_{ab} (t) = \frac{1}{16\pi^2}\,
\bigg\{1 - \frac{\Delta_{ab}}{t} \ln\frac{M_a}{M_b} +
\frac{\Sigma_{ab}}{\Delta_{ab}}\ln\frac{M_a}{M_b}
+
\frac{M_a M_b}{\Sigma_{ab} - M_a M_b \left[ \chi_{ab} (t) + \chi^{-1}_{ab} (t) \right]} 
\left[ \chi_{ab} (t) - \frac{1}{\chi_{ab} (t)} \right] \ln \chi_{ab} (t)
\bigg\}
,
\end{equation}
with
\begin{equation}
\chi_{ab} (t) =
\left\{
\begin{array}{cl}
\displaystyle{\frac{\sqrt{(M_a + M_b)^2 - t} - \sqrt{(M_a - M_b)^2 - t}}{\sqrt{(M_a + M_b)^2 - t} + \sqrt{(M_a - M_b)^2 - t}}} &
\qquad [t < (M_a - M_b)^2]
\\
&
\\
\displaystyle{\frac{\sqrt{(M_a + M_b)^2 - t} - i \sqrt{t - (M_a - M_b)^2 }}{\sqrt{(M_a + M_b)^2 - t} + i \sqrt{t - (M_a - M_b)^2}}} &
\qquad [(M_a - M_b)^2 < t < (M_a + M_b)^2]
\end{array}
\right.  
,
\end{equation}
so that $0\le\chi_{ab}(t)\le 1$ when $t<(M_a - M_b)^2$ and $\vert \chi_{ab}(t)\vert = 1$ for $(M_a - M_b)^2 < t < (M_a + M_b)^2$.
Notice also the identity
\begin{equation}
t\,=\,M_a^2 + M_b^2 - M_a M_b \left[ \chi_{ab} (t) + \frac{1}{\chi_{ab} (t)}\right]
.
\end{equation}
The possibility to perform the required integrals hinges on finding a function ${\cal F}_{ab} (x)$ that satisfies
\begin{equation}
 \frac{d}{d x} {\cal F}_{ab} (x) = \frac{1}{x} \ln \frac{M_a}{M_b} \cdot
\frac{M_a M_b \left(x - \displaystyle{\frac{1}{x}}\right)}{\Sigma_{ab} - M_a M_b\left(x + \displaystyle{\frac{1}{x}}\right)}
- \frac{\ln x}{x} \frac{\Delta_{ab}}{\Sigma_{ab} - M_a M_b\left(x + \displaystyle{\frac{1}{x}}\right)}
.
\end{equation}
Such a function can indeed be found and reads
\begin{equation}
 {\cal F}_{ab} (x) = H_{1,0} \left( \frac{M_a}{M_b} x \right) - H_{1,0} \left( \frac{M_b}{M_a} x \right) +
\ln \frac{M_a}{M_b} \ln x
,
\end{equation}
where
\begin{equation}
H_{1,0} (x) = - {\rm Li}_2 (x) - \ln x \ln (1-x)
\label{H_10_<1}
\end{equation}
belongs to the family of functions known as harmonic polylogarithms \cite{polylogs}.
The function $H_{1,0} (x)$ is defined in the complex plane, with a cut along
the negative real axis. Notice that, despite what the expression (\ref{H_10_<1}) might suggest,
there is no problem along the positive real axis for $x > 1$: the discontinuities of
the functions ${\rm Li}_2 (x)$ and $\ln x \ln (1-x)$ along this line compensate 
each other exactly.
This can also be inferred from the equivalent expression 
\begin{equation}
H_{1,0} (x) = {\rm Li}_2 \left( \frac{1}{x} \right) - \ln x \ln (x-1) + \frac{1}{2} \ln^2 x - \frac{\pi^2}{3}
,
\label{H_10_>1}
\end{equation}
which follows from the properties of the dilogarithm function.

With these definitions, we obtain (see also App.~A of \cite{DescotesGenon:2012gv}):
\begin{eqnarray}
 16 \pi^2 {\cal J}_{ab}^{(2)} (t) &=& \left( \frac{4}{3} + \frac{\Sigma_{ab}}{\Delta_{ab}} \ln \frac{M_a}{M_b} \right) \frac{t^3}{3}
- \left( \Delta_{ab}  \ln \frac{M_a}{M_b} + \frac{\Sigma_{ab}}{6} \right) \frac{t^2}{2}
- \left( M_a^4 + M_b^4 + 10 M_a^2 M_b^2 \right)  \frac{t}{6}
\nonumber\\
&&
+ \frac{M_a M_b}{6} \left[ 2 t^2 - \Sigma_{ab}  t - \left( M_a^4 + M_b^4 + 10 M_a^2 M_b^2 \right) \right] 
\left( \chi_{ab} (t) - \frac{1}{\chi_{ab} (t)} \right) \ln \chi_{ab} (t) 
\nonumber\\
&&
+ M_a^2 M_b^2 \Sigma_{ab} \ln^2 \chi_{ab} (t)
,
\\
\nonumber\\
 16 \pi^2 {\cal J}_{ab}^{(1)} (t) &=& \left( \frac{3}{2} + \frac{\Sigma_{ab}}{\Delta_{ab}} \ln \frac{M_a}{M_b} \right) \frac{t^2}{2} -
\left( \frac{\Sigma_{ab}}{2} + \Delta_{ab}  \ln \frac{M_a}{M_b} \right) t +
\frac{M_a M_b}{2} \left( t - \Sigma_{ab} \right) \left( \chi_{ab} (t) - \frac{1}{\chi_{ab} (t)} \right)  \ln \chi_{ab} (t) 
\nonumber\\
&&
+ M_a^2 M_b^2 \ln^2 \chi_{ab} (t)
,
\\
\nonumber\\
 16 \pi^2 {\cal J}_{ab} (t) &=&
\left(
2 + \frac{\Sigma_{ab}}{\Delta_{ab}} \ln \frac{M_a}{M_b}
\right) t +
M_a M_b \left( \chi_{ab} (t) - \frac{1}{\chi_{ab} (t)} \right)  \ln \chi_{ab} (t) + \frac{\Sigma_{ab}}{2} \ln^2 \chi_{ab} (t)
\nonumber\\
&&
+ \Delta_{ab} \left[
H_{1,0} \left( \frac{M_a}{M_b} \chi_{ab} (t) \right) - H_{1,0} \left( \frac{M_b}{M_a} \chi_{ab} (t) \right) +
\ln \frac{M_a}{M_b} \ln \chi_{ab} (t)
\right]
,
\\
\nonumber\\
 16 \pi^2 {\cal K}_{ab} (t) &=& \frac{\Delta_{ab}}{2} \left\{
\frac{\Delta_{ab}}{t} \ln \frac{M_a}{M_b}
- M_a M_b \left( \chi_{ab} (t) - \frac{1}{\chi_{ab} (t)} \right)  \frac{\ln \chi_{ab} (t)}{t} - \frac{1}{2} \ln^2 \chi_{ab} (t) 
\right\}
\nonumber\\
&&
- \frac{\Sigma_{ab}}{2} \left[
H_{1,0} \left( \frac{M_a}{M_b} \chi_{ab} (t) \right) - H_{1,0} \left( \frac{M_b}{M_a} \chi_{ab} (t) \right) +
\ln \frac{M_a}{M_b} \ln \chi_{ab} (t)
\right]
,
\\
\nonumber\\
 16 \pi^2 {\cal M}_{ab} (t) &=&
\left( \frac{2}{9} + \frac{1}{12} \frac{\Sigma_{ab}}{\Delta_{ab}} \ln \frac{M_a}{M_b} \right) t +
\frac{M_a M_b}{12} \left( 1 + 4 \frac{\Sigma_{ab}}{t} - 2 \frac{\Delta_{ab}^2}{t^2} \right) \left( \chi_{ab} (t) - \frac{1}{\chi_{ab} (t)} \right) \ln \chi_{ab} (t)
\nonumber\\
&&  
+ \frac{\Sigma_{ab}}{8} \ln^2 \chi_{ab} (t)
- \frac{\Delta_{ab}^2}{6 t} \left( 1 + 3 \frac{\Sigma_{ab}}{\Delta_{ab}} \ln \frac{M_a}{M_b}  \right) +
\frac{\Delta_{ab}^3}{6t^2} \ln \frac{M_a}{M_b}
\nonumber\\
&&
+ \frac{\Delta_{ab}}{4} \left[
H_{1,0} \left( \frac{M_a}{M_b} \chi_{ab} (t) \right) - H_{1,0} \left( \frac{M_b}{M_a} \chi_{ab} (t) \right) +
\ln \frac{M_a}{M_b} \ln \chi_{ab} (t)
\right]
.
\end{eqnarray}

As far as the ranges of the integrations in the expressions (\ref{f_0_+-}) and (\ref{f_0_00})
are concerned, we may note that 
$0 < (M_{K^0} - M_{\eta})^2 < {t_-^c} (s,s_\ell) \le {t_+^c} (s,s_\ell) \le (M_K - M_\pi)^2 < (M_{K^0} - M_{\pi^0})^2$, whereas
$0 < (M_K - M_\eta)^2 < {t_-^n} (s,s_\ell) \le {t_+^n} (s,s_\ell) \le (M_{K^0} - M_\pi)^2 < (M_K - M_{\pi^0})^2$.

\section{Numerical representation}\label{app:approximate}
\setcounter{equation}{0}

The previous appendices as well as the main part of this article provide all the elements needed to compute
the isospin-breaking corrections to the difference $\delta_0^0-\delta_1^1$ as measured in $K^+\to \pi^+\pi^- \ell^+\nu_\ell$. It is also useful to provide a numerical approximation of the (lengthy) expression for the inputs described in Sec.~\ref{sec:IBtwoloops}. 
$\Delta_{\mbox{\scriptsize{IB}}}(s,s_\ell)$ is approximated at the level of 0.03 mrad for $a_0^0$ between 0.18 and 0.30,
$a_0^2$ between -0.06 and -0.03 and $s_\ell$ between 0 and $(M_K-\sqrt{s})^2$ by the following expression
\begin{equation}
\Delta_{\mbox{\scriptsize{IB}}}(s,s_\ell)=\sum_{i,j,k,l} c_{ijkl} \left(\frac{a_0^0}{0.22}\right)^i  \left(\frac{a_0^2}{-0.045}\right)^j
   \left(\sqrt{\frac{s}{4 M_\pi^2}-1}\right)^k \left(\frac{s_\ell}{4 M_\pi^2}\right)^l\,,
\end{equation}
with the values of the coefficients $c_{ijkl}$ given in Tab.~\ref{Tab:coeffnumdelta}.

\begin{table}[t]
\begin{tabular}{ccc}
\begin{tabular}{c|c|c|c|c}
$i$ & $j$ & $k$ & $l$ & $c_{ijkl}$\\
\hline
0 & 0 & 0 & 0 & -0.0928488\\
0 & 0 & 0 & 1 & -0.000489348\\
0 & 0 & 1 & 0 & 1.53009\\
0 & 0 & 1 & 1 & 0.0170472\\
0 & 0 & 2 & 0 & -0.714631\\
0 & 0 & 2 & 1 & -0.104036\\
0 & 0 & 3 & 0 & 2.01142\\
0 & 0 & 3 & 1 & 0.265872\\
0 & 0 & 4 & 0 & -1.21606\\
0 & 0 & 4 & 1 & -0.293487\\
0 & 0 & 5 & 0 & 0.306475\\
0 & 0 & 5 & 1 & 0.115913\\
0 & 1 & 0 & 0 & 4.5098\\
0 & 1 & 0 & 1 & 0.0524669\\
0 & 1 & 1 & 0 & -4.56129\\
0 & 1 & 1 & 1 & -0.691058\\
0 & 1 & 2 & 0 & 42.4981\\
0 & 1 & 2 & 1 & 4.31352\\
0 & 1 & 3 & 0 & -63.0096\\
0 & 1 & 3 & 1 & -10.4222\\
0 & 1 & 4 & 0 & 43.5422\\
0 & 1 & 4 & 1 & 11.1704\\
0 & 1 & 5 & 0 & -11.6411\\
0 & 1 & 5 & 1 & -4.33411\\
0 & 2 & 0 & 0 & 0.0163751\\
0 & 2 & 0 & 1 & 0.000172447\\
0 & 2 & 1 & 0 & 0.653006\\
0 & 2 & 1 & 1 & 0.0023838\\
0 & 2 & 2 & 0 & -0.0434267\\
0 & 2 & 2 & 1 & -0.0110707\\
0 & 2 & 3 & 0 & 1.77494\\
0 & 2 & 3 & 1 & 0.0349048\\
0 & 2 & 4 & 0 & -2.32223\\
0 & 2 & 4 & 1 & -0.0382942\\
0 & 2 & 5 & 0 & 0.557611\\
0 & 2 & 5 & 1 & 0.0144076
\end{tabular}
&
\begin{tabular}{c|c|c|c|c}
$i$ & $j$ & $k$ & $l$ & $c_{ijkl}$\\
\hline
1 & 0 & 0 & 0 & 19.2585\\
1 & 0 & 0 & 1 & 0.210249\\
1 & 0 & 1 & 0 & -77.115\\
1 & 0 & 1 & 1 & -2.71633\\
1 & 0 & 2 & 0 & 173.301\\
1 & 0 & 2 & 1 & 16.9493\\
1 & 0 & 3 & 0 & -223.67\\
1 & 0 & 3 & 1 & -41.3346\\
1 & 0 & 4 & 0 & 150.903\\
1 & 0 & 4 & 1 & 44.2302\\
1 & 0 & 5 & 0 & -41.0664\\
1 & 0 & 5 & 1 & -17.1066\\
1 & 1 & 0 & 0 & -0.0584899\\
1 & 1 & 0 & 1 & 0.00208991\\
1 & 1 & 1 & 0 & 1.83986\\
1 & 1 & 1 & 1 & -0.0214383\\
1 & 1 & 2 & 0 & -0.0524592\\
1 & 1 & 2 & 1 & 0.137195\\
1 & 1 & 3 & 0 & -5.35591\\
1 & 1 & 3 & 1 & -0.331961\\
1 & 1 & 4 & 0 & 3.03996\\
1 & 1 & 4 & 1 & 0.35232\\
1 & 1 & 5 & 0 & -1.03269\\
1 & 1 & 5 & 1 & -0.135533\\
1 & 2 & 0 & 0 & -0.041962\\
1 & 2 & 0 & 1 & -0.000228634\\
1 & 2 & 1 & 0 & 0.000847618\\
1 & 2 & 1 & 1 & 0.00462323\\
1 & 2 & 2 & 0 & -0.382409\\
1 & 2 & 2 & 1 & -0.0277188\\
1 & 2 & 3 & 0 & 0.471365\\
1 & 2 & 3 & 1 & 0.0687998\\
1 & 2 & 4 & 0 & -0.361924\\
1 & 2 & 4 & 1 & -0.0743238\\
1 & 2 & 5 & 0 & -0.0169208\\
1 & 2 & 5 & 1 & 0.0288682
\end{tabular}
&
\begin{tabular}{c|c|c|c|c}
$i$ & $j$ & $k$ & $l$ & $c_{ijkl}$\\
\hline
2 & 0 & 0 & 0 & -0.973318\\
2 & 0 & 0 & 1 & -0.00886144\\
2 & 0 & 1 & 0 & -0.142484\\
2 & 0 & 1 & 1 & 0.18153\\
2 & 0 & 2 & 0 & -9.73948\\
2 & 0 & 2 & 1 & -1.09236\\
2 & 0 & 3 & 0 & 17.7732\\
2 & 0 & 3 & 1 & 2.66039\\
2 & 0 & 4 & 0 & -12.8517\\
2 & 0 & 4 & 1 & -2.88117\\
2 & 0 & 5 & 0 & 3.21743\\
2 & 0 & 5 & 1 & 1.12552\\
2 & 1 & 0 & 0 & -0.149598\\
2 & 1 & 0 & 1 & -0.000761696\\
2 & 1 & 1 & 0 & -0.0112002\\
2 & 1 & 1 & 1 & 0.0154114\\
2 & 1 & 2 & 0 & -1.23163\\
2 & 1 & 2 & 1 & -0.092462\\
2 & 1 & 3 & 0 & 0.849874\\
2 & 1 & 3 & 1 & 0.229646\\
2 & 1 & 4 & 0 & -0.029795\\
2 & 1 & 4 & 1 & -0.248227\\
2 & 1 & 5 & 0 & -0.129968\\
2 & 1 & 5 & 1 & 0.0964606\\
2 & 2 & 0 & 0 & -0.00190432\\
2 & 2 & 0 & 1 & $-8.87957\cdot 10^{-6}$\\
2 & 2 & 1 & 0 & -0.000665274\\
2 & 2 & 1 & 1 & 0.000179941\\
2 & 2 & 2 & 0 & -0.0128156\\
2 & 2 & 2 & 1 & -0.00108143\\
2 & 2 & 3 & 0 & -0.0208564\\
2 & 2 & 3 & 1 & 0.00269032\\
2 & 2 & 4 & 0 & 0.0307737\\
2 & 2 & 4 & 1 & -0.00291208\\
2 & 2 & 5 & 0 & -0.00645683\\
2 & 2 & 5 & 1 & 0.00113295
\end{tabular}
\end{tabular}
\caption{Coefficients for the numerical approximation of the  central value for the isospin-breaking correction $\Delta_{\mbox{\scriptsize{IB}}}$.}
\label{Tab:coeffnumdelta}
\end{table}

The uncertainty $\delta\Delta_{\mbox{\scriptsize{IB}}}(s,s_\ell)$ induced by the variation of the inputs 
is approximated at the level of 0.01 mrad for the same range of $a_0^0$, $a_0^2$ and $s_\ell$ by the following expression
\begin{equation}
\delta\Delta_{\mbox{\scriptsize{IB}}}(s,s_\ell)=\sum_{i,j,k,l} d_{ijkl} \left(\frac{a_0^0}{0.22}\right)^i  \left(\frac{a_0^2}{-0.045}\right)^j
   \left(\sqrt{\frac{s}{4 M_\pi^2}-1}\right)^k \left(\frac{s_\ell}{4 M_\pi^2}\right)^l\,,
\end{equation}
with the values of the coefficients $d_{ijkl}$ given in Tab.~\ref{Tab:coeffnumdeltaerr}. 
We recall that $s_\ell$ is constrained to lie within 0 and $(M_K-\sqrt{s})^2\leq (M_K-2M_\pi)^2$, so 
that $0 \leq s_\ell/(4M_\pi^2)\leq 0.6$. The size of the coefficients $c_{ijkl}$ and $d_{ijkl}$ indicate that the central value $\Delta_{\mbox{\scriptsize{IB}}}$ is insensitive to the value of $s_\ell$ in the range considered here, whereas $\delta\Delta_{\mbox{\scriptsize{IB}}}$ exhibits a very mild dependence.

\begin{table}[t]
\begin{tabular}{ccc}
\begin{tabular}{c|c|c|c|c}
$i$ & $j$ & $k$ & $l$ & $d_{ijkl}$\\
\hline
0 & 0 & 0 & 0 & 0.0360932\\
0 & 0 & 0 & 1 & 0.0132719\\
0 & 0 & 1 & 0 & -0.157746\\
0 & 0 & 1 & 1 & -0.0995196\\
0 & 0 & 2 & 0 & 0.235716\\
0 & 0 & 2 & 1 & 0.455559\\
0 & 0 & 3 & 0 & -0.0501353\\
0 & 0 & 3 & 1 & -0.645175\\
0 & 0 & 4 & 0 & 0.0663443\\
0 & 0 & 4 & 1 & 0.331087\\
0 & 1 & 0 & 0 & 0.0261216\\
0 & 1 & 0 & 1 & -0.0280331\\
0 & 1 & 1 & 0 & 0.534865\\
0 & 1 & 1 & 1 & 0.354181\\
0 & 1 & 2 & 0 & 2.03114\\
0 & 1 & 2 & 1 & -0.708541\\
0 & 1 & 3 & 0 & -3.85332\\
0 & 1 & 3 & 1 & 0.692502\\
0 & 1 & 4 & 0 & 1.80675\\
0 & 1 & 4 & 1 & -0.347696\\
0 & 2 & 0 & 0 & -0.000588477\\
0 & 2 & 0 & 1 & -0.00036999\\
0 & 2 & 1 & 0 & 0.0545846\\
0 & 2 & 1 & 1 & -0.0332993\\
0 & 2 & 2 & 0 & -0.310945\\
0 & 2 & 2 & 1 & -0.0629525\\
0 & 2 & 3 & 0 & 0.464568\\
0 & 2 & 3 & 1 & 0.141621\\
0 & 2 & 4 & 0 & -0.155224\\
0 & 2 & 4 & 1 & -0.068107
\end{tabular}
&
\begin{tabular}{c|c|c|c|c}
$i$ & $j$ & $k$ & $l$ & $d_{ijkl}$\\
\hline
1 & 0 & 0 & 0 & 0.0888772\\
1 & 0 & 0 & 1 & -0.084117\\
1 & 0 & 1 & 0 & 0.139356\\
1 & 0 & 1 & 1 & 0.136112\\
1 & 0 & 2 & 0 & 1.18191\\
1 & 0 & 2 & 1 & -1.19902\\
1 & 0 & 3 & 0 & -2.60141\\
1 & 0 & 3 & 1 & 1.40357\\
1 & 0 & 4 & 0 & 1.78954\\
1 & 0 & 4 & 1 & -0.526573\\
1 & 1 & 0 & 0 & 0.0230242\\
1 & 1 & 0 & 1 & 0.0151728\\
1 & 1 & 1 & 0 & -0.574102\\
1 & 1 & 1 & 1 & -0.532437\\
1 & 1 & 2 & 0 & -1.22998\\
1 & 1 & 2 & 1 & 1.52903\\
1 & 1 & 3 & 0 & 2.8769\\
1 & 1 & 3 & 1 & -2.264\\
1 & 1 & 4 & 0 & -1.18219\\
1 & 1 & 4 & 1 & 1.15153\\
1 & 2 & 0 & 0 & 0.00594686\\
1 & 2 & 0 & 1 & 0.0041983\\
1 & 2 & 1 & 0 & -0.00200251\\
1 & 2 & 1 & 1 & -0.0176096\\
1 & 2 & 2 & 0 & 0.430117\\
1 & 2 & 2 & 1 & 0.318301\\
1 & 2 & 3 & 0 & -0.617296\\
1 & 2 & 3 & 1 & -0.550716\\
1 & 2 & 4 & 0 & 0.310329\\
1 & 2 & 4 & 1 & 0.276654
\end{tabular}
&
\begin{tabular}{c|c|c|c|c}
$i$ & $j$ & $k$ & $l$ & $d_{ijkl}$\\
\hline
2 & 0 & 0 & 0 & 0.0800327\\
2 & 0 & 0 & 1 & 0.0170832\\
2 & 0 & 1 & 0 & 0.0248212\\
2 & 0 & 1 & 1 & -0.0947344\\
2 & 0 & 2 & 0 & 0.322594\\
2 & 0 & 2 & 1 & 0.531996\\
2 & 0 & 3 & 0 & 0.260986\\
2 & 0 & 3 & 1 & -0.699728\\
2 & 0 & 4 & 0 & -0.261437\\
2 & 0 & 4 & 1 & 0.282638\\
2 & 1 & 0 & 0 & 0.0172326\\
2 & 1 & 0 & 1 & -0.00298951\\
2 & 1 & 1 & 0 & 0.198811\\
2 & 1 & 1 & 1 & 0.181986\\
2 & 1 & 2 & 0 & 0.407994\\
2 & 1 & 2 & 1 & -0.632245\\
2 & 1 & 3 & 0 & -0.684285\\
2 & 1 & 3 & 1 & 1.00794\\
2 & 1 & 4 & 0 & 0.28593\\
2 & 1 & 4 & 1 & -0.515946\\
2 & 2 & 0 & 0 & 0.00167104\\
2 & 2 & 0 & 1 & -0.00163906\\
2 & 2 & 1 & 0 & -0.0199531\\
2 & 2 & 1 & 1 & 0.00855621\\
2 & 2 & 2 & 0 & -0.0973739\\
2 & 2 & 2 & 1 & -0.113327\\
2 & 2 & 3 & 0 & 0.182151\\
2 & 2 & 3 & 1 & 0.194166\\
2 & 2 & 4 & 0 & -0.0931676\\
2 & 2 & 4 & 1 & -0.0986163\\
\end{tabular}
\end{tabular}
\caption{Coefficients for the numerical approximation of the  uncertainty for the isospin-breaking correction  $\delta\Delta_{\mbox{\scriptsize{IB}}}$.}
\label{Tab:coeffnumdeltaerr}
\end{table}

\end{document}